\documentclass[12pt,preprint,epsfig]{aastex}
\usepackage{epsfig}
\usepackage{color}
\begin{document}
\topmargin=0in
\shorttitle{Cosmogenic Lyman alpha}
\shortauthors{Kollmeier et al.}
\newcommand{\msun}{M_{\odot}}
\newcommand{\kms}{\, {\rm km\, s}^{-1}}
\newcommand{\cm}{\, {\rm cm}}
\newcommand{\gm}{\, {\rm g}}
\newcommand{\erg}{\, {\rm erg}}
\newcommand{\kel}{\, {\rm K}}
\newcommand{\kpc}{\, {\rm kpc}}
\newcommand{\mpc}{\, {\rm Mpc}}
\newcommand{\seg}{\, {\rm s}}
\newcommand{\kev}{\, {\rm keV}}
\newcommand{\hz}{\, {\rm Hz}}
\newcommand{\etal}{et al.\ }
\newcommand{\yr}{\, {\rm yr}}
\newcommand{\mpyr}{{\rm mas}\, {\rm yr}^{-1}}
\newcommand{\lya}{Ly$\alpha$}
\newcommand{\lyb}{Ly$\beta$}
\newcommand{\hmpc}{$h^{-1}$ {\rm Mpc}}
\newcommand{\hkpc}{$h^{-1}$ {\rm kpc}}
\newcommand{\gyr}{\, {\rm Gyr}}
\newcommand{\asec}{^{\prime \prime}}
\newcommand{\fluxunits}{\erg\seg^{-1}\cm^{-2}}
\newcommand{\eq}{eq.\ }
\def\arcsec{''\hskip-3pt .}
\definecolor{purple}{rgb}{0.31,0,0.79}
\definecolor{green}{rgb}{0,0.7,0.3}
\definecolor{wine}{rgb}{0.7,0.1,0.2}
\definecolor{red}{rgb}{1,0.0,0.0}
\definecolor{blue}{rgb}{0,0,1}
\def\jak#1{{\bf[{\textcolor{purple}{JK: #1}}]}} 
\def\zz#1{{\bf [{\textcolor{green}{ZZ: #1}}]}}
\def\dw#1{{\bf [{\textcolor{red}{DW: #1}}]}}
\def\jm#1{{\bf [{\textcolor{blue}{JM: #1}}]}}
\def\ag#1{{\bf [{\textcolor{wine}{AG: #1}}]}}

\title{Lyman-$\alpha$ Emission From Cosmic Structure I: Fluorescence}
\author{Juna A. Kollmeier \altaffilmark{1}, Zheng Zheng \altaffilmark{2,3}, Romeel Dav{\'e} \altaffilmark{4}, Andrew Gould \altaffilmark{5}, Neal Katz \altaffilmark{6}, \\ Jordi Miralda-Escud{\'e} \altaffilmark{7,8}, \& David H. Weinberg \altaffilmark{5} }

\altaffiltext{1}{Observatories of the Carnegie Institution of Washington, 813 Santa Barbara Street, Pasadena, CA 91101}
\altaffiltext{2}{Institute for Advanced Study, Einstein Drive, Princeton, NJ 08540}
\altaffiltext{3}{John Bachall Fellow}
\altaffiltext{4}{Steward Observatory/TAP, Univ. of Arizona, 933 N. Cherry Ave.
Tucson, AZ 85721}
\altaffiltext{5}{Dept. of Astronomy, The Ohio State University,
140 W. 18th Ave, Columbus, OH 43210}
\altaffiltext{6}{Dept. of Astronomy, University of Massachusetts, Amherst, MA 01003}
\altaffiltext{7}{Instituci{\'o} Catalana de Recerca i Estudis Avan{\c{c}}ats, Barcelona, Spain}
\altaffiltext{8}{Institut de Ci{\`e}ncies del Cosmos, Universitat de Barcelona}

\begin{abstract}
  We present predictions for the fluorescent \lya\ emission signature
  arising from photoionized, optically thick structures in Smoothed
  Particle Hydrodynamic (SPH) cosmological simulations of a
  $\Lambda$CDM universe using a Monte Carlo \lya\ radiative transfer
  code. We calculate the expected \lya\ image and 2-dimensional
  spectra for gas exposed to a uniform ultraviolet ionizing background
  as well as gas exposed additionally to the photoionizing radiation
  from a local quasar, after correcting for the self-shielding of
  hydrogen.  As a test of our numerical methods and for application to
  current observations, we examine simplified analytic structures that
  are uniformly or anisotropically illuminated.  We compare these
  results with recent observations.  We discuss future observing
  campaigns on large telescopes and realistic strategies for detecting
  fluorescence owing to the ambient metagalactic ionization and in
  regions close to bright quasars.  While it will take hundreds of
  hours on the current generation of telescopes to detect fluorescence
  caused by the Ultraviolet Background (UVB) alone, our calculations
  suggest that of order ten sources of quasar-induced fluorescent
  \lya\ emission should be detectable after a 10 hour exposure in a 10
  arcmin$^2$ field around a bright quasar.  These observations will
  help probe the physical conditions in the densest regions of the
  intergalactic medium as well as the temporal light curves and
  isotropy of quasar radiation.

\end{abstract}

\section{Introduction \label{sec:intro}}

A cornerstone of the current picture of galaxy formation and evolution
is the existence of filaments of non-uniform gas that form the
backbone of cosmic structure.  The presence of this material has been
inferred via neutral hydrogen absorption line studies of background
quasars for nearly a half century \citep[e.g.,][]{bahcall65,
bahcall66, bps66, lynds71}, and comparison of these types of
observations with hydrodynamic cosmological simulations led to a major
breakthrough in understanding how this observable material relates to
the underlying dark matter distribution \citep[Cen et al. 1994, Bi \&
Davidsen 1997]{zhang95,miralda96,hernquist96,hg97, croft98}.  But
these observations only provide information along one-dimensional (1D)
lines-of-sight through the matter distribution, which is compared to
similar 1D cuts through theoretical models.  Owing to the rarity of
close quasar pairs, information transverse to the line of sight is
difficult to obtain in absorption.  As a result, the 3D geometry,
contents, and specific relation of intergalactic gas to galaxies remain
among the most important outstanding questions in galaxy formation.
 
It has been long recognized that exploiting the emission in the strong
$1s$ --- $2p$ (\lya) transition of hydrogen could prove helpful for
eventually observing the 3D intergalactic medium (IGM) {\it directly}
\citep{hogan87,gw96}, allowing us to test models of the structure of
the IGM and determine the role of the IGM in the process of galaxy
formation as well as the effects of galaxy formation on the IGM. Apart
from the potential \lya\ emission from intergalactic stars, there are
two mechanisms for generating \lya\ emission that dominate in the
overdense regions of the intergalactic medium probed by current
technologies: recombination radiation following photoionization
(fluorescence) and cooling radiation.  A third mechanism, scattering
of photons emitted by any source and redshifted into \lya\ is
important at low column density as discussed below.  Photoionization
of intergalactic neutral hydrogen followed by recombination yields
fluorescent emission of \lya\ photons from the recombining gas at an
efficiency of approximately 0.66 \lya\ photons for each ionizing
photon \citep{osterbrock, spitzer, gw96}.  The ionizing radiation that
keeps the \lya\ forest highly ionized (the metagalactic ultra-violet
background (UVB)) probably originates from galaxies and quasars.  Some
of these sources are very luminous, implying large-scale fluctuations
in the radiation intensity.  Another important source of intergalactic
\lya\ emission is cooling radiation.  As gas settles into galactic
potential wells, it radiates its gravitational potential energy, and a
significant fraction of this energy emerges in the \lya\ line because
much of the cooling gas has temperature $T \sim 10^4-10^5$ K even when
the halo virial temperature is higher
\citep{binney77,katz92,fardal01,haiman00}.  Fluorescence stimulated by
the UVB, fluorescence stimulated by local sources (e.g., nearby
quasars), and cooling radiation all have potential to reveal the
structure of the IGM and the mechanisms of gas accretion by forming
galaxies.

The original predictions for the fluorescent \lya\ emission signal
from the uniform UVB concluded that it is faint, requiring many hours
of integration on 10m class telescopes even with optimistic
assumptions for the UV background radiation field
\citep[e.g.,][]{gw96}.  These studies indicated that if one is
searching for the fluorescence signature from the UVB alone, only
optically thick systems, corresponding to dense patches of the IGM or
the outer regions of galaxies, could be realistically probed with
current technology.  Here ``optically thick'' refers to the Lyman
continuum, implying neutral hydrogen column densities in excess of
$2\times 10^{17} {\rm cm}^{-2}$.  An optically thick cloud, in the
absence of all ionizing sources except the photoionization from a
uniform UVB, should glow with a maximum surface brightness of roughly
50\% of the intensity of the ionizing background (or approximately
$1.4 \times 10^{-19} [(1+z)/3]^{-4} {\rm erg\ s^{-1}\ cm^{-2}\
  arcsec^{-2}})$.  Precise measurements of this signal can probe the
actual {\it value} of the UVB itself, and the large ${\it O}$(100
hour) programs that are necessary to reach this low signal level are
currently underway but have yet to reach these faint levels (e.g., Rauch et al. 2008).

A substantially enhanced \lya\ surface brightness may be produced,
however, in optically thick systems in the vicinity of luminous sources
that increase their rate of recombinations and \lya\ emission when they
are exposed to the radiation of a local source. There is good reason
to search for the glow of Lyman limit systems near the most luminous
quasars, which could be detected in much less observing time than the
glow owing to the metagalactic background. These observations may reveal
the masses, sizes, and kinematics of the absorption systems.

Recent observations have discovered significant numbers of extended
``\lya\ blobs'' whose sizes and surface brightnesses demand
explanation \citep[e.g.,][]{steidel00, matsuda06, dey05} and seem too
large to be consistent with \lya\ emission from star formation alone.  To
interpret the results from increasingly larger samples of observed
extended \lya\ emission, accurate predictions for the \lya\ emission
signal from the processes described above are necessary to
understand the physical origin of the luminosity of these
systems. Furthermore, as increasingly ambitious surveys are planned to
look for this faint emission, it is important to have accurate
theoretical expectations from which these surveys can optimize their
observing strategy and telescope resources.

Because \lya\ is a resonant line, it is non-trivial to estimate the
detailed \lya\ emission from cosmological simulations. Fluorescent
\lya\ photons are typically generated at an ionizing optical depth,
$\tau_{\rm ion} \sim 1$.  This optical depth corresponds to $\tau_{\rm
  Ly\alpha} \sim 10^4$ at the \lya\ line center, for a typical
temperature of $\sim 10^{4}$K.  Therefore, the photon will be absorbed
and re-emitted a number $\tau_{ \rm Ly\alpha} \sim 10^4$ times,
undergoing a random walk in frequency until it is scattered into the
line wing by a high-velocity atom, where the optical depth is of order
unity, at which point the photon can emerge from the gas toward the
observer. Simulating the \lya\ emission signature, therefore, requires
computationally expensive radiative transfer calculations that cannot
currently be performed self-consistently at runtime in cosmological
simulations. \citet{ZM02a} (hereafter ZM02) demonstrated that one can
obtain accurate line-transfer results by employing a Monte Carlo
technique.  This technique makes it possible to predict \lya\ emission
from arbitrary gas-density, temperature, and velocity distributions.

In this work, we combine the Monte Carlo line transfer method with
the outputs of large-scale hydrodynamic simulations and examine the
\lya\ emergent from structures that form in an $\Lambda$CDM universe.  In Paper I we focus on fluorescent
\lya\ from the uniform UVB and also from local ionizing sources.  We will
address cooling radiation in Paper II (Kollmeier et. al, in
preparation) and will refer to it here only briefly. These computations have two objectives: 1)
to allow a comparison of simulations with observations, which will
reveal successes and failures of the treatment of gas physics in the
current generation of hydrodynamic cosmological models and 2) to serve
as a guide to future large observational programs by providing
theoretical benchmarks from specific simulations.

Several studies have improved on the predictions of \cite{gw96} and
have analyzed the \lya\ emission signature from cosmological
simulations without including detailed line radiative transfer
\citep{fardal01, furlanetto05}.  More recently, several authors have
used the method of ZM02 to include line transfer for a variety of
applications ranging from fluorescence \citep{cantalupo05} to cooling radiation \citep{dijkstra06a} to \lya\ emitters \citep{dijkstra06b,hansen06,tasitsiomi06}

For purposes of this paper we {\it define} fluorescent \lya\ emission
to be that produced by recombinations that directly follow
photoionizations by the UVB or a quasar source.  Specifically, this
means that the fluorescent emissivity of a gas element is 0.66 times
its photoionization rate \citep{osterbrock, spitzer}.  Although the
observations cannot tag photons separately as fluorescent emission and
cooling radiation, we treat them separately in our studies for two
reasons.  First, they are physically distinct mechanisms, and it is
interesting to investigate them separately and see whether they have
different observational signatures (source sizes, velocity widths,
etc.).  Second, the cooling radiation predictions are sensitive to the
gas temperatures, and the simulations do not compute these
self-consistently because they do not include self-shielding during
dynamical evolution.  We will devote considerable attention in the
next paper in this series to correcting the gas temperatures for
self-shielding and to understanding the sensitivity of the cooling
radiation predictions to these corrections, but here we circumvent the
issue by focusing on fluorescent emission alone.  We still have to
worry about the effect of gas temperature on neutral fractions, but
the effect is smaller and we discuss this in Appendix C.  Finally,
we note that photoionization also induces \lya\ emission by {\it heating}
the gas, and that this effect is similar in magnitude to the direct
recombinations.  We also treat the \lya\ emission induced by
photoionization heating as cooling radiation.

Finally, the third process of scattered \lya\ photons from all distant
sources becomes important at low column densities in the \lya\
forest. Continuum photons emitted between \lya\ and Ly$\beta$ can be
scattered when they are redshifted to the \lya\ resonance line. The
brightness of this scattered \lya\ emission relative to the
fluorescent \lya\ emission discussed here can be easily computed in the limit
when both the \lya\ optical depth, $\tau_{\alpha}$, and the optical depth
at the Lyman limit, $\tau_{LL}$, are small. The scattered intensity
$I_s$, compared to the fluorescent brightness, $I_f$, is given by
\begin{equation}
 I_s \sim I_f  \frac{J_{\alpha}}{J_{LL}}  \frac{f_{\alpha}} {\bar g_{\nu}}  2 (\beta+3)   \frac{ \rm max(1,\tau_{LL})}{ \rm max(1,\tau_{\alpha})}  
\end{equation}
where $J_{\alpha}$ and $J_{LL}$ are the background intensities at the
\lya\ and Lyman limit frequencies, the spectral index of the ionizing
background is $J_{\nu} \sim \nu^{-\beta}$,$f_{\alpha}=0.416$ is the
oscillator strength of \lya\, and $\bar g_{\nu}\simeq 0.9$ is the
average Gaunt factor of the ionization cross section. The expression
at the end of this equation provides a rough approximation of what is
expected for the case when the optical depths are not small. The
scattered radiation dominates in the Lya forest, but is small when
$\tau_{LL}$ becomes close to 1 unless the decline of the background
intensity from the \lya\ to the Lyman limit frequencies is extremely
large (note that $\tau_{\alpha}/\tau_{LL} \sim 10^{3.5}$ for the typical
\lya\ forest velocity dispersion).

The current paper will present both our methods and our first results.
We will present the results from idealized models of fluorescent gas
clouds and from cosmological hydrodynamic simulations.  The use of
idealized models is complementary to results from hydrodynamic
simulations and serves two main functions.  First, we use these cases
as illustrations and tests of our numerical machinery, checking that
it functions properly when we have analytic results with which to
compare.  Second, owing to the freedom we have in modeling idealized
cases, they are relevant to and can be compared with current
observations of individual systems.  The complementary role of the
hydrodynamic simulations is to provide realistic predictions for large
samples of emitters in arbitrary patches of the universe in a
cosmological model.  These predictions are useful for future surveys
in which {\it ensembles} of systems are being examined.

In \S\ref{sec:method}, we review the method developed in ZM02 and
describe how we adapt their algorithm to work in conjunction with
generalized particle distributions, and with the output of smoothed
particle hydrodynamic (SPH) cosmological simulations in particular.
We present our results for the \lya\ emission signature from a simple
spherical geometry in \S\ref{sec:results1} and for two cosmological
simulations in \S\ref{sec:cosmo}.  We discuss these results in the
context of currently available observational facilities and recent
observations in \S\ref{sec:observables}.  We summarize our results and
present our conclusions in \S\ref{sec:discussion}.  For the interested
reader, we provide more information about the computations carried out
in this study in the Appendices.

\section{Method \label{sec:method}}

\subsection{Overview of Machinery}

The machinery we develop includes three parts. First, we apply
self-shielding corrections given the distribution of SPH particles,
either from the output of cosmological simulations or from model
structures. Then, the distributions of gas density, temperature,
velocity, and emissivity, represented by SPH particles, are put onto a
grid. Finally, we apply the Monte Carlo \lya\ radiative transfer
algorithm of ZM02 to the grid and obtain \lya\ images and spectra.
The ZM02 algorithm can be applied to systems with arbitrary geometry
and arbitrary distributions of gas density, temperature, and velocity.
It is modified to work in conjunction with the gas distribution
prepared by the first two parts of the machinery. We give a brief
review of the ZM02 algorithm below and describe the first two parts of
the machinery in detail in the next two subsections. 

In the ZM02 Monte Carlo algorithm, for each photon, the scattering
process is generally described by three steps: 1) the initial position
of each \lya\ photon is generated based upon the emissivity
distribution in the gas while its initial direction is randomly drawn;
2) the optical depth, $\tau$, through which the photon will travel
before scattering is drawn from an exponential distribution ${\rm
  exp}{(-\tau)}$, and the spatial location for the scattering at this
optical depth is determined along the initial direction from the
neutral hydrogen distribution (density, temperature, and velocity) and
the scattering cross section; 3) the thermal velocity of the
scattering atom is determined, and the new frequency and direction of
the photon are calculated.  In general, we use the term ``scattering''
to refer to this process of absorption and re-emission.  In the rest
frame of the absorbing atom, the \lya photon is re-emitted with an
unchanged frequency, except for the recoil effect (which the code
accounts for but is negligibly small for our applications).  The
calculation of scattering is performed in the restframe of the atom
and the frequency and direction of the scattered photon are
transferred back to the laboratory frame.  When calculating the photon
free path, the bulk motion (fluid velocity) of the medium is taken
into account by using the frequency in the fluid frame to compute the
(thermally broadened) scattering cross section.  With the new
frequency and direction, steps 2) and 3) are repeated until the photon
escapes the system. 

To generate the image of the \lya\ emission, a fixed direction of
observation is chosen, and the output of the computation is stored in
a 3D array containing the observed \lya\ spectrum at each projected
spatial position.  At each photon scattering, the probability that the
photon escapes along the chosen direction of observation is
calculated, and this probability is added to the pixel in the 3D array
corresponding to the projected position and frequency of the photon.
The scattering of \lya\ photons can be divided into two
regimes. Around the line center, the scattering cross section has a
thermal core with high amplitude, and at large frequency offsets, the
cross section follows the Lorentz wing.  For \lya\ scatterings in a
medium with high optical depth, the frequency of a \lya\ photon
changes back and forth around the line center (``core'' scatterings)
with little change in its spatial position, until it suffers a
scattering that leads to a large frequency jump that shifts it out of
the core regime. To avoid spending excessive computational time
performing the core scatterings with little spatial diffusion, we
introduce a numerical acceleration scheme to skip the core
scatterings.  In the fluid frame, if the absolute value of the
frequency offset from the line center $\nu_0$ is within $q$ times
$(\sigma/c)\nu_0$ before scattering, where $\sigma$ is the 1D thermal
velocity dispersion of the hydrogen atoms and $q$ is a positive
number, we draw a frequency offset directly from a distribution to
assign the frequency after scattering.  This distribution is a
Gaussian distribution of width $(\sigma/c)\nu_0$ with the central $\pm
q(\sigma/c)\nu_0$ part excluded. The photon then travels with this new
frequency until the next scattering. In our applications, the
acceleration scheme is invoked only if the line-center optical depth
across the grid cell (see \S 2.3) exceeds $10^3$. We take $q=3$ and
find that choosing it to be smaller does not affect the final
spectra. We also tested the using the acceleration scheme advocated by
\citet{tasitsiomi06}, which assumes an optical-depth dependent core
width.  We find that adopting that scheme does not have noticeable
effects in the results of our application here, and we therefore use
our constant core width approach.

As discussed in \S\ref{sec:intro}, we apply our machinery to 
analytically specified, isolated gas clouds and to gas distributions
extracted from SPH simulations.
The only
difference between our two configurations (in terms of code operation)
is the setup of the gas particle properties.  Once these are
determined from the cosmological or analytic density field, we proceed
in exactly the same manner for both cases.  We describe our procedure
for determining the gas properties below.

\subsection{SPH Particles and the Self-Shielding Correction \label{sec:ssc}}

In the SPH technique, the density field is represented by discrete
particles with an extent determined by a 3D kernel or smoothing length
\citep{monaghan85}. The smoothing lengths of the particles are chosen
to overlap a fixed number of particles to ensure accurate
representation of the fluid.  Each particle has an associated
temperature and velocity that capture the (continuous) properties of
the fluid. 

In our cosmological SPH simulation, particles are exposed to a uniform
photoionizing background in the approximation that all of the gas is
optically thin \citep{katz96}.  In reality, however, some of the gas
is optically thick and should be self-shielded.  Accounting for the
self-shielding effect is particularly important for the \lya\ emission
signature, a signature that critically depends on the recombination
rate and, therefore, on the distribution of ionized and neutral gas.
We introduce an algorithm to perform the self-shielding correction
{\it a posteriori} to the neutral fractions of particles of gas
experiencing illumination by either the uniform UV background or a
local ionizing source.

We note that because of this self-shielding correction, if simulation
temperatures were directly adopted, they would also be too high in
general, but specifically in dense regions.  We make the following
correction for this effect: for particles with hydrogen number
densities $n_H > 1\times 10^{-3}\,$cm$^{-3}$ and temperatures $T<5
\times 10^4$K, we set the particle temperature to $T_{\rm
  corr}=10^4$K.  Particles with temperatures in excess of $T=5\times
10^4$K have been shock-heated, and these high temperatures are thought
to be more robust, so we do not modify them.  We calculate neutral
fractions using these revised temperatures including photo- and
collisional ionization of the gas.  We do not alter the simulation
temperatures when running the scattering calculation.  We discuss the
effects of adopting the simulation temperatures directly in Appendix
C.  The self-shielding correction is performed directly on the
particles, rather than on a grid, to retain the full resolution in the
SPH gas distribution.

For all of our calculations we adopt a power-law spectrum for the
ionizing background.  To make predictions for fluorescence in
the presence of both the UVB and a local ionizing source, we have run
additional cases in which we have placed a bright quasar at different
locations relative to the gas distribution.  We assume the quasar also
has a power-law ionizing spectrum and emits isotropically.

We correct for the effect of self-shielding on a particle-by-particle
basis by computing for each particle the optical depth contributed by
all particles that lie within 3 smoothing lengths, $h_{s}$, of the
sightline from each particle along the 6 principal directions ($\pm
x,\pm y$ and $\pm z$) of the box (plus the additional quasar direction
when the quasar is present).  For computational convenience, we use
the equivalent Gaussian form of the kernel for our calculations,
however the simulation is run with a cubic spline kernel.  The
contribution of a particle's density to the optical depth outside of 3
smoothing lengths is negligible.  Because we are primarily interested
in the transition layers between optically thick and thin regimes in a
given structure, we must further correct the optical depth to account
for the density gradient across a particle.  We describe our procedure
and tests for this in detail in Appendix A for the interested reader;
including the density gradient makes a critical difference to the
accuracy of the result.  The average of the attenuated UVB intensity
over the six directions is adopted as the mean intensity at the center
of the particle. At this position, the neutral hydrogen fraction is
determined through photoionization equilibrium.  For the case of
illumination by a quasar, which can be put at any reasonable position,
the attenuated ionizing flux from the quasar's direction is also
calculated and added to the photoionization term for the neutral
hydrogen determination.  Since each particle's neutral fraction
changes via this procedure, we carry it out iteratively until
achieving a fractional convergence of $10^{-2}$ between the old and
new neutral fractions for the most discrepant particle in the region.
The code for the self-shielding correction can thus deal with the
general case of photoionization from UVB and/or a local ionizing
source.

For particles with ionizing $\tau \sim 1$ we are particularly
sensitive to the resolution of our underlying simulation.  We correct
the optical depths for particles as described in Appendix A.  This
correction ensures an accurate representation of both the density
field and the neutral fractions of particles throughout the box.
However, in the case of quasar-induced illumination, the resolution of
our simulations is simply not sufficient as detailed in Appendix A.
These calculations should be regarded as lower limits to the possible
detectable emission. For the purposes of applying the \lya\ transfer
code, we represent the corrected gas distributions from particles with
a regular grid as we describe below.

\subsection{From Particles to a Grid}

To conveniently apply the scattering code, we resample the SPH output
at a fixed redshift onto a 3-dimensional grid.  In each grid cell, the
quantities to be determined from the particle distribution are the
neutral hydrogen density, the emissivity, the temperature, and the
fluid velocity.

For the {\it neutral} density $\rho_{\rm cell}$ and emissivity $\epsilon_{\rm cell}$
in a cell, we determine the fraction of neutral mass and \lya\ luminosity of
each particle that falls into the cell according to the SPH profile of
each particle and add contributions from all relevant particles. We
have
\begin{equation}
\rho_{\rm cell} =\frac{\sum_{i=1}^{N_p} m_{{\rm neut},i}\  K_i}{V_{\rm cell}}
\end{equation}
and
\begin{equation}
\epsilon_{\rm cell} =\frac{\sum_{i=1}^{N_p} l_i\  K_i}{V_{\rm cell}},
\end{equation}
where $N_p$ is the number of particles contributing to the cell under
consideration, $V_{\rm cell}$ is the volume of the cell, $m_{{\rm
neut},i}$ and $l_i$ are the neutral hydrogen mass and \lya\ luminosity of
the $i$-th particle, and $K_i$ is the fraction of the particle that
overlaps the cell based on its SPH kernel (eq.~[14] of \citealt{katz96}). The
luminosity of a particle is determined from the emissivity at the
particle position (defined as 0.66 times the ionization rate at the
location of the particle) and its volume.

Only the distribution of neutral hydrogen is important for \lya\ scattering.  Therefore, for the temperature or fluid velocity in each cell, we calculate the neutral-mass-weighted average, that is
\begin{equation}
Q_{\rm cell} =\frac{\sum_{i=1}^{N_p} m_{{\rm neut},i}\  K_i Q_i}{\sum_{i=1}^{N_p} m_{{\rm neut},i}\  K_i},
\end{equation}
where $Q$ is either the temperature $T$ or one of the three components
$(v_x,v_y,v_z)$ of the bulk velocity. The bulk velocity of each
particle is the sum of its peculiar and the Hubble flow velocity ${\bf
v}_{H} = H{\bf {r}}$ (referenced to the center of the box).

To perform the \lya\ scattering calculation, we must represent the
emissivity with a finite number of photons and then ``launch'' these
photons in the gas distribution.  We could do this in a variety of
ways: for example, we could launch photons with a number in proportion
to $\epsilon$, or launch a single photon per cell and weight these
photons by $\epsilon$.  We choose an intermediate course to
efficiently sample the emissivity distribution while minimizing
computation time.  We map the emissivity, $\epsilon$, of a cell to the
number, $N_\gamma$, of photons launched from the cell through the
monotonic function $G(\epsilon)$,
\begin{equation}
G(\epsilon) =
 \left\{  
    \begin{array}{ll}
          f \epsilon/\epsilon_{\rm crit}, & \mbox{if $\epsilon/\epsilon_{\rm crit} \leq N_{\rm tr}$},\\
          f N_{\rm tr} \log_{N_{\rm tr}}(\epsilon/\epsilon_{\rm crit}), & \mbox{if $\epsilon/\epsilon_{\rm crit} > N_{\rm tr}$}.
        \end{array}
 \right. 
\end{equation} 
The values of $\epsilon_{\rm crit}$ and $f$ determine the number of
photons launched given the gas distribution and grid size.  In
practice, we choose $\epsilon_{\rm crit}$ such that we draw a
sufficient number of photons for a fiducial grid resolution
(e.g. $32^3$) with $f=1$. We scale $f$ in proportion to the grid
resolution, $N_{\rm grid}$, as $N_{\rm grid}^{-3}$ for other
resolutions. We choose $N_{\rm tr}=10$ in our calculation for
convenience.  Adopting this scheme, the total number of photons
launched from the entire grid is approximately independent of the grid
resolution. We note that the 2D spatial resolution of the \lya\ image is
always matched to the the 3D resolution of the grid.

We weight the photons such that we recover the correct luminosity for
each cell. The number $N_\gamma$ of launched photons from each cell is
forced to be an integer. If $G(\epsilon) \geq 1$, we round it to the
nearest integer $N_\gamma=[G(\epsilon)]$ and assign a weight
$\epsilon V_{\rm cell}/N_\gamma$. If $G(\epsilon)<1$ for a cell, we draw a uniform
random deviate between 0 and 1. If the random deviate is greater than
$G(\epsilon)$, no photon is launched from the cell. If it is smaller
than $G(\epsilon)$, a single photon is launched with a weight of
$\epsilon V_{\rm cell}/G(\epsilon)$. That is, for these undersampled cells
[$G(\epsilon)<1$], the single photon drawn carries the luminosity
corresponding to $1/G(\epsilon)$ cells of similar emissivity.

\section{Case I: SPH Singular Isothermal Sphere \label{sec:results1} }

We first turn our attention to the case of a singular isothermal 
sphere (SIS) in rotation represented by SPH particles.  The simple case of a
rotating spherical cloud can help to develop physical intuition for
what we should expect for images and line profiles of fluorescent
clouds.  This will prove useful for studying and interpreting the more
complex results from the 3D simulations.

This case has a well-understood solution and therefore acts as a
benchmark test of our machinery.  The SIS has the added benefit of
being analogous and easily adaptable to specific
high-surface-brightness configurations (e.g., a cloud irradiated by a
nearby quasar) that may be observed with substantially reduced
telescope time.  The case of fluorescence from a singular isothermal sphere is
discussed for a range of physical parameters in ZM02a, and we compare
with their results as appropriate.  We further anisotropically
illuminate our SIS by a luminous local source (a quasar), which we
will discuss in \S\ref{sec:iso_qso}. We fix the temperature of our
sphere to $2\times10^4$ K, and neglect collisional ionization to
compute the ionized fraction.

\subsection{A SIS in a Uniform UV Background \label{sec:iso_uni}}
  
We first examine a $z=3$ singular isothermal sphere exposed to a
uniform ionizing background. In our calculations we assume a UVB
intensity of the form $I_{\nu}=3 \times 10^{-22} \left(
  {\nu_L}/{\nu}\right){\rm erg\ s^{-1}\ cm^{-2}\ Hz^{-1}\ sr^{-1}}$
where $\nu_L$ is the frequency at the Lyman limit.  This is close to
the spectrum computed by, e.g., Haardt \& Madau (1996) in shape and
over the frequency range that matters, and is consistent with recent
measurements of the cosmic UVB \citep[e.g.,][]{kirkman05}.  We further
assume a $\Lambda$CDM cosmology with Hubble constant $H_0 = 65 \kms$,
$\Omega_m=0.3$ and $\Omega_\Lambda$=0.7.  The sphere has total mass of
$10^{11}\msun$ and a 5\% gas fraction.  The virial radius $R_{\rm
  vir}=37.4\;\kpc$ and virial velocity $V_{\rm vir}=107\;\kms$ are set
by the total halo mass (e.g., Padmanabhan 1993).  The velocity
dispersion (from both thermal and turbulent contributions) of the
system is set to be $51\kms$.  The cloud is rotating with a flat
rotation curve with a circular velocity equal to $V_c ^2=V_{\rm vir}^2
- 2 \sigma^2$.  We ignore the ellipticity this rotation would
induce.  We set the temperature of the sphere to be $2\times 10^4$ K
throughout. For gas in high-density shielded regions, the cooling
times are very short and the gas is likely to have the indicated low
temperature given the available cooling and heating processes.  The
density of the sphere is represented by particles of fixed mass and
with smoothing lengths chosen to enclose 12 neighboring particles.  We
distribute the mass according to the SPH kernel.

\begin{figure}
\centering
\begin{tabular}{c}
\small
\epsfig{file=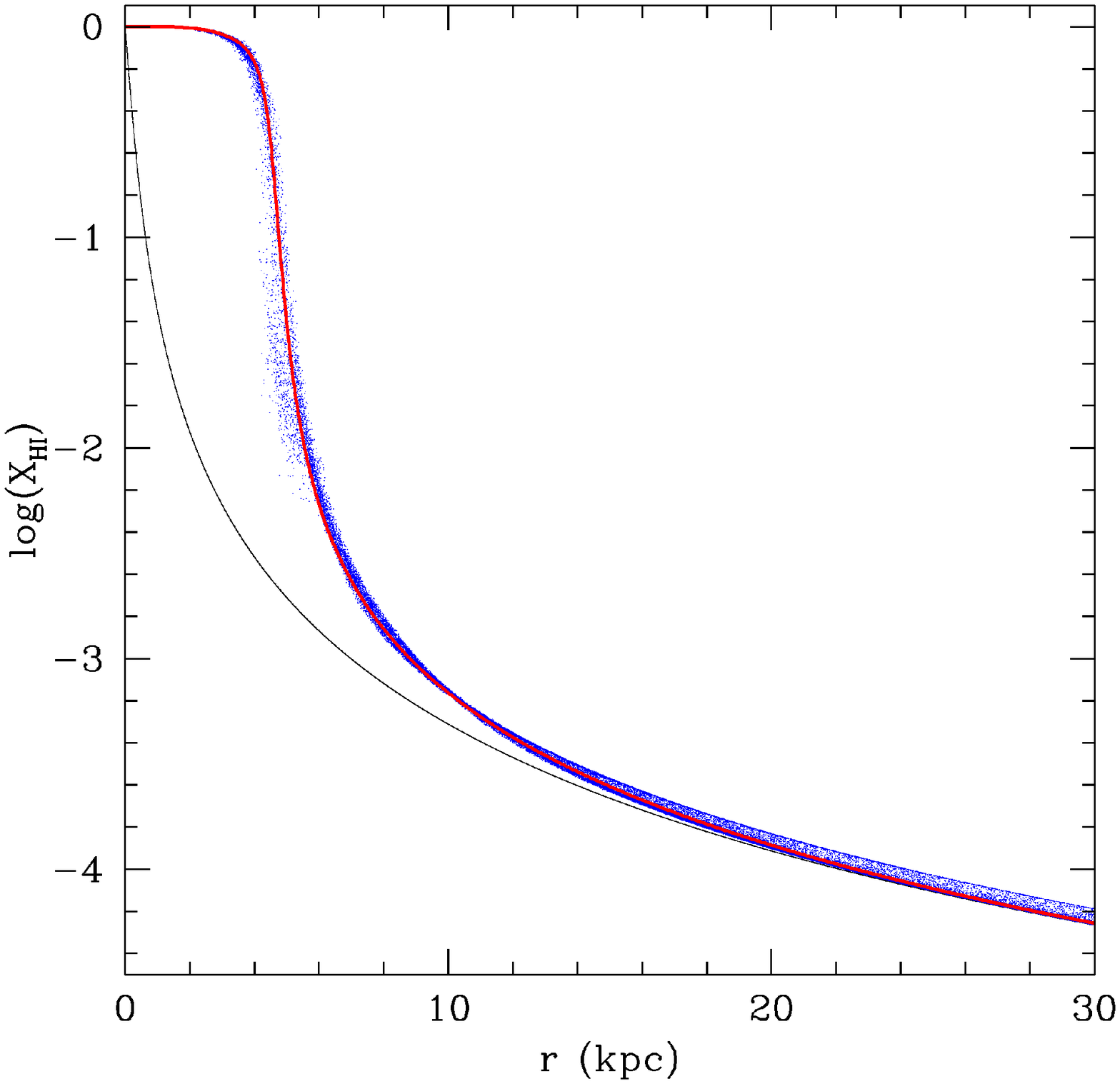,width=0.4\linewidth,clip=} \\
\epsfig{file=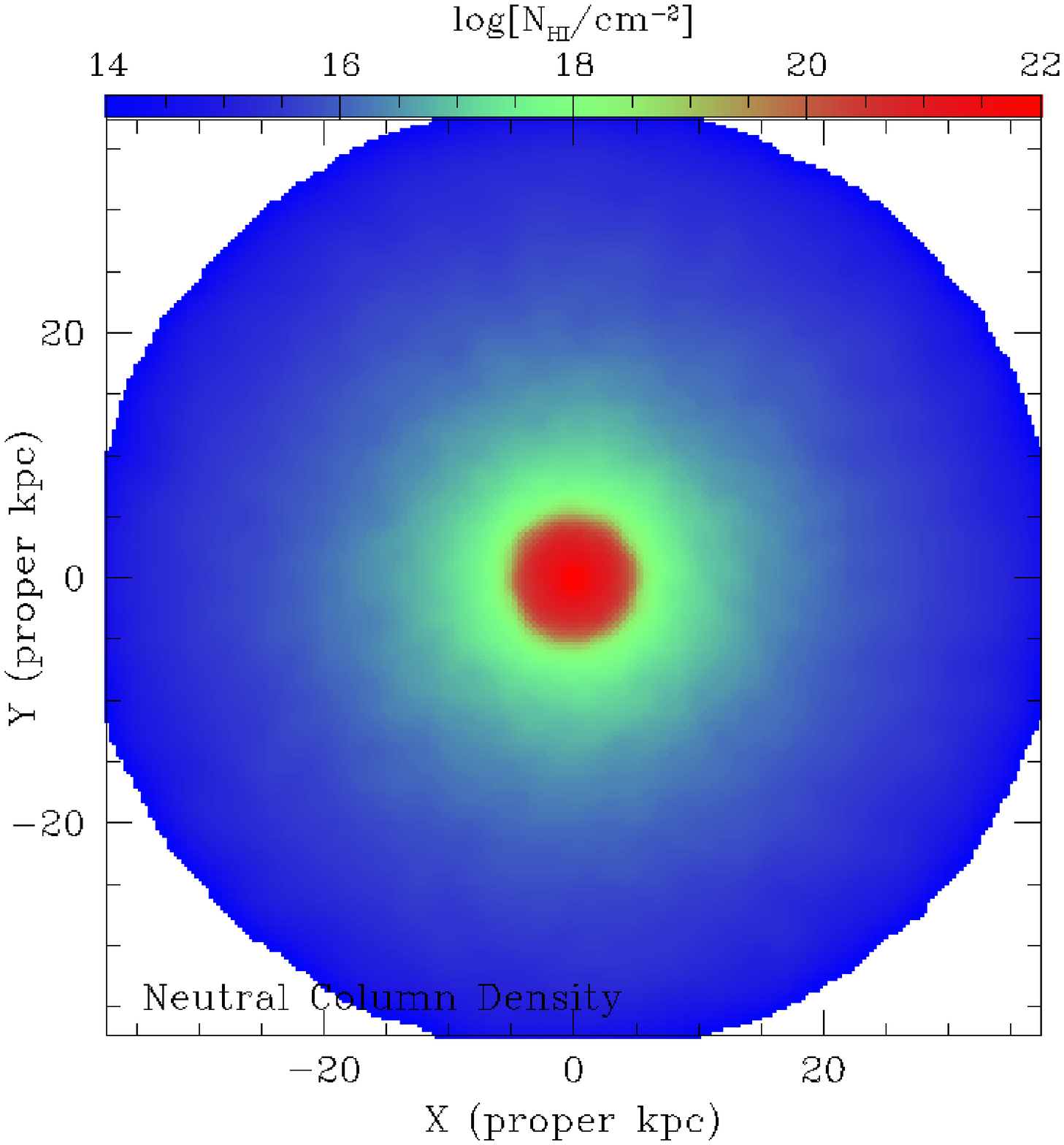,width=0.4\linewidth,clip=} \\
\epsfig{file=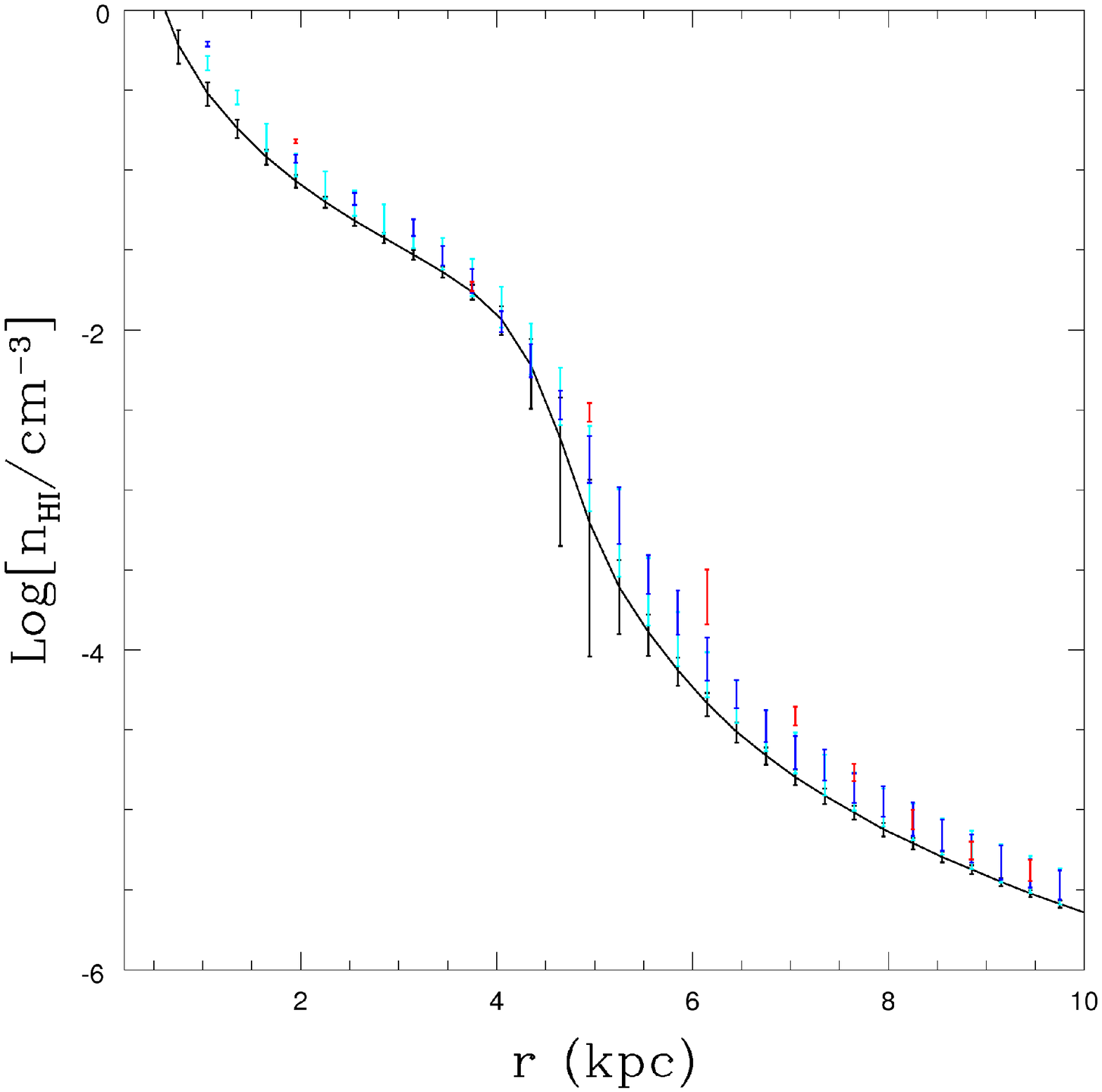,width=0.4\linewidth,clip=} 
\end{tabular}
\caption{\singlespace \scriptsize Properties for a singular isothermal
  sphere.  Top: Neutral fraction of particles for a $z=3$ isothermal
  sphere.  Black points show the optically thin case in which all
  particles are exposed to the same ionizing flux.  Blue points show
  include the effect of self-shielding.  The red curve shows the
  results of ZM02b.  At the self-shielding radius ($\sim 5\,$kpc) the
  cloud rapidly changes from nearly transparent to nearly opaque.
  Middle: Projected column density distribution for the $z=3$
  isothermal sphere.  The cloud would be viewed as a DLA over a total
  region $\sim 10\kpc$ in diameter.  Bottom: Comparison of the neutral
  density profile between particles and gridded cells.  The black curve
  shows the mean and 2-$\sigma$ variation for particles
  as a function of radius. Red, green, and cyan correspond to the mean
  and 2-$\sigma$ variation for the $32^3,64^3$, and $128^3$ cells,
  respectively.}
\label{fig:isoproperties} 
\end{figure}

As a key component, our self-shielding code should calculate the
correct value for the neutral fraction of each particle.  This
determines the photoionization rate, and therefore, the emission rate
of \lya\ photons. In the top panel of Figure~\ref{fig:isoproperties}
we show the neutral fraction, X$_{\rm HI}$, of particles in the sphere
as a function of radius.  The effect of self-shielding is clear in
this diagram.  The black points show the optically thin case, in which
we have exposed each particle in the cloud to the same ionizing flux
(corresponding to a photoionization rate of $9.5 \times 10^{-13} {\rm
  s^{-1}}$). The blue points show the neutral fraction of particles
after we have applied our self-shielding correction.  At the center of
the cloud, the gas becomes completely neutral owing to the shielding
layer, which is recombining rapidly enough to keep the inner cloud
completely neutral: no photoionizing photons are able to penetrate to
this depth.  This shielding layer is very thin --- it is effectively a
skin of about only 2 kpc separating nearly completely ionized from
completely neutral gas.  It is from this thin layer, in addition to
the extended emission from the larger optically thin regions, from
which \lya\ emission emerges.  Our results for the ``SPH'' version of the
singular isothermal sphere are in good agreement with the results of
ZM02 (shown by the red line in the figure). Of interest for absorption
line studies is the projected neutral hydrogen column density of this
cloud, which is shown in the middle panel of
Figure~\ref{fig:isoproperties}.  If there were a quasar directly
behind this system, it would be considered a Damped \lya\ system (DLA)
over the $\sim 10$\,kpc central region.  We will return to this column
density distribution below.  The fluorescent emissivity of \lya\
photons at each position is computed as 66\% of the photoionization
rate.

Once we have determined the emissivity, neutral density, temperature
and velocity at each location in the gas distribution, we put these
quantities on a uniform grid with length $2 R_{\rm vir}$ on a side and
run our radiative transfer code.  We first ensure that our grid
resolution is fine enough to resolve the self-shielding layer in our
cloud. We show in the bottom panel of Figure~\ref{fig:isoproperties}
the neutral density profile of the sphere (directly from the
particles) and compare it to the density profile generated from grids
with resolutions of $32^3$, $64^3$, and $128^3$.  The SPH smoothing lengths
of the particles are 0.7 kpc on average, while the smoothing length in
the transition region (from optically thin to thick) is ~0.25 kpc. There is a slight offset between the particle distribution and the gridded distribution, however, this is simply due to the the subtle difference between the density as determined at a given particles' position in the sphere and the density as determined from the sum of overlapping particle mass profiles.  This figure demonstrates that the $128^3$ grid recovers the density
profile with sufficient accuracy for our calculations, and we adopt it
for subsequent calculations of the SIS.

\begin{figure*}
\plotone{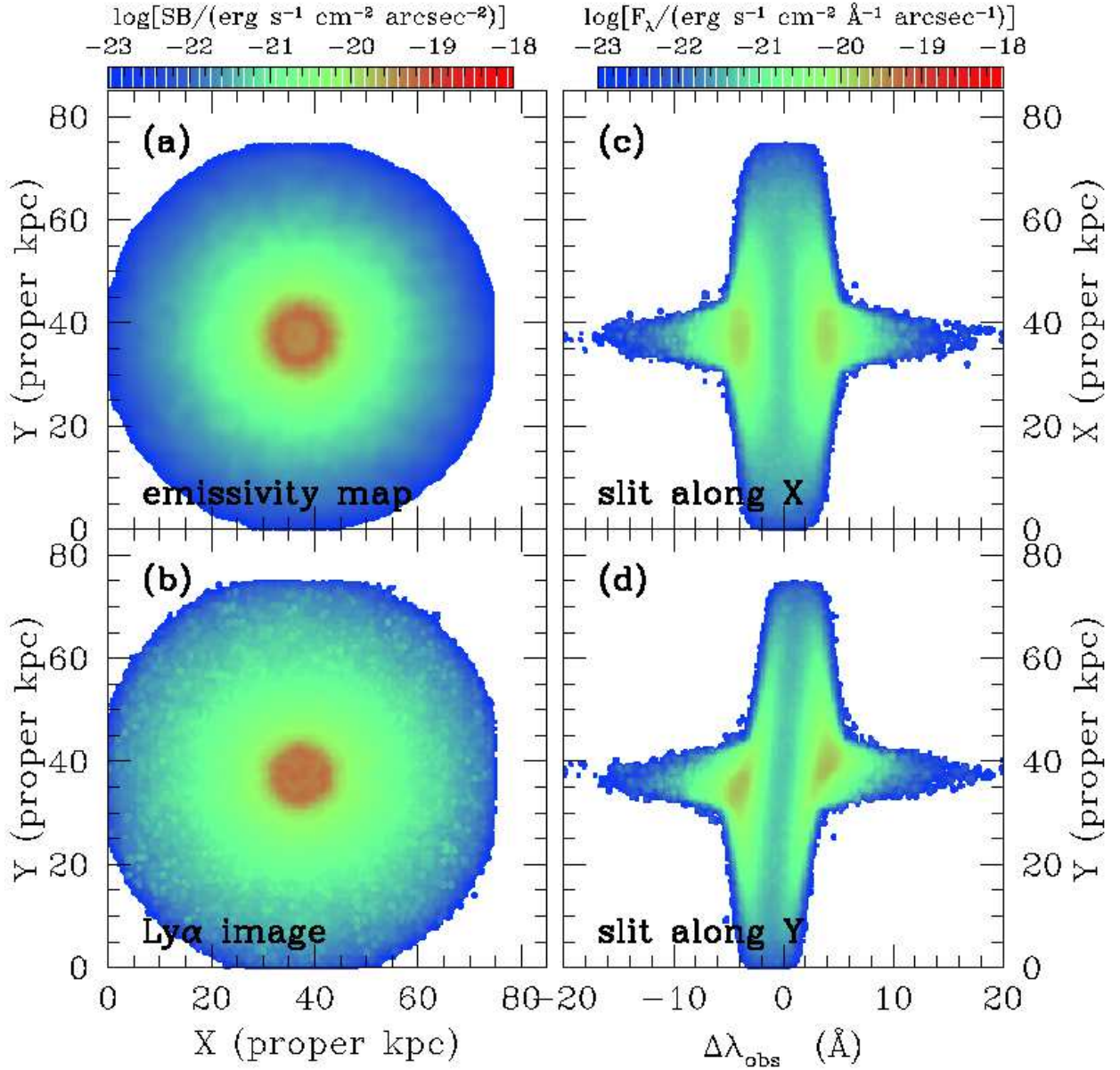}
\caption[\lya\ emission from uniformly illuminated
S.I.S.]{\singlespace Prediction of \lya\ emission for the
$z=3$ isothermal sphere induced
by a uniform UVB.  Panels (counter-clockwise from upper left) are
column emissivity, \lya\ image, 2D spectrum when slit is placed along
the y-axis, 2D spectrum when the slit is placed along the x-axis. The cloud
is rotating around the x-axis in this projection.  The rotation is clear
in the 2D spectrum when the slit is placed along the y-axis.  The
\lya\ image looks slightly smeared compared to the emissivity image.  At this
column density, however, the photons diffuse little in space, but rather shift
in frequency to emerge from the cloud.}
\label{fig:lya_z3iso}
\end{figure*}

In Figure~\ref{fig:lya_z3iso} we show the results of the radiative
transfer calculation for the isothermal sphere described above.  The
upper left panel in this figure shows the projected \lya\ emissivity
map, which we obtain by integrating the \lya\ emissivity along the
line of sight and assuming that \lya\ photons isotropically escape the
cloud without scattering.  In this sense it is a ``column emissivity
per solid angle'' map.  This emissivity image is the surface
brightness one would measure if the \lya\ photons underwent no
scattering and streamed directly out of the cloud over the 4$\pi$
solid angle from where they were physically produced.  This is similar
to what one would observe if seeing this object in non-resonant line
radiation such as H$\alpha$ or H$\beta$, although of course these
would be at much lower intensities.  The lower left panel shows the
emergent scattered \lya\ image.  A comparison of the true (i.e.,
scattered) \lya\ image to the ``column emissivity'' map, illustrates
the effects of spatial diffusion of the photons.  Note the graininess
in the \lya\ image at low surface brightness is due to the finite
number of photons we run.  With respect to the emissivity map, \lya\
emission seen in the scattered image shows spatial diffusion caused by
the scattering, although the effect is small.

\begin{figure*}
\plotone{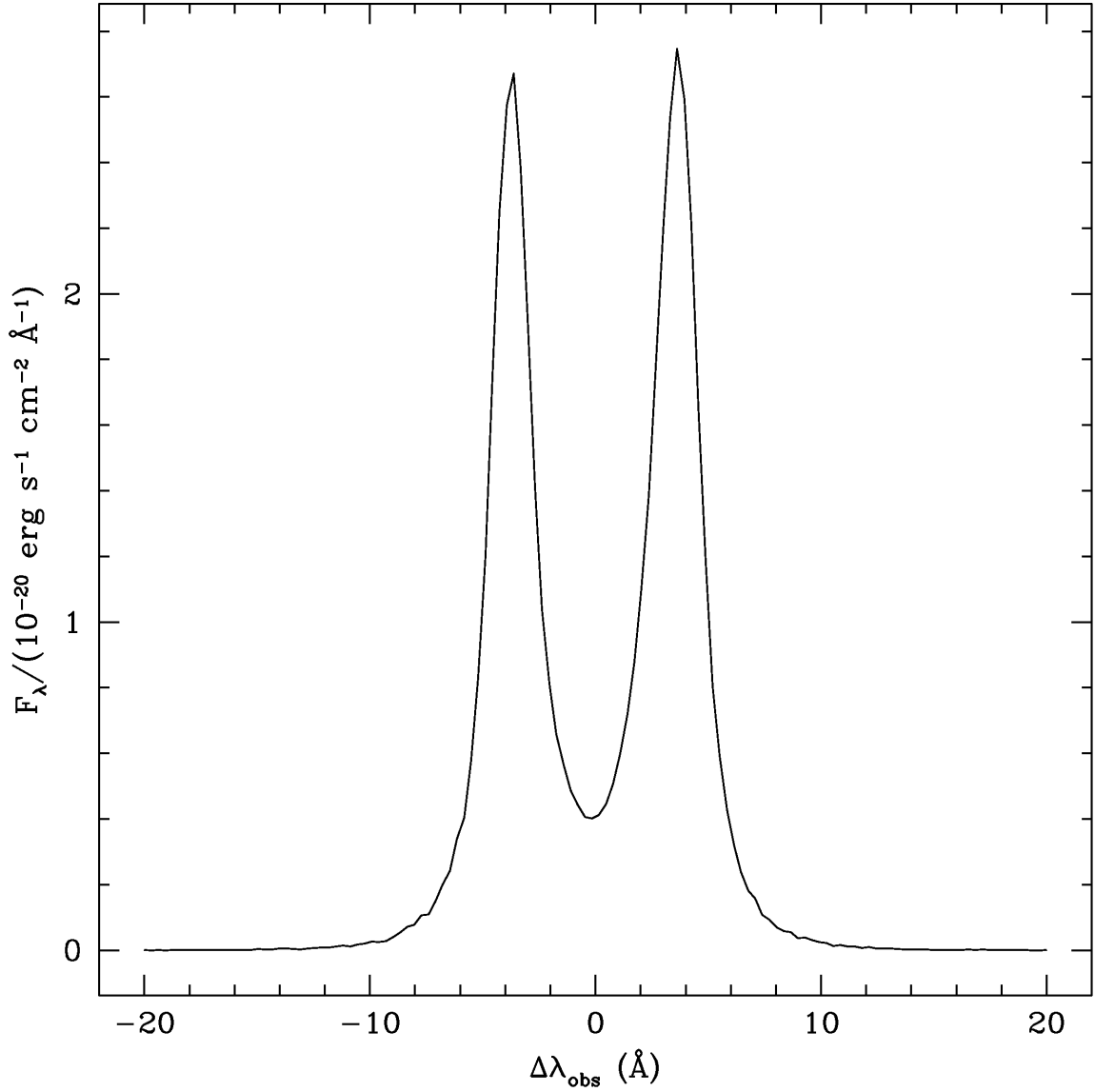}
\caption[1D Spectrum for S.I.S.]
{\singlespace 1D spectrum of $z=3$ isothermal sphere induced by a
uniform UVB.  The spectrum is the equivalent of one that would be
observed if a single fiber were placed on the sphere. The peak
separation is $\sim 7$\AA, with positions that approximately
correspond to $\pm 4\sigma / c \lambda_{\rm Ly\alpha}$ where $\sigma$
is the velocity dispersion of this cloud ($51 \kms$).}
\label{fig:lya_z3iso_1d}
\end{figure*}

The right hand panels of Figure~\ref{fig:lya_z3iso} show the 2D
spectra of the cloud.  These spectra are generated by orienting a wide
slit (over the entire cloud) along the $x$-axis (upper-right) and
$y$-axis (lower-right).  The rotation curve for the sphere is clearly
seen in the lower-right panel.  Since the sphere is set to rotate
around the x-axis, there is no effect of rotation in the upper-right
panel, which shows the characteristic double-peaked \lya\ profile.
For clarity, we show the 1D spectrum of the whole cloud in
Figure~\ref{fig:lya_z3iso_1d}.  The velocity profile becomes more
apparent in the 1D diagram.  It is clear from these figures that the
photons are escaping the cloud primarily by substantial shifts from
the line-center frequency, so that the scattering cross section
becomes sufficiently low to allow the photon to escape.  The peaks are
separated by $\approx 7$ \AA, which corresponds approximately to the
width given by $\pm 4\sigma\times \lambda_{\rm Ly\alpha}/c$ where
$\sigma$ is the 1D thermal velocity dispersion \citep{gw96}.  ZM02
also present the \lya\ images and spectrum for this case, and we find
the agreement is, as expected, excellent.  The brightest pixel in the
\lya\ image corresponds to a surface brightness of $\sim 6.0\times
10^{-20}\ {\rm erg\ s^{-1}\ cm^{-2}\ arcsec^{-2}}$, which is 30\%
higher than expected under the "simple mirror" approximation (GW96)
from our adopted ionizing background at the redshift of the SIS in the
absence of heating (eq.~[5] of GW96).  As we show in
Appendix A, this is in accordance with the expectations from an exact
solution for this system based on ZM02.  The excess flux over the
simple mirror expectation arises from a simple
limb-brightening effect.  Even so, at these flux levels detecting such
a system is a challenge for modern 10m-class telescopes.  The maximum
source surface brightness should be compared with the $B$-band sky
brightness of $\sim 2.9 \times 10^{-17}\ { \rm erg\ s^{-1}\ cm^{-2}\
  arcsec^{-2} \AA^{-1}}$, corresponding to ${ \rm B}=22.2 { \rm\ mag \,
  arcsec^{-2}}$.  At the specified redshift, $z=3$, this relatively
bright region has size $\sim 10 \kpc$ that would correspond to a
diameter of $\sim 1.3 \asec$.  For our adopted UVB at this redshift,
detecting this object with a signal-to-noise ratio (S/N) of 1 would
require approximately 100 hours on a 10m telescope assuming a 10 \AA
\ filter, 30\% telescope efficiency and 80\% atmospheric transparency.
The detectability is not, however, this remote.  For this calculation
we have ignored two important effects: 1) heating of the gas from
high-energy photoelectrons and 2) cooling radiation. We have not
explicitly included the first effect in our calculation, but one can
estimate its magnitude from equation (13) of GW96: it would basically
double the observed surface brightness. We address the 2nd effect more
completely in Paper II, as its amplitude is entirely dependent on the
adopted temperature of the gas that, in this case, is physically
motivated but otherwise arbitrarily chosen.

Even given more optimistic estimates for the surface brightness, it
remains a major observational undertaking to detect \lya\ emission
caused by the UVB alone --- for example, larger structures of size
$10\; {\rm arcsec}^{2}$ could be detected in $\sim26$ hours with a 10m
telescope should they exist at these redshifts.  More promising at
present is the possibility of detecting \lya\ emission from clouds
exposed to an enhanced ionizing field.  We turn our attention to this
case.

\subsection{A SIS in a Quasar Radiation Field \label{sec:iso_qso}}

\begin{figure*}
\plotone{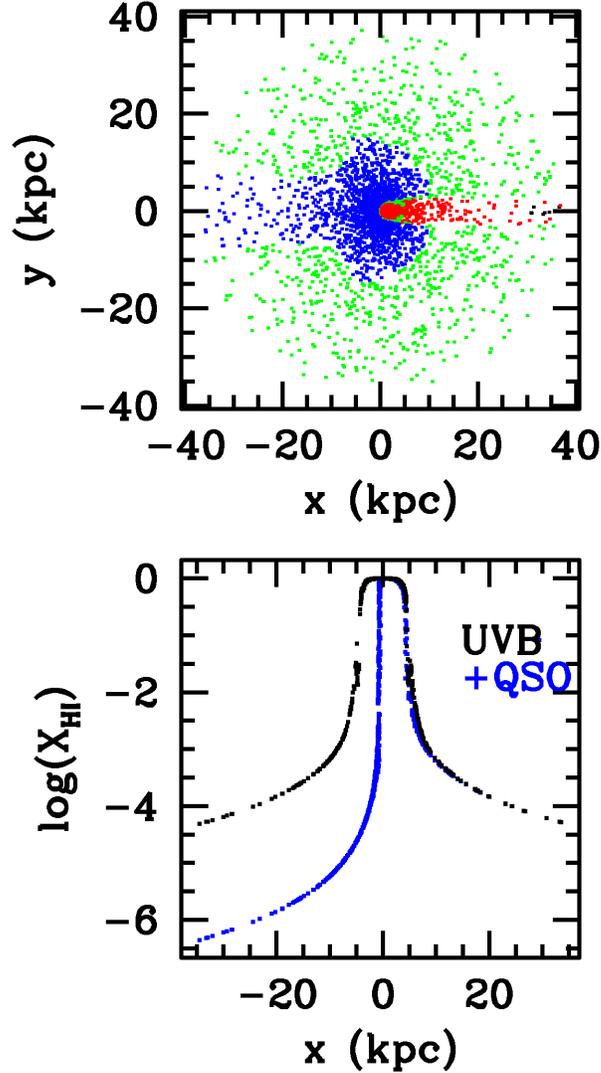}
\caption[effect of quasar on neutral fractions for S.I.S.]
{\singlespace The effect of quasar flux on particle neutral fractions.  
Top panel: The particle distribution through a slice near the
$z=0$ plane in the SIS color-coded according to neutral fraction {\it
relative} to the uniform UVB case.  Black, red, green, and blue points
show particles for which the ratio of the neutral fraction
($R_{HI}$) in the presence of the QSO to the neutral fraction in the
uniform UVB case is unchanged, $1>R_{HI}>0.1$, $0.1>R_{HI}>0.01$, and
less than 0.01, respectively.  Bottom Panel: The neutral Fraction of
particles near the $x$-axis in the presence of bright QSO.  The
quasar is located to the left in this figure at approximately $-500$ kpc 
from the center of the sphere.}
\label{fig:z3qso_effect}
\end{figure*}

We now investigate the case of a singular isothermal cloud,
constructed as described above, that is irradiated by a local bright
quasar.  The UVB + quasar case is of particular interest because of
the potential surface brightness enhancement and therefore
detectability of these systems.  It is timely to analyze this case
because of the recent detection of \lya\ fluorescence in a DLA system
irradiated by a bright quasar \citep{adelberger06}.  As noted
previously, our code for performing the self-shielding correction is
equipped to deal with an anisotropic radiation field to study
anisotropically irradiated gas fields, novel with respect to the cases
studied in ZM02.  We place the quasar at $(x,y,z)=(-500,0,0)$ kpc
(physical) from the center of the sphere.  The quasar is assumed to
emit isotropically and have a power-law continuum shortward of 912
\AA\ of $L_\nu=L_{\nu_L}(\nu/\nu_L)^{\alpha}$ where $\alpha$ is taken
to be $-1.57$ in accord with observations \citep{telfer02}.  We assume
a specific 912 \AA\ luminosity, $L_{\nu_L}$ at the Lyman limit $\nu_L$
of $1.0\times 10^{31}\; {\rm erg s^{-1} Hz^{-1}}$ \citep{liske01}
unless otherwise specified.  When we turn on the quasar and examine
the resulting neutral fraction for particles near the $y=z=0$ line, we
see in the bottom panel of Figure~\ref{fig:z3qso_effect} that the
quasar has indeed ionized the outer edges of the cloud facing it as
expected from this configuration.  At the cloud, the quasar intensity
corresponds to an enhancement in ionizing photon flux per unit area
above the UVB of approximately 60.

In the top panel of Figure~\ref{fig:z3qso_effect} we show a particle
representation of the cloud near the $z=0$ plane, where particles are
color coded according to their relative neutral fractions, i.e., their
neutral fractions in the presence of the quasar relative to their
neutral fractions when exposed to the UVB alone.  The blue points in
the diagram show those particles that have been most strongly affected
by the quasar.  The quasar has the most dramatic effect on the nearly
neutral innermost particles and particles on direct sightlines to the
quasar radiation.  Particles on the opposite side of the cloud from
the QSO experience a less dramatic reduction of their neutral fractions
because the quasar radiation is attenuated or entirely blocked from
view by the central optically thick structure.  The half moon
illumination caused by the quasar is apparent from this figure.  In
fact, it looks more like a keyhole, but the highly ionized outer
layers of the cloud contribute very little to the \lya\ surface
brightness owing to their low density.  We now explore how this
translates into \lya\ emissivity and, ultimately, scattered radiation.

\begin{figure*}
\plotone{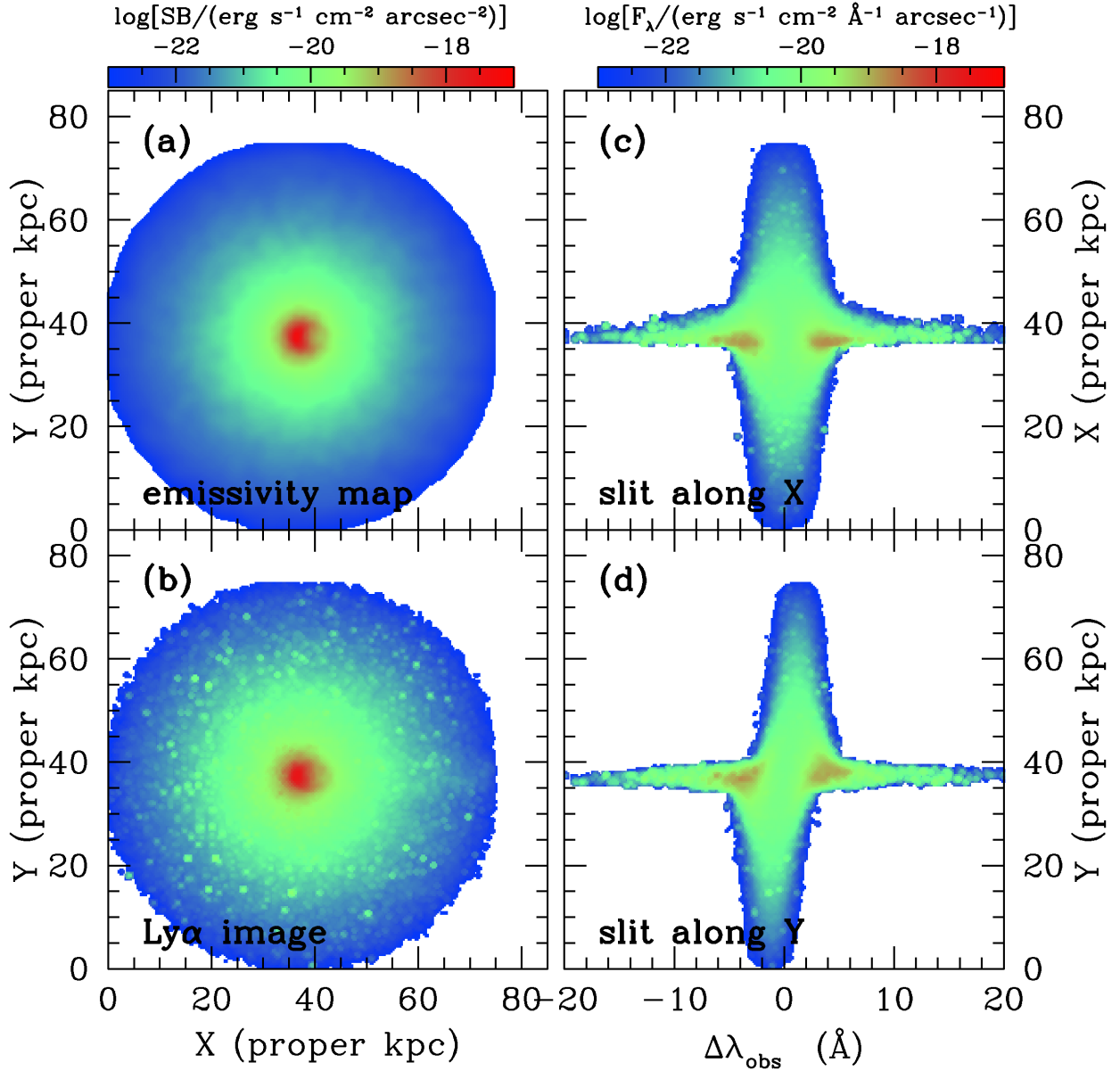}
\caption[\lya\ emission for anisotropically illuminated S.I.S.]
{\singlespace Predictions for \lya\ emission from a $z=3$ isothermal sphere
induced by both a uniform UVB and a bright quasar.  Panels are as in
Figure~\ref{fig:lya_z3iso}.  The quasar is located to the left at a distance $500\; \kpc$ from the center of the isothermal sphere (off the panels), and has a power law continuum with
slope $-1.57$ and luminosity $L_{\nu_L}=1.0\times 10^{31}\; {\rm erg\
s^{-1} Hz^{-1}}$ at the Lyman limit.  Note the difference in the color scale from Fig.~\ref{fig:lya_z3iso}.
}
\label{fig:lya_z3qiso}
\end{figure*}

Based on the properties of the quasar, we can determine
order-of-magnitude expectations for the surface brightness of \lya\
emission for the case in which the DLA acts simply as a ``mirror'', converting $66\%$ of the quasar's ionizing radiation into \lya\ fluorescence\footnote{We use the term ``mirror'' to mean an optically thick surface that converts 66\% of impinging ionizing photons to the \lya\ frequency and re-emits these photons at a random angle.}. At the distance of the
cloud, these photons should emerge as:

\begin{eqnarray}
\Gamma_{\rm mirror} &=& \frac{0.66\ r_{SS}^2\ \dot{N}_{\rm ionizing}}{4 d_{q}^2},\\
\dot{N}_{\rm ionizing} &=& \int_{\nu_L}^{\infty} \frac{L_{\nu}}{h\nu} d\nu ,
\end{eqnarray}
Where $r_{SS}$ is the self-shielding radius of the cloud, $d_{q}$ is
the cloud-quasar distance, The prefactor of 0.66 comes from the
fraction of ionizing photons that eventually cascade to the \lya\
transition \citep[][GW96]{gw96}.  For the systems we examine, $d_{q}$ is
$500\;\kpc$ and $r_{SS}$ depends on the quasar flux as we show below.  For the quasar spectrum we adopt, the peak surface brightness should go roughly as $\sim 1.01 \times 10^{-17} [(1+z)/4]^{-4} [L_{\nu_L,31}][ d_q/(500 {\kpc})]^{-2} {\rm erg\ s^{-1}\ cm^{-2}\
  arcsec^{-2}}$ where  $L_{\nu_L,31}$ is the quasar luminosity at the Lyman limit in units of $10^{31} {\rm erg s^{-1} Hz^{-1}}$.
Figure~\ref{fig:lya_z3qiso} shows the \lya\ images and spectra when
we include the radiation field of the quasar. We see in 
the upper left hand panel that the
effect of the quasar on the \lya\ emission is to produce a half-moon
region of high surface brightness gas.  The half-moon illumination
reflects the higher ionization rate and, therefore, a higher recombination rate
in the portion of the cloud facing the quasar.  This should be
compared with the uniform emission shown in Figure~\ref{fig:lya_z3iso}
for the case in which the only source of illumination is the UVB.  In
the \lya\ image itself, this half-moon shape is preserved.  Such a
feature serves as a diagnostic of this configuration because it
implies either a specially arranged gas density distribution, or an
anisotropic illumination similar to the case discussed here.  The spectra
are also modified in the presence of the quasar.  The amplitude of
emission is higher as expected, and the peak separation is smaller
than the case with UVB illumination alone.

\begin{figure*}
\plotone{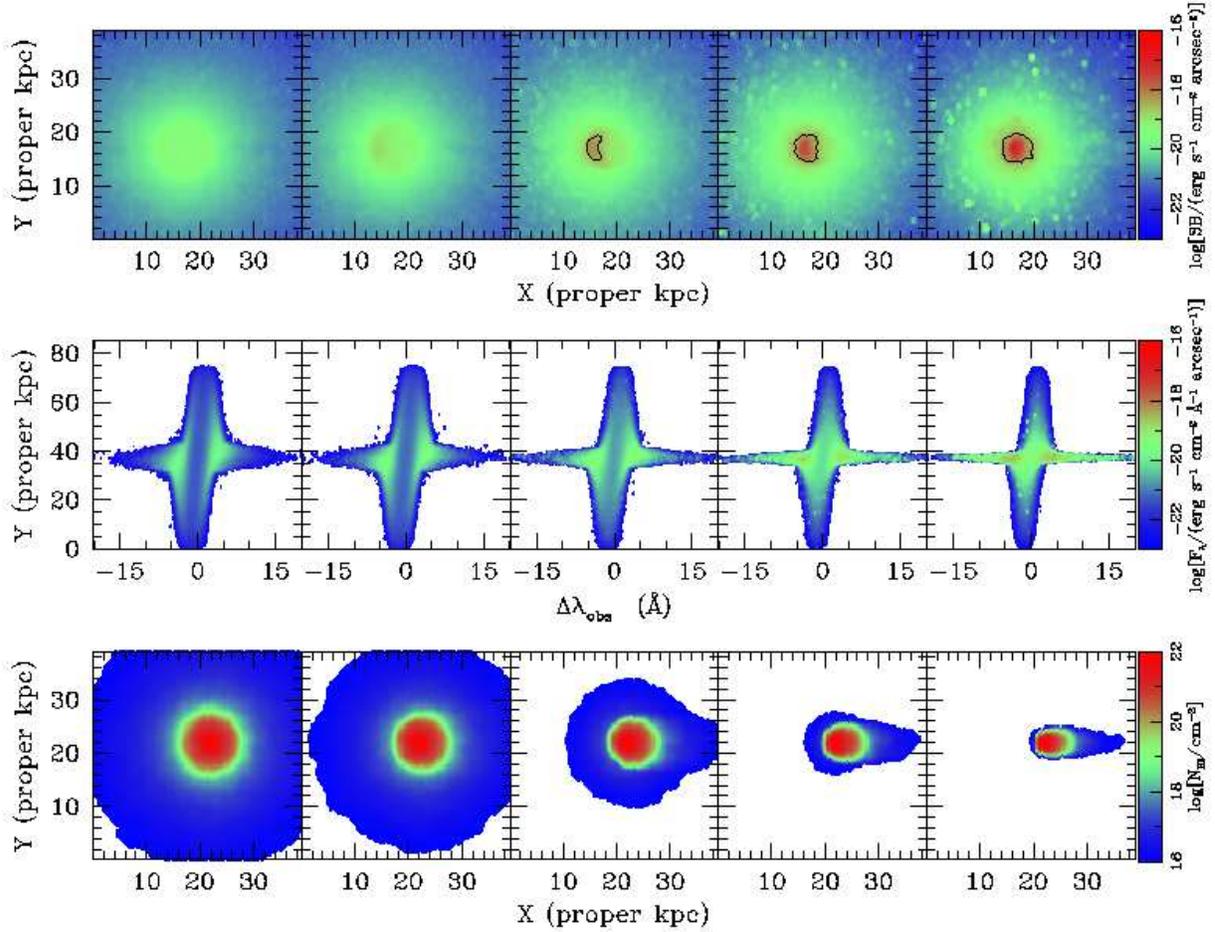}
\caption[Effect of quasar luminosity on \lya\ surface brightness and
neutral column density]{\singlespace Sequences in Ly$\alpha$ surface
  brightness (top), 2D spectrum (middle), and neutral column density
  (bottom) for an isothermal sphere as a function of the illuminating
  quasar.  The quasar is turned off in the leftmost column and the
  cloud is exposed to the UVB only.  The quasar is located to the left
  at a distance $500\; \kpc$ from the center of the isothermal sphere.
  The quasar is turned on from specific luminosity at the Lyman limit
  of $L_{\nu_L}=1.0\times 10^{29} {\rm erg\ s^{-1} Hz^{-1}}$ to a
  maximum value of $L_{\nu_L}=1.0\times 10^{32} {\rm erg\ s^{-1}
    Hz^{-1}}$ in increments of factors of 10. The characteristic half
  moon illumination pattern is most pronounced in the middle panel,
  where the quasar's radiation further ionizes the exposed area of the
  cloud.  The black contour in the upper panel shows a constant
  surface brightness level of $3\times10^{-19} {\rm erg\ s^{-1}\
    cm^{-2}\ arcsec^{-2}}$. The bright emission comes from gas that
  would have high neutral density in the absence of the quasar flux,
  and has a high recombination rate once the quasar radiation impinges
  upon it.  This demonstrates the increasing tendency of the quasar to
  fully ionize the outer edges of the cloud, and to shrink the highly
  neutral regions of the cloud from the left.  }
\label{fig:crank_qso} 
\end{figure*}

The brightest feature in the QSO-irradiated cloud is 53 times brighter
than the brightest feature seen from exposure to the UVB alone. The
quasar itself contributes $\sim$200 times more ionizing flux than the
UVB alone at the center of the cloud.  Since the UV-absorbing and
\lya\-emitting surface is finite and curved, as opposed to an infinite
flat surface, \lya\ photons are emitted into $2\pi-4\pi$
steradians. This geometric effect leads to a factor of 2-4 reduction
in the surface brightness with respect to the expectation from an
infinite flat ``mirror''. The calculated increased surface brightness
relative to the UVB-only case is in line with this expectation, given
the strength of the quasar radiation field.  As we show in Appendix A,
the reason for any small discrepancy is that the neutral fraction
profile is not being faithfully represented in our 100,000 particle
case.  When we run this case with 500,000 particles, we get closer to
the expected value, but even this is not fully adequate.  Since
high-resolution cosmological simulations typically have at most
1,000,000 particles in their most well-resolved structures, current
SPH simulations are simply unable to faithfully capture the steep
transition from optically thick to thin that occurs in these systems.
In the limit of a very bright quasar, this half-moon shape is
eliminated as the ionizing flux propagates further into the cloud. We
illustrate this effect in Figure~\ref{fig:crank_qso} in which we show
a sequence of surface brightness and neutral column density images as
the quasar luminosity $L_{\nu_L}$ at the Lyman limit is increased from
0 (the UVB-only case) in the left-most panel, and then from
$1.02\times10^{29}$ to $ 1.02\times 10^{32}\; {\rm erg\ s^{-1}
  Hz^{-1}}$.  We show the impact of the quasar on the projected
neutral column density of the system in the bottom panels of
Figure~\ref{fig:crank_qso}.  The sequence of neutral column density
shows that, as expected, the neutral layers are progressively blasted
away, leaving only a very small, dense core in the case of very strong
quasar illumination.

There are three features to note in these figures that, in
conjunction, can provide constraints on the physical situation of
individual optically thick systems when observed.  The first is simply
the \lya\ surface brightness.  Given the quasar luminosity and
distance, it is straightforward to calculate how bright (in the
absence of dust) the optically thick cloud will glow.  Depending on
the impinging flux, this anisotropic illumination will create a
half-moon or a ``pac-man'' type structure.  The second is the
frequency distribution of the photons. Most importantly, fluorescence
will manifest itself as a double peaked profile with a peak separation
approximately equal to $8 \sigma$ where $\sigma$ is the cloud's
velocity dispersion, as demonstrated here.  In the absence of other
bulk flows, the peak separation, together with an estimate of the
object's size, directly constrains the mass of the object.  Rotation
within the cloud will be apparent in the spectrum depending on the
orientation of the slit to the rotation axis of the cloud.  In the
case we show in Figure~\ref{fig:lya_z3qiso}, a ``double'' rotation
curve can be clearly seen in the 2D \lya\ spectrum.  The third
diagnostic is the size of the ``absorber''. For a given impinging
flux, halo mass, and density profile, there is a specific size over
which the cloud will appear as a LLS or DLA.  Combining emission
observations with absorption line studies to measure the column
density, one can set limits on this size and thereby set constraints
on simple models such as those presented here.  In
\S\ref{sec:observables}, we demonstrate how these diagnostics work
together by comparing these simple models in detail to a recently
discovered system \citep{adelberger06}.

\section{Case II: A Cosmological Volume \label{sec:cosmo}}
We now turn our attention from simplified cases with well-understood
geometries to predictions for the variety of structures produced in
cosmological hydrodynamic simulations.  For this work, we use two
cosmological simulations that have complementary strengths.  Our
primary simulation is a 5.555 \hmpc\ (comoving) box at $z=3$ with
cosmological parameters $\Omega_m =0.4, \Omega_{\Lambda}=0.6,
\Omega_{b}=0.0473, \sigma_8=0.8, H_0= 100h \kms {\rm Mpc^{-1}}$ with
$h=0.65$ (hereafter L5).  The simulation has $128^3$ dark matter and
$128^3$ gas particles and the gravitational forces are softened using
a cubic spline kernel with radius of 1.25 \hkpc\ (comoving).  The mass
per gas particle in the simulation is $1.7\times 10^6 \msun$. We
supplement this box with a second larger, but lower resolution, box of
22.222 \hmpc\ (comoving) at $z=2$ with the same cosmology and the
same number of particles (hereafter L22). The softening radius and
mass resolution for the L22 box are exactly 4 and 64 times larger than
in the L5 box respectively.  We use the lower-resolution simulation
primarily to illustrate the redshift dependence of the
\lya\ emission signature in amplitude and morphology. 

The simulations make use of the parallel version of TREESPH
\citep{hernkatz89, katz96, katz99, dave97} that combines smoothed particle
hydrodynamics \citep{lucy77, gm77} and a hierarchical tree algorithm
for the computation of gravitational forces \citep{bh86,
  hernquist87}. The calculation is described extensively in
\citet{katz96} and \citet{keres05}, and we refer the interested reader
to that work for more details.

\subsection{ Large Scale Structure in the UVB}

\begin{figure*}
\epsscale{0.9}
\plotone{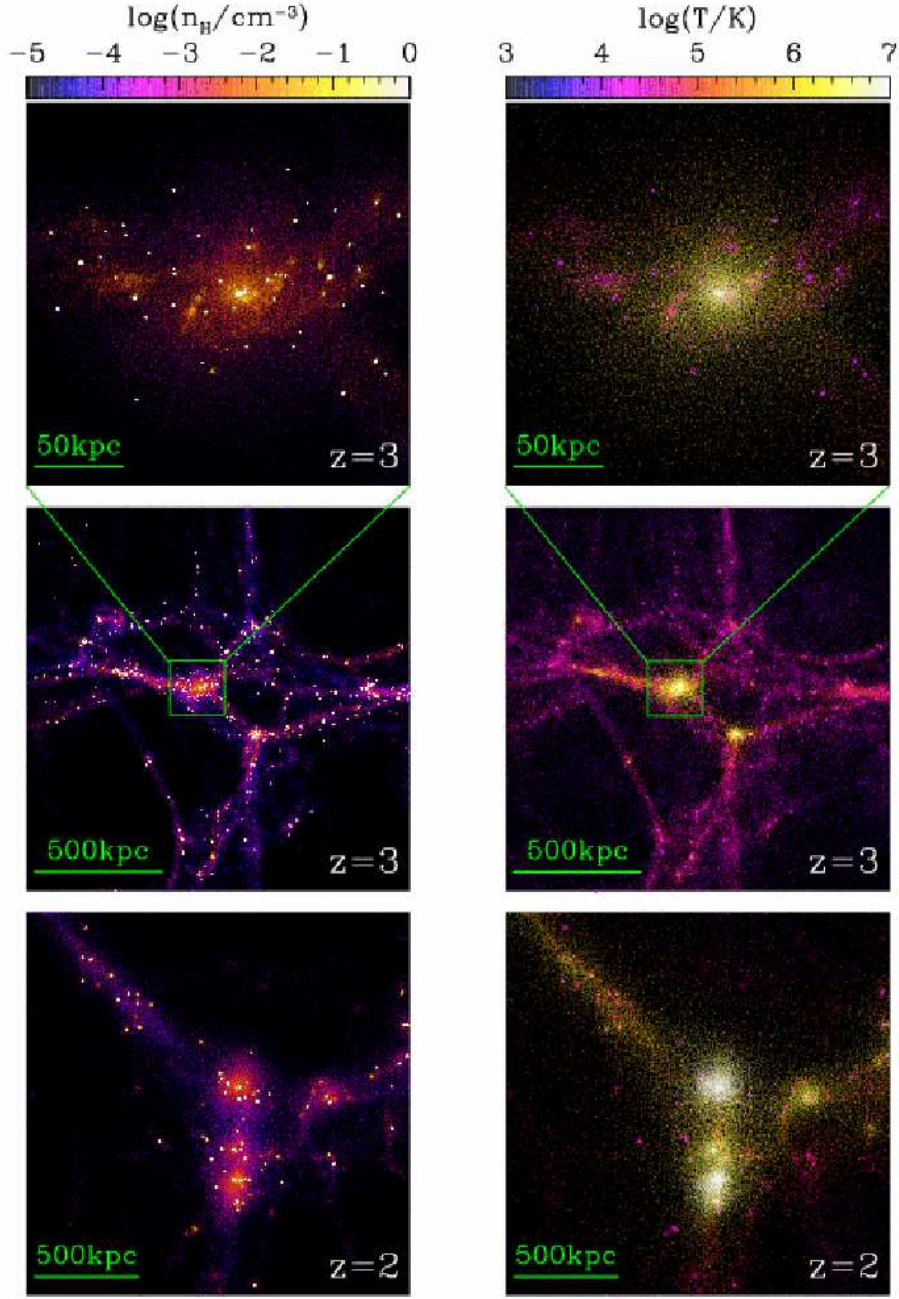}
\caption{\singlespace Physical properties of the cosmological volumes analyzed.
Gas density is shown on the left and gas temperature on the right.
The middle panels are for the 1.5 Mpc (physical) sub-region of the
5.555 \hmpc\ (comoving) simulation box at $z=3$.  Upper panels show a
200 kpc region extracted from within the region shown in the middle panels.
Bottom panels show a 1.8 Mpc (physical) sub-region extracted from the
22.222 \hmpc\ (comoving) simulation box.}
\label{fig:simpics}
\end{figure*}

\begin{figure*}
\plotone{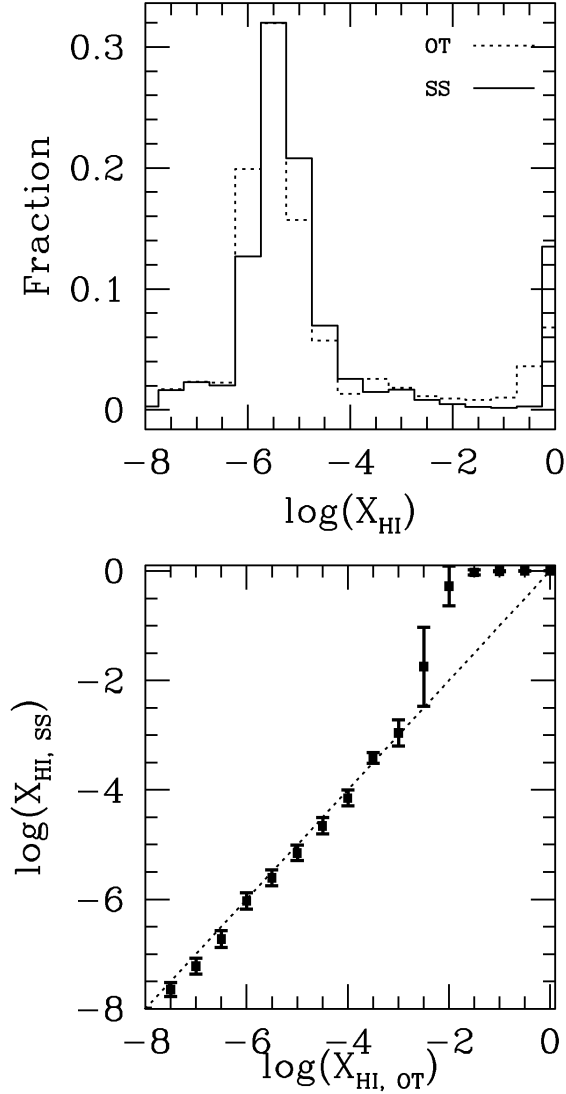}
\caption[Distribution of neutral fractions for the L5 simulation particles]{\singlespace The upper panel shows the distribution of neutral fraction ($X_{\rm HI}$) for self-shielded particles compared to the optically thin approximation in the L5 simulation.  The bottom panel shows the optically-thin (OT) versus self-shielded (SS) neutral fraction particle-by-particle. The effect of self-shielding is to move dense particles to higher neutral fractions, which is particularly important at large values of the neutral fraction. }
\label{fig:cosmo_neutralfrac} 
\end{figure*}

We have selected a 1.5 Mpc (physical) region and a smaller 200 kpc
(physical) region from the L5 simulation as well as a 1.8 Mpc
(physical) region from the L22 simulation for which we make
predictions for fluorescent \lya\ emission.  We show the total gas
density and temperature from these simulations in
Figure~\ref{fig:simpics}. Irradiating these gas structures with a
uniform ionizing background \footnote{We adopt the same UVB for $z=2$
and $z=3$.  The UVB is not expected to vary significantly between
these epochs \citep{haardt96}}, we determine the neutral fractions
for all of the particles within these sub-regions of the simulations
using the well-tested algorithms of our self-shielding correction,
which we describe in \S 2.2.  Since we now have an arbitrary geometry
and gas density distribution, we can only compare our particle neutral
fractions before and after we correct them for the effects of
self-shielding: we do not have a simple reference such as a
self-shielding radius.  We show in Figure~\ref{fig:cosmo_neutralfrac}
the resultant neutral fraction distribution after we perform the
self-shielding correction on the SPH particles compared to the case in
which all particles are exposed to a uniform UVB for the L5
simulation.  The effect of the self-shielding is seen clearly by the
shift toward higher neutral fractions in dense regions, with a
substantial number of particles becoming completely neutral. We note
that there is almost no change in the low density regions that are
optically-thin to \lya.

To preserve the high resolution achievable with SPH
simulations, we would ideally have grid sizes that were smaller than
the smallest smoothing lengths in the box.  The smallest physical
scale resolved in this simulation (i.e., the smallest particle
smoothing length) is 0.07[0.38] kpc for the L5[L22] simulation.  For a
1.5[1.8] Mpc region, this would correspond to an unrealistically high
resolution grid of $(2.1\times10^4)^3$[$(4.7\times10^3)^3$] cells.
However, because they are shielded, extremely dense regions play no
role in generating photons, and the requirement of grid resolution
should be much less stringent in these regions. For our applications,
the boundaries of any dense region will simply act as ``mirrors'' for
the incoming photons. We, therefore, need to resolve dense regions as a
whole, not individual particles inside them.  We have tested the
effect of grid resolution and based upon these experiments, have
adopted a resolution of $300^3$ cells (corresponding to 5kpc and 6kpc
spatial resolution for the large regions of the L5 and L22 boxes
respectively) to make our predictions.  We adopt a spatial resolution
of $128^3$ cells (corresponding to 1.6kpc resolution) for the small
sub-sub region of the L5 box.  The details of these tests can be found
in Appendix B.

\begin{figure*}
\plotone{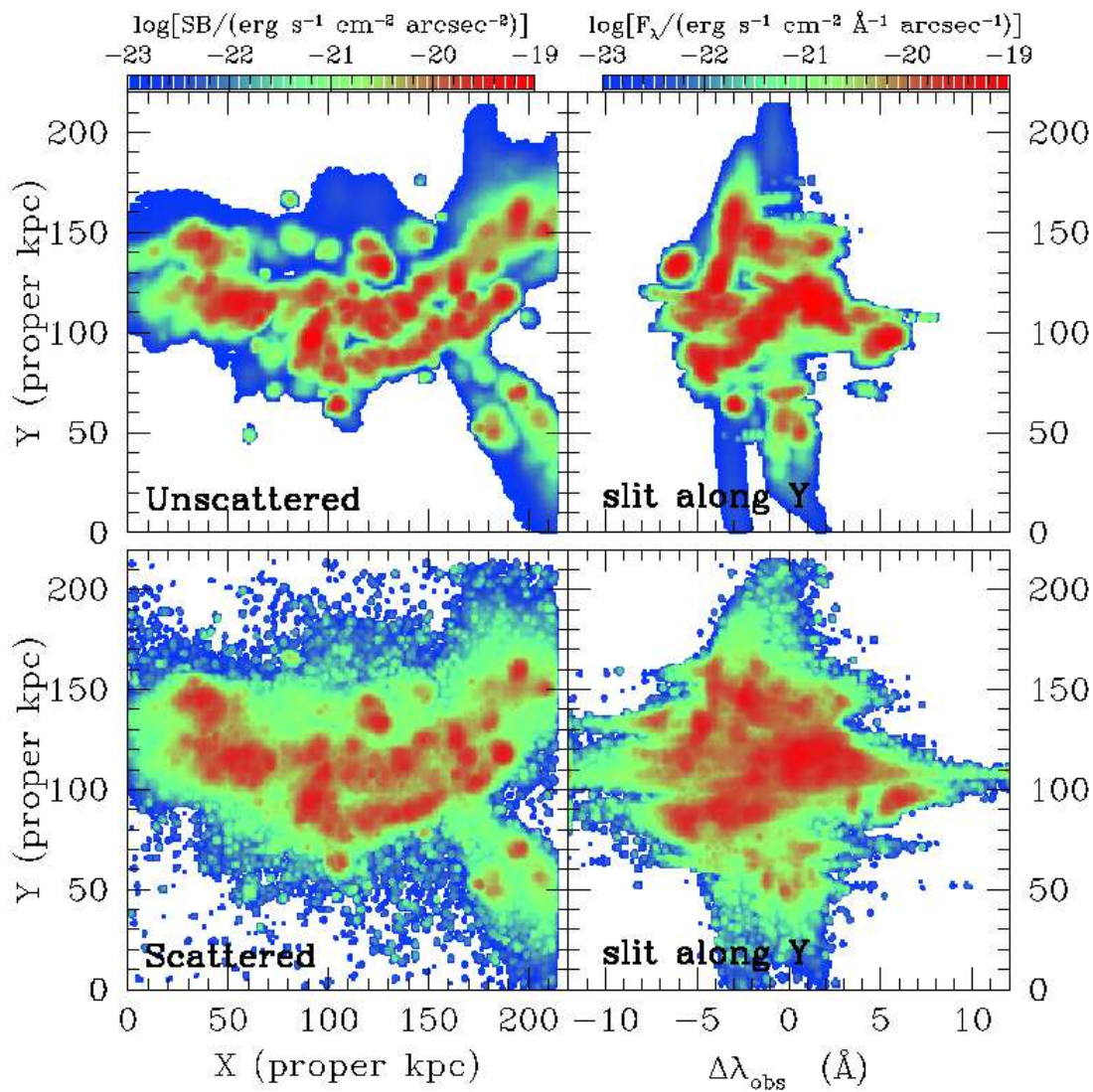}
\caption[\lya\ fluorescence at $z=3$ from 200kpc subregion of cosmological simulation 
L5.]{\singlespace \lya\ map of the central region of the L5 simulation.  This
projection corresponds to the upper panels in
Fig.~\ref{fig:simpics}. Left panels show \lya\ surface brightness and
right panels show the 2D spectrum with slit along the $y$-direction.  The upper
panels show the image and spectrum one obtains without radiative
transfer.  Bottom panels show the post-radiative transfer image and spectrum. }
\label{fig:lya_L5zoom}
\end{figure*}

We first examine the 200 kpc sub-sub region of the L5 simulation that
contains multiple optically-thick structures.  To explicitly examine
the effect of frequency diffusion, we generate a \lya\ map with the
radiative transfer turned {\it off}, which is the same as the
emissivity map.  The top panels of Figure~\ref{fig:lya_L5zoom} show
the unscattered \lya\ image and 2D spectrum extracted along the
$y$-direction.  The bottom panels in this figure show the image and
spectrum from the full radiative transfer calculation.  The connection
between gas density and emission is striking in this zoomed region as
one can see by comparing the upper left panel of
Figure~\ref{fig:simpics} with the \lya\ image in the bottom left panel
of Figure~\ref{fig:lya_L5zoom} . Very dense knots of gas produce a
substantially higher emission signature in both images and spectra.  

A comparison of the top and bottom panels of
Figure~\ref{fig:lya_L5zoom} directly shows the effect of the resonant
scattering.  The image in the scattered case is smeared compared to
the non-scattered case owing to the spatial diffusion of the photons.
The differences between these images shows the modest amount of
spatial diffusion that occurs in these systems.  The comparison
between the 2D spectra for both cases shows that photons diffuse
primarily in frequency space, in contrast to a spatial random walk.
In the non-scattered image, each blob gives rise to a narrow bright
line near line-center in the 2D spectrum in contrast to the
characteristic broad double-peaked line profiles.  The peculiar
velocity and Hubble flow in the gas cause frequency shifts away from
zero. When we examine the scattered image, however, we see that each
structure now gives rise to a more diffuse line profile in frequency.
While this realistic case is more complex than the idealized cases
presented in \S\ref{sec:results1}, we can still see the fingerprints
of resonant scattering on this scale.  With the appropriate scaling,
the top panel of Figure~\ref{fig:lya_L5zoom} could represent an image
in an optically thin recombination line such as H$\alpha$.  Taking the
ratio of the top and bottom panels yields an estimate of the relative
morphology for lines that are optically thin and thick, respectively.
While in high-emissivity locations, the difference is modest, in
low-emissivity regions the difference between, e.g. an H$\alpha$ image
and a \lya\ image would be quite substantial.

\begin{figure}
\centering
\small
\plotone{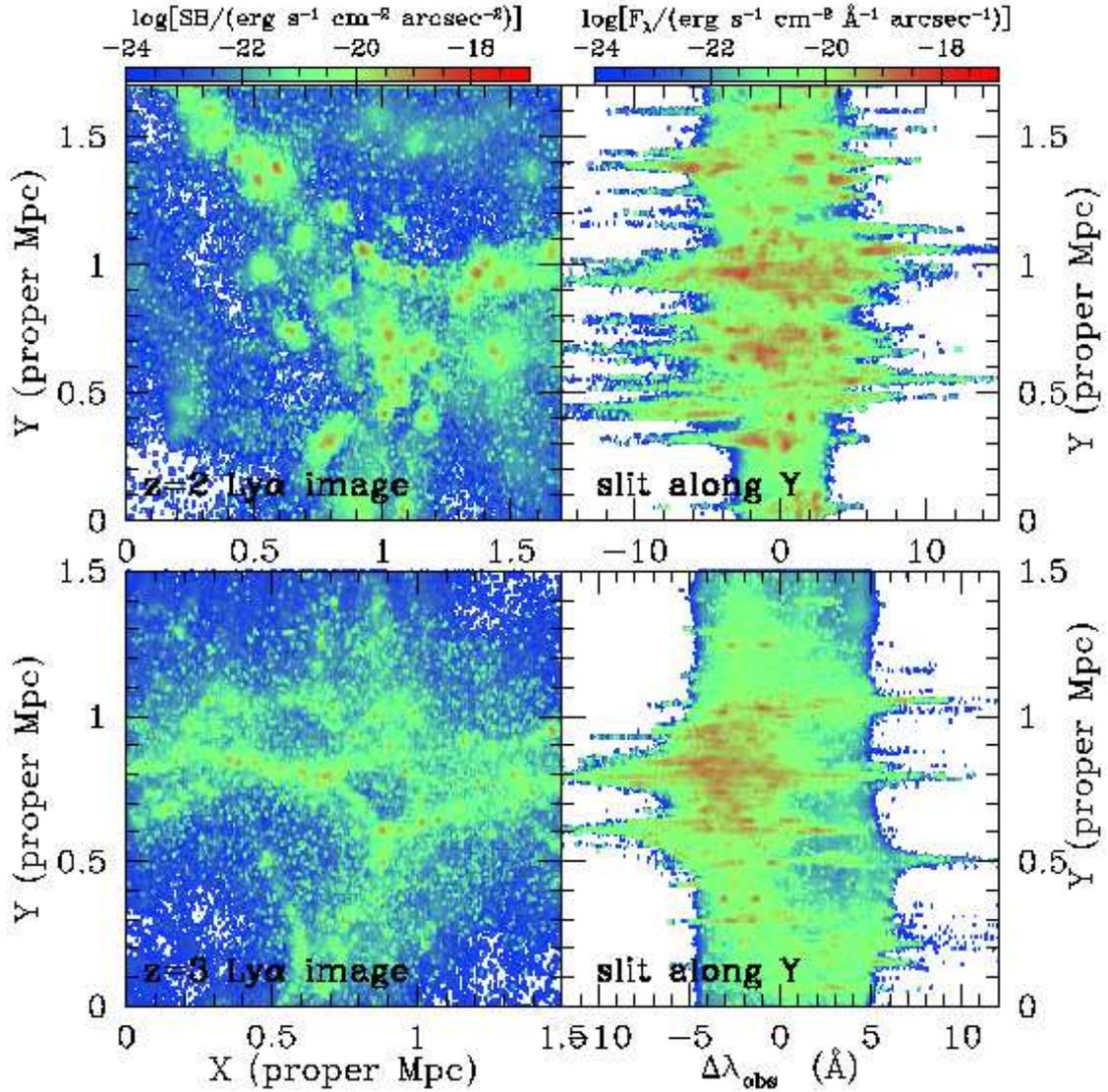}
\caption[Tada!  Fluorescence from cosmological simulations]
{\lya\ fluorescence from cosmological simulations.  
Top:  Emission at $z=2$ from the gas distribution in a
sub-region of size 1.8 Mpc from cosmological simulation L22. Left
panel shows the \lya\ surface brightness and right panel shows the 2D
spectrum with slit along the $y$-direction.
Bottom: Emission at $z=3$ from a sub-region of size 1.5 Mpc from cosmological
simulation L5. Note the surface brightness scale is different from 
Figure~\ref{fig:lya_L5zoom}.}
\label{fig:lya_sims} 
\end{figure}

We now examine the larger regions of the L5 and L22 simulations in
which blobs of the size shown in Figure~\ref{fig:lya_L5zoom} are only
a small portion.  We show in Figure~\ref{fig:lya_sims} the \lya\
images and 2D spectra generated from the large L5 and L22 simulation
regions shown in Figure~\ref{fig:simpics} (with different surface
brightness scale).  We show the results for both
simulation boxes in Figure~\ref{fig:lya_sims} to facilitate
comparison between the $z=2$ (top panels) and $z=3$ (bottom panels)
cases from the L22 and L5 simulations, respectively.  In contrast to
the isothermal-sphere case and owing to the highly disturbed gas
density and velocity structure, we do not have clean diagnostics of
characteristic radii and analytic expectations for the separations of
spectral features.  However, a comparison of the panels of
Figure~\ref{fig:lya_sims} and the bottom panels of
Figure~\ref{fig:simpics} shows that the emission primarily originates from
dense knots of material, as one would expect, since the emission
comes from the rapidly recombining skins of optically thick cores,
which should occur in regions of high density.  Since we have adopted
the same UVB for both redshifts, the difference in emission intensity
between the images is caused primarily by the cosmological surface
brightness dimming, which is $\propto (1+z)^{-4}$.  

The 2D spectra are quite complex as can be seen in the right panels of
Figure~\ref{fig:lya_sims}.  As we showed in detail in
Figure~\ref{fig:lya_L5zoom}, each blob of optically thick gas gives
rise to a disturbed version of the characteristic double-peaked
profile.  Since we have a large number of such blobs, the resultant
spectra are a superposition of many of these profiles, each modified
by the fluid velocity field and the geometry (e.g. the frequency
distribution may not be symmetric about the line center).  The
profiles shift with respect to each other because of the Hubble flow
and the peculiar velocity of the gas.  The complex structure seen in
the 2D spectra is, therefore, generically expected owing to transfer
effects.  It is also worth noting that, as one approaches sufficiently
low surface brightness limits, the ``forest'' of Lyman Limit systems
illuminated by the UVB emerges.  That is, many systems become visible
at a similar surface brightness level --- once this threshold in
surface brightness is achieved, the filamentary structure is evident
i.e., once one LLS is observable, many others are also visible.  

\begin{figure*}
\plotone{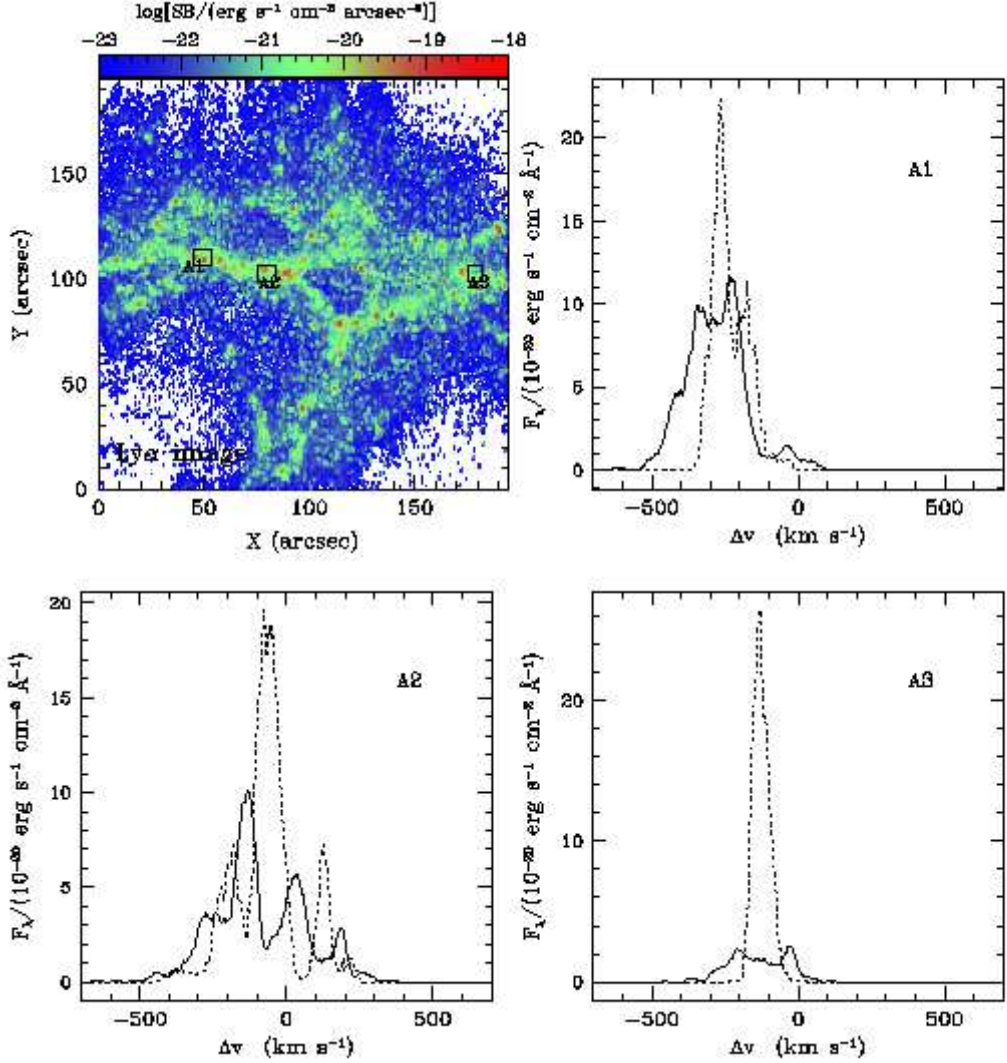}
\caption[1D spectra from \lya\ emitting clumps in cosmological
simulation]{\singlespace1D spectra from multiple apertures throughout the $z=3$ structure in Figure~\ref{fig:lya_sims}.  The top-left panel shows the \lya\ image (as in Figure~\ref{fig:lya_sims}) with 3 square apertures ($7.5\asec \times 7.5 \asec$) overlaid.  The other three panels show the 1D spectra from the three apertures, respectively. Solid lines in the figure show the post-transfer spectra and dotted lines show the case in which the photons are not scattered. }
\label{fig:lya_cosmo1d} 
\end{figure*}

In Figure~\ref{fig:lya_cosmo1d}, we show several 1D spectra from blobs
of gas in different regions of the structure in the L5 box. This is
analogous to obtaining a narrow-band image for a particular field and
follow-up spectroscopy of the identified sources. Solid lines in the
figure show the post-transfer spectra and dotted lines show the
``unscattered'' spectra.  The overall double peaks owing to the
transfer of photons are clear in some cases (e.g. A3 and A1) but
unlike the isothermal case, they are not symmetric owing to the bulk
velocity of the gas in the simulations.  While comparison of any
individual system requires detailed modeling, characteristic
line-widths and morphologies from large-volume calculations such as
these could be derived.  Observational campaigns to obtain deep
narrow-band imaging and follow-up spectroscopy are currently underway
and comparison with calculations such as these will prove helpful for
understanding the origin of the \lya\ emission.  We discuss the
requirements for observing such fields in \S\ref{sec:observables}.

\subsection{Large Scale Structure in a Quasar Radiation Field}

\begin{figure*}
\plotone{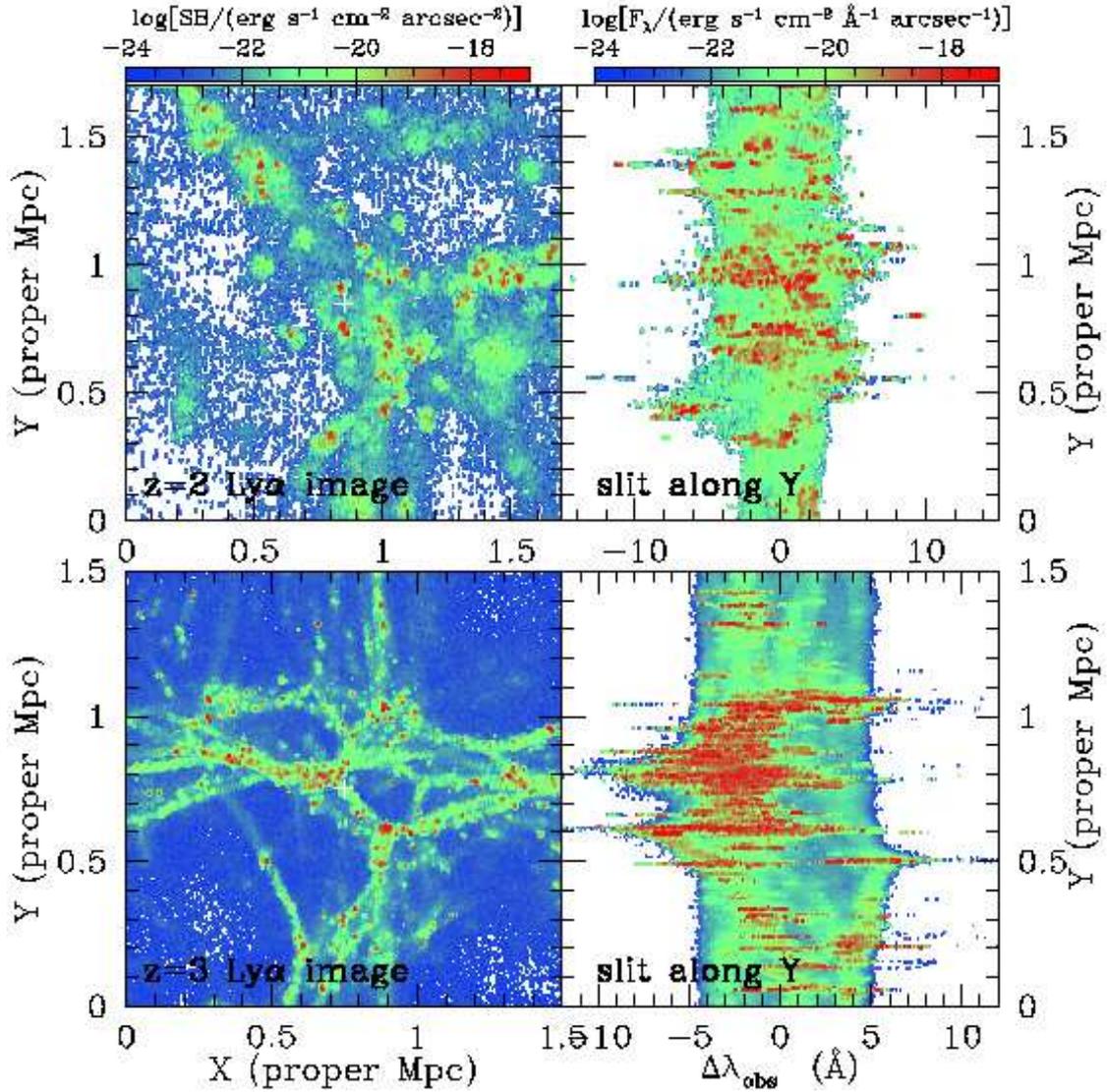}
\caption[Effect of quasar on \lya\ emission from cosmological simulation region]
{\singlespace \lya\ maps including a quasar source.  Panels are as in
Fig.~\ref{fig:lya_sims} but we have now placed a bright quasar with Lyman
limit luminosity  $L_{\nu_L}=1.0\times 10^{32}\; {\rm erg\
s^{-1} Hz^{-1}}$ at the center of the region (marked with crosses).  Left panels show the
\lya\ surface brightness on the sky for redshifts $z=2$ (upper panels)
and $z=3$ (lower panels).  Right panels show the 2D spectrum with slit along the
$y$-direction from the images shown in the left. }
\label{fig:lya_simsQ}
\end{figure*}

Similar to the case discussed in \S\ref{sec:iso_qso}, we now place a
bright quasar with $L_{\nu_L}=1.0\times 10^{32}\; {\rm erg\ s^{-1}
  Hz^{-1}}$ (10 times brighter than for the previous SIS case) in the
center of our cosmological sub-regions to examine the effect on the
\lya\ emission strength and morphology and to show the expectations
for observing a field containing a bright quasar.  The quasar will
have little effect on the emission from gas that is already highly
ionized by the UVB, but will have a substantial effect on the emission
from the dense optically thick clumps that are not already
significantly ionized by the UVB.  We see this in
Figure~\ref{fig:lya_simsQ} where we show the resulting \lya\ image and
2D spectra for the L22 (top panels) and L5 (bottom panels) simulations
including a quasar.  Comparing with Figure~\ref{fig:lya_sims} one can
see that in the dense knots, the emission is brighter by more than an
order of magnitude relative to the UVB-only case.  We note that even
in the outer reaches of this system, there is a substantial {\it
  relative} increase in the surface brightness -- the quasar is
sufficiently powerful to reach the edges of this region.

\begin{figure*}
\plotone{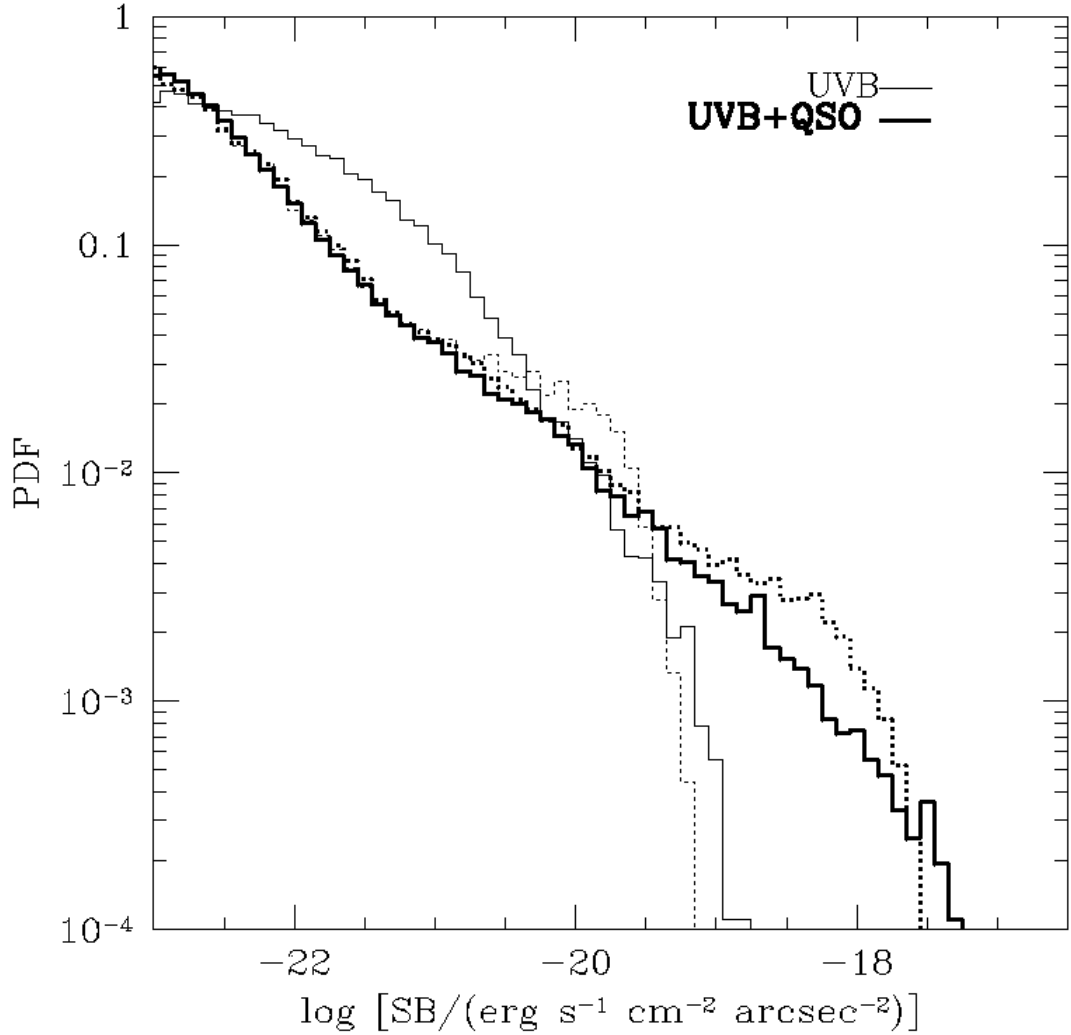}
\caption[Surface brightness distributions for UVB and UVB+QSO cases]
{\singlespace Distributions of resultant \lya\ surface brightness of
  pixels for the UVB case and the UVB+QSO case.  Solid (dotted) thin
  lines show the UVB-only case and solid (dotted) thick lines show the
  UVB+QSO pixel distribution after (before) \lya\ radiative transfer.
  The shift toward higher surface brightness pixels results directly
  from the quasar radiation impinging on the dense, optically thick
  clouds in the simulation. }
\label{fig:pixels_cosmo_qso} 
\end{figure*}

In Figure~\ref{fig:pixels_cosmo_qso} we quantify the shift to higher
surface brightness in the distribution of pixels in the resulting
\lya\ map caused by the QSO illumination.  In the UVB-only case, the
brightest pixels were set by the intensity of the UVB.  Now, the
brightest pixels are determined by the quasar flux impinging on the
densest regions.  The faintest pixels come from low column-density
material that is highly ionized.  There is, therefore, little
difference at low emissivities between the UVB and UVB+QSO case since
these systems already emit near maximum.  The brightest systems are
those that are able to remain optically thick in the presence of the
vastly increased ionizing flux of the quasar.  The UV photon
enhancement caused by the additional ionizing flux of the quasar is
$\approx 500 (d_q/1 {\rm Mpc})^{-2}$.  In our calculations,
the brightest pixel is now $\sim 2\times10^{-17}  {\rm erg\ s^{-1}\ cm^{-2}\ arcsec^{-2}}$ , a factor of $\sim$400 over the UVB-only case.
The enhancements in the average pixel surface brightness relative to the UVB-only case are shown in Figure~\ref{fig:pixels_cosmo_qso}.  On average, the resultant enhancement over the UVB is more modest.  This is due to the geometry of the emission which can result in a factor of 2-4 suppression of the expected \lya\ surface brightness. 

Most striking is that the presence of the quasar highlights the
morphology of the densest knots in the large scale density
distribution.  This is the ``meatball'' topology of \lya\ emission
referred to in GW96.  The quasar brings the contrast between the
emission from optically thick and optically thin sources into sharp
relief since only the densest knots are able to reprocess the
increased ionizing radiation.  The lower surface brightness material,
which appears spatially extended and fluffy, becomes sinewy in the
presence of the quasar.  This corresponds to material that was
previously partly neutral, and has become fully ionized in the
quasar's radiation field.  Hence, its fluorescent emissivity has
decreased since the emissivity of the gas is proportional to the
photoionization rate, which is small for gas with very low neutral
fractions.

The 2D spectra for the QSO case also highlights the densest systems.
The brightest knots have narrower frequency distributions compared
with their counterparts in the UVB-only case.  The QSO has completely
photoionized many low-emissivity structures, eliminating their
contribution in both the image and the spectra.  The highest
column-density systems are relatively smaller and brighter
and, because the quasar radiation has generally lowered the neutral
column-density of the gas, the spectral pattern is narrower in
frequency since the photons undergo smaller frequency diffusion at lower
column-densities.  The double-peaks of the highest density systems
become much more prominent in the presence of the quasar's radiation.
For relatively isolated blobs, one can see clearly the double-peaked
spectral feature associated with the object in the 2D spectrum
(e.g. the systems located at $(X,Y)=(0.8,0.3)$ and $(0.8,1.3)$).

Fluorescent \lya\ emission in quasar fields should have a very
different morphology as a function of luminosity compared to
fluorescence from the UVB alone.  Bright knots of emission from
high-column density dominate over the general emission from the IGM.

\section{Observables \label{sec:observables}}

We now quantify the results presented in \S\ref{sec:cosmo} to demonstrate
how such predictions can be explicitly compared with current narrow-band
imaging surveys \citep[e.g.][]{steidel00, ouchi05}.  We will first go over some
analytic expectations and see how these relate to our more detailed
calculations.  Then we present simulated images and extract sources from those
images.

\subsection{Analytic Considerations \label{sec:observability}}

What are the realistic prospects for observing \lya\ emission from the
IGM from the ground and from space?  A useful figure
of merit is the amount of observing time required to reach a fixed
S/N.  We gain some insight into the practical difficulty of this
problem by considering the simple mirror approximation as cast by
GW96 who obtained the following expression for S/N as a function of
observing time:

\begin{eqnarray}
 S/N & = & \xi \frac{\Phi_{\rm obs}}{(1+z)^{1/2}} \left[ \frac{\pi^{1/2} D^2 f T \Delta\Omega} {16 (\sigma/c) \phi_{\rm sky} \lambda_{ Ly\alpha }} \right]^{1/2} 
\\
  & \sim & 7.5 \left( \frac{1+z}{3.2}\right)^{-3.5} \left( \frac{f}{0.25}\right)^{1/2}\left( \frac{D}{10 {\rm m}} \right) \left(\frac{T}{20 {\rm hr}} \right)^{1/2} \left ( \frac {\Delta\Omega}{10 {\rm arcsec^2}} \right )^{1/2} \left(\frac{\sigma}{35 {\rm km\; s^{-1}}} \right)^{-1/2} 
\end{eqnarray}
where $T$ is the integration time, $D$ is the telescope diameter, $f$
is the telescope efficiency, $\phi_{sky}$ is the flux from the sky,
$\Phi_{\rm obs}$ is the source flux, $\sigma$ is the velocity
dispersion of the source, $\xi$ is the atmospheric transmission, and
$\Delta\Omega$ is the source size.  In equation~(9) we have assumed
$\xi=0.9$ and $\phi_{\rm sky}=1.85 \times 10^{-2} ({\rm \gamma\ s^{-1}
  m^{-2} arcsec^{-2} \AA^{-1}})$, corresponding to a $B$-band surface
brightness $B=22.2$.  The strong $(1+z)$ scaling in equation~(9)
arises from the redshift dependence of $\Phi_{\rm obs}$, assuming that
the UVB intensity is constant with redshift. Using their values for
the ionizing background, telescope setup and source size, they
inferred that the IGM \lya\ fuorescence from the UVB would be
marginally detected in $\sim$ 20 hours.

However, this prediction may be optimistic in several respects.  The
lower amplitude and flatter shape of the UVB we adopt, cause our value
of  $\Phi_{\rm obs}$ 
to be lower than the GW96 value by a factor of 3.11,
which leads to an increase in observation time by a factor of 9.7.
The typical source sizes for the brightest fluorescent sources in the
simulation are not $10~ \rm arcsec^2$, but rather more like $4 \rm
arcsec^2$, causing a factor of 2.5 increase in observing time. This
already implies typical observing programs of ~500 hours instead of
the 20 hours GW96 obtained for marginal detection.  GW96 also adopted
a matched filter that is approximately 3 times narrower than even a
very narrow 10 \AA\ filter, causing another factor of ~3 increase in
observing time.  GW96 further assume that sources have a Gaussian profile and that
observations would reach a frequency resolution of $\sim 1$ \AA. These
observations would be very powerful for detecting low-level \lya\ emission.  If the source sizes in our
simulation and our adopted UVB intensity are correct, then, in the
presence of the terrestrial night sky, it will require $\sim$ 1500
hour exposures to detect the typical sources of fluorescence from the
uniform UVB with current ground based telescopes.  

This probably explains why fluorescence from the general IGM has been
difficult to detect with the large 100 hour programs currently
completed \citep{rauch08}.  In space, the sky background is 1
magnitude fainter and integration times can be longer.  Future
dedicated space-based and ground-based facilities will be ideal for
detecting the glow of the IGM.  In the near term, however,
observations in quasar fields where the ionizing flux can easily be
1500 times the uniform UVB are feasible in only hours on 10m class
telescopes.  As we showed in \S\ref{sec:cosmo}, the morphology of
emission near quasars highlights the ``forest'' of Lyman limit
systems.

We now simulate observational maps of \lya\ fluorescent emission from
our cosmological simulations.

\subsection{Lyman $\alpha$ Maps}

To mimic narrow band \lya\ observations, we add a background
of sky photons and Poisson noise to our theoretical predictions for a
specific observational setup.  We fix the telescope aperture,
integration time, narrow band filter width and telescope efficiency to
make ``exposures'' of our theoretical structures.  We subtract the
background from these frames --- simply taken to be the minimum pixel
count --- to create ``sky-subtracted'' images from our predicted \lya\
images. We then convert our image files to standard observational
image format and use the Source Extractor program \citep{Bertin96} to
identify sources from our image, just as would be done for an observed
narrow-band image.  We use a 10\AA\ filter on a 10m telescope aperture
with $30\%$ efficiency and integration times of 10 and 1500 hours to
generate our observed maps. 

\begin{figure}
\centering
\begin{tabular}{cc}
\epsfig{file=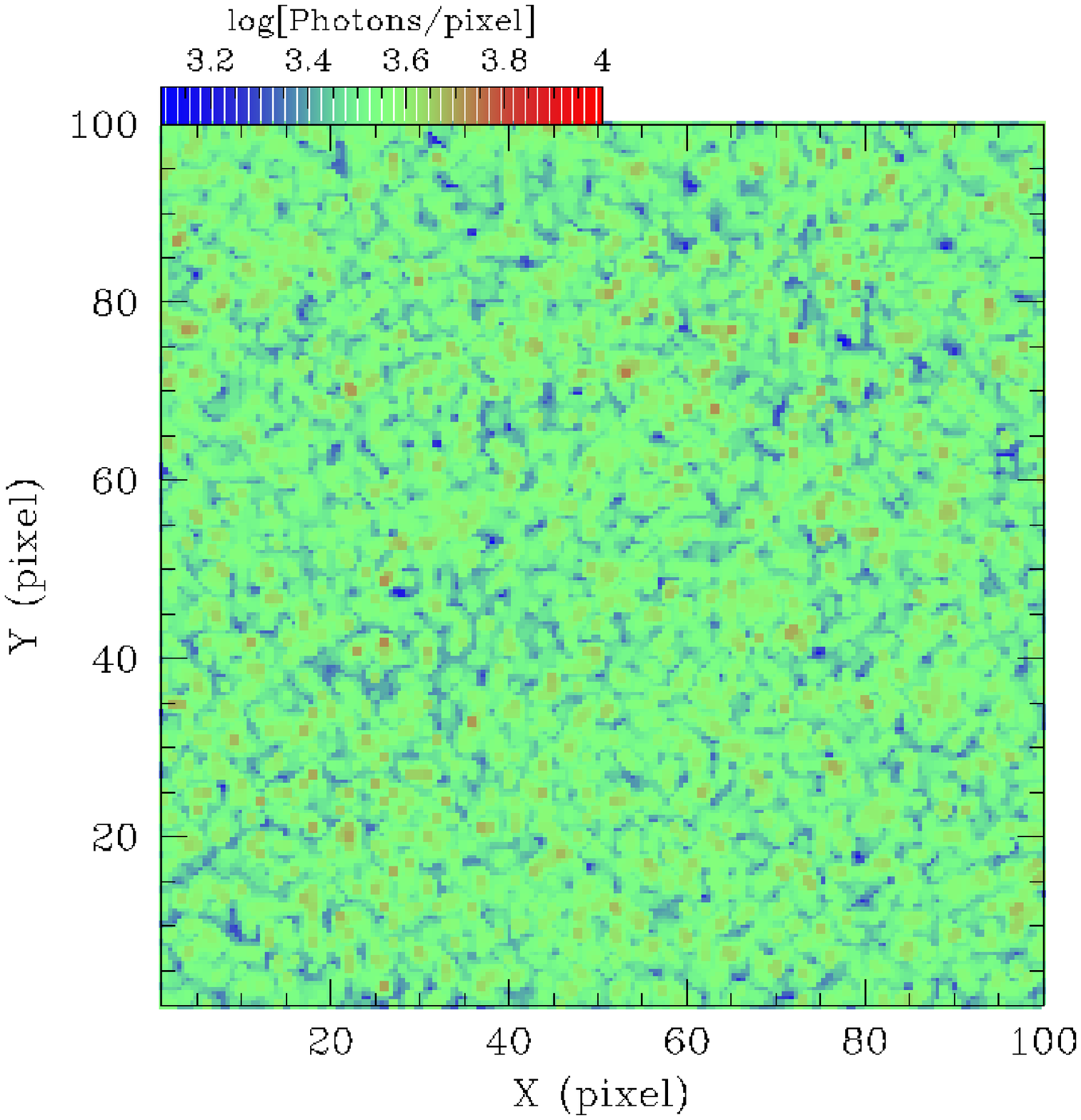,width=0.38\linewidth,clip=} & \epsfig{file=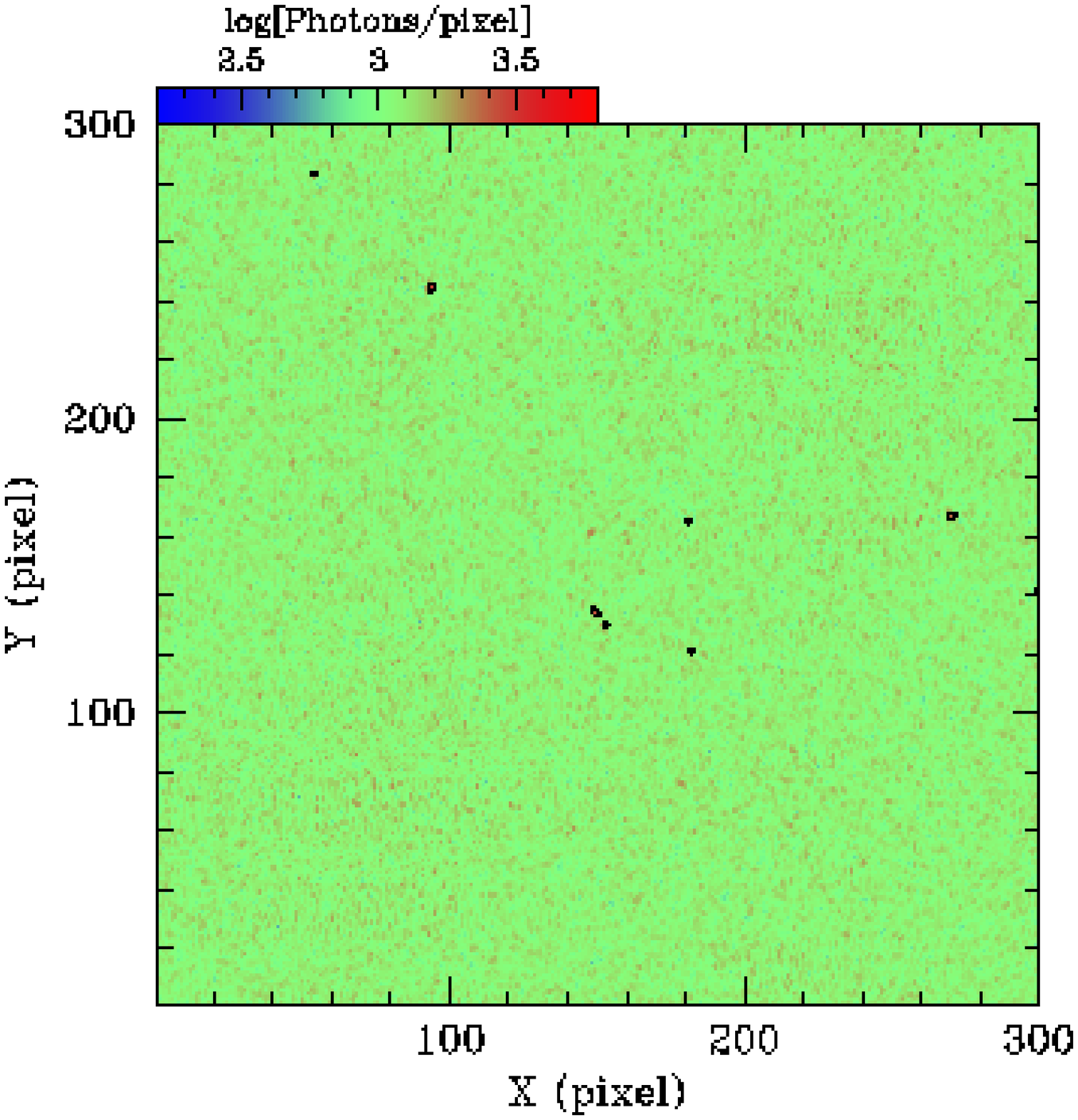,width=0.38\linewidth,clip=} \\
\epsfig{file=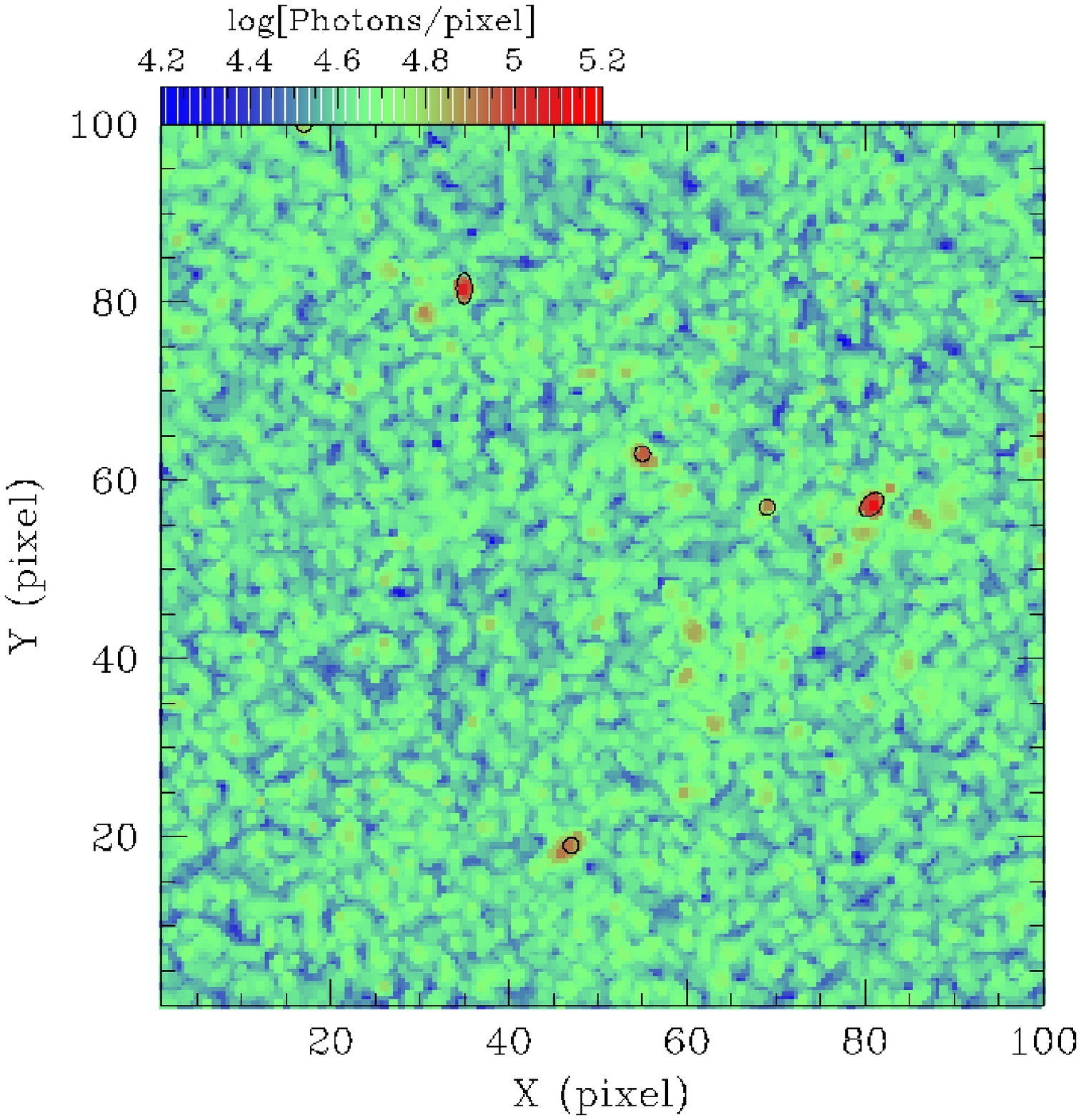,width=0.38\linewidth,clip=} & \epsfig{file=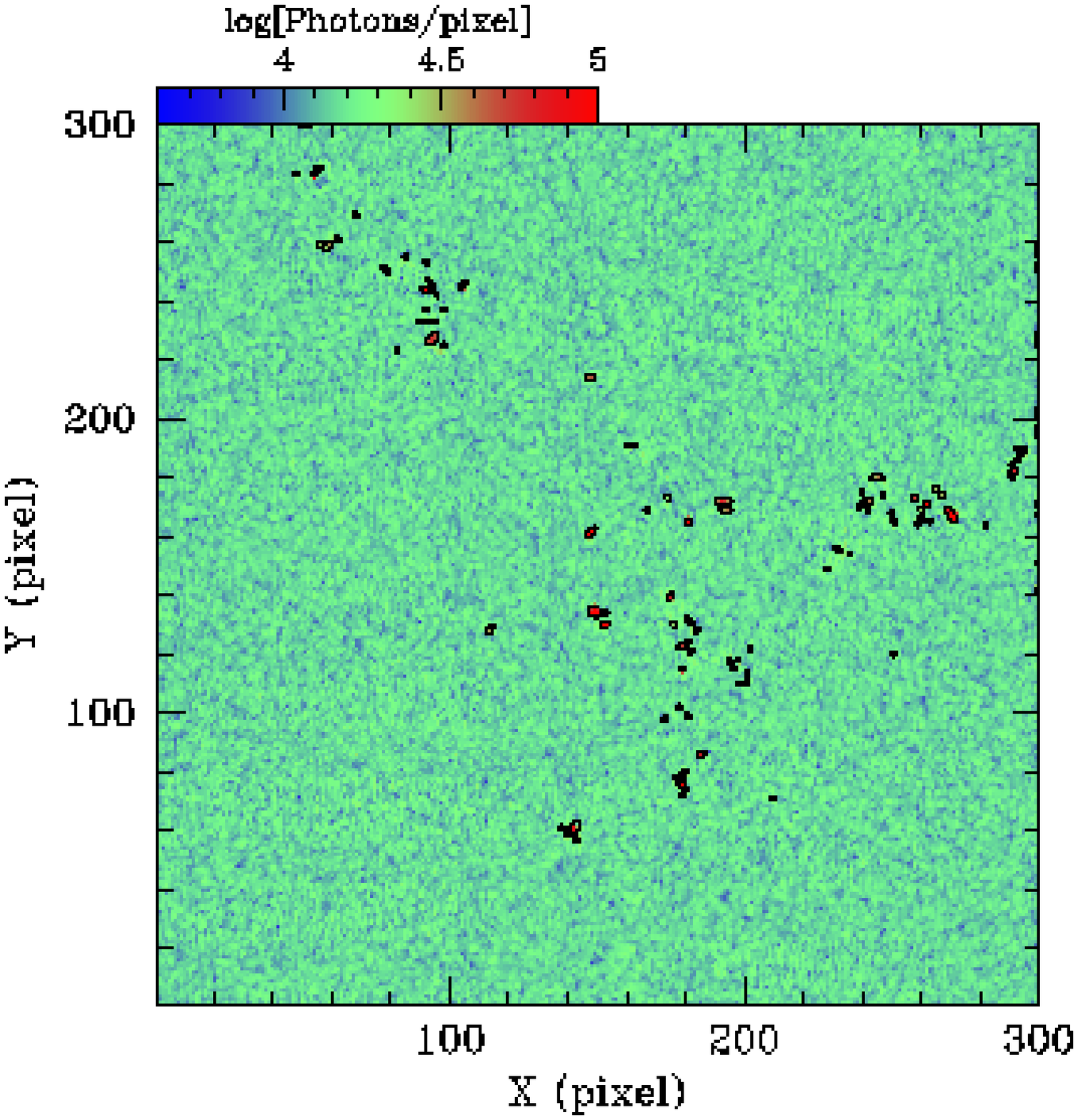,width=0.38\linewidth,clip=} \\
\epsfig{file=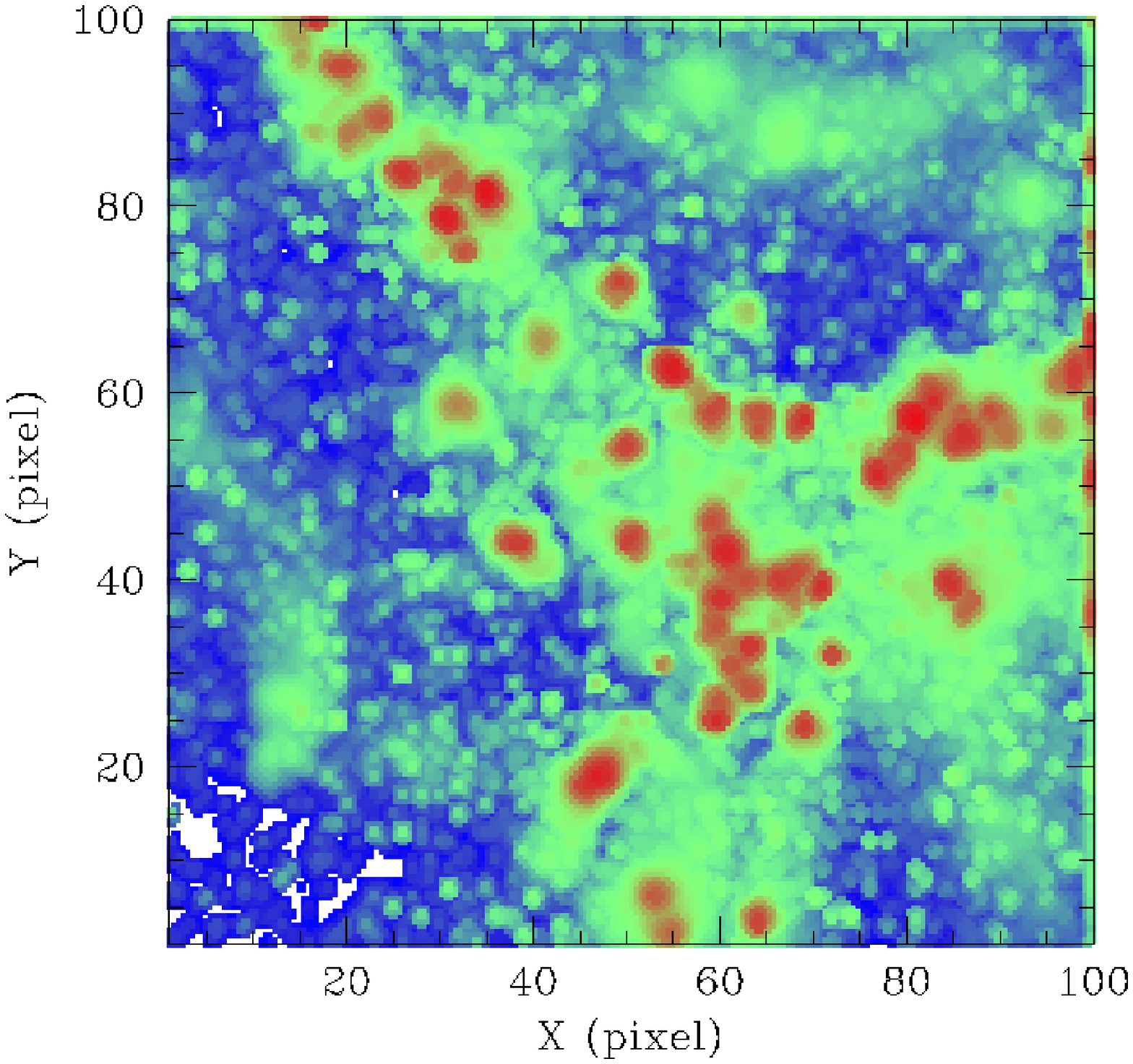,width=0.38\linewidth,clip=} & \epsfig{file=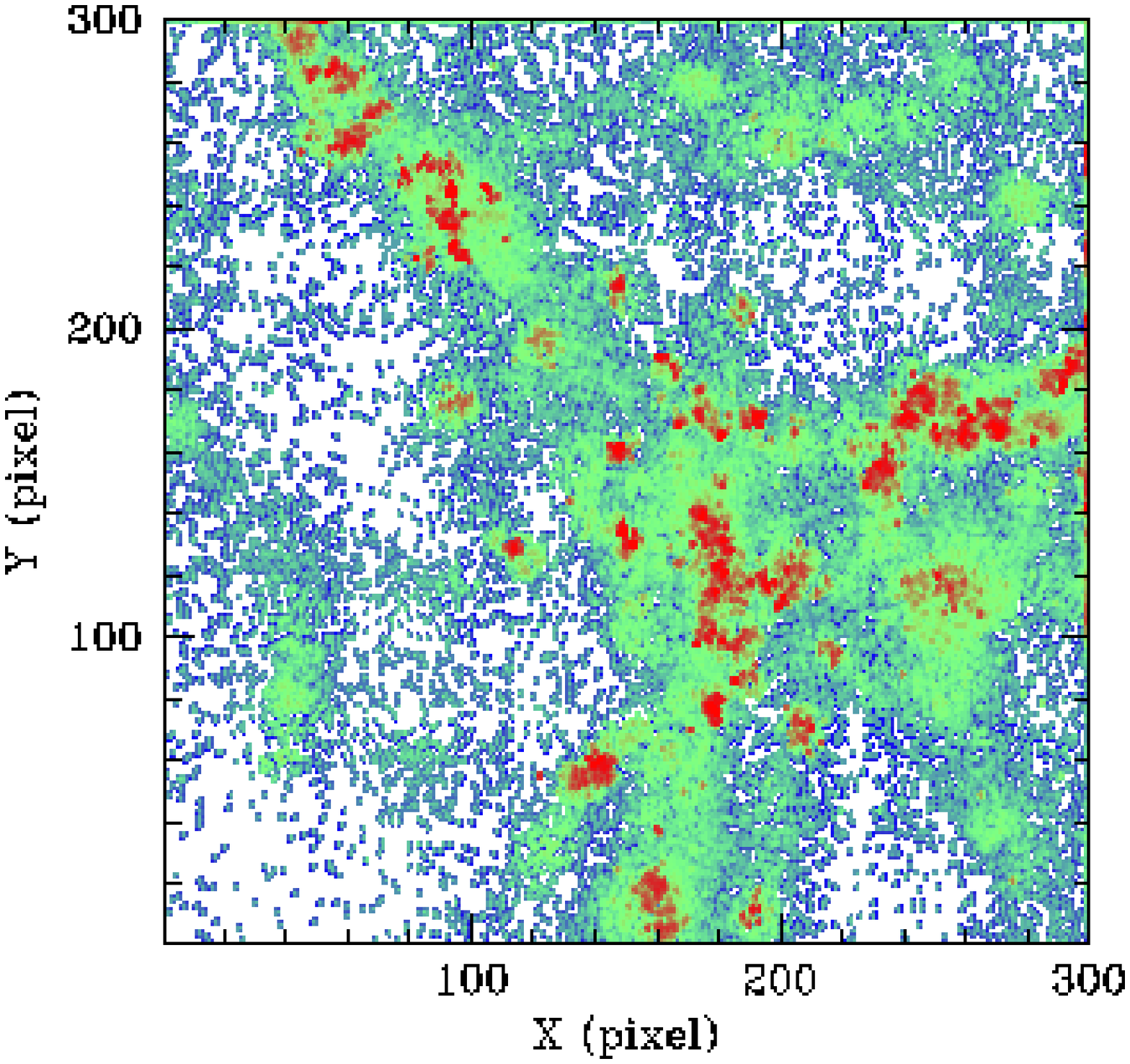,width=0.38\linewidth,clip=} \\
(a) & (b)
\end{tabular}
\caption{\singlespace \small Simulated observed maps of \lya\ fluorescence in the L22 region at
$z=2$.  Top panels show a 10 hour observation using a 10 \AA\,
filter on a 10m telescope with 30\% efficiency.
Middle panels show a 1500 hour integration.
Bottom panels show the noiseless image to aid with identifying features in the 
noisy maps.  
Left panels are the case for fluorescence from the UVB only.
Right panels are for fluorescence boosted by the presence of a bright quasar with $L_{\nu_L}=1.0\times 10^{32}\; {\rm erg\
s^{-1} Hz^{-1}}$ at the center of the region.
The pixels in these images are roughly 2x2 and 1x1 ${\rm arcsec^2}$ for the left and right panels respectively.
Ellipses in the maps show sources identified by SExtractor.}
\label{fig:noisymapsL22} 
\end{figure}

\begin{figure}
\centering
\begin{tabular}{cc}
\small
\epsfig{file=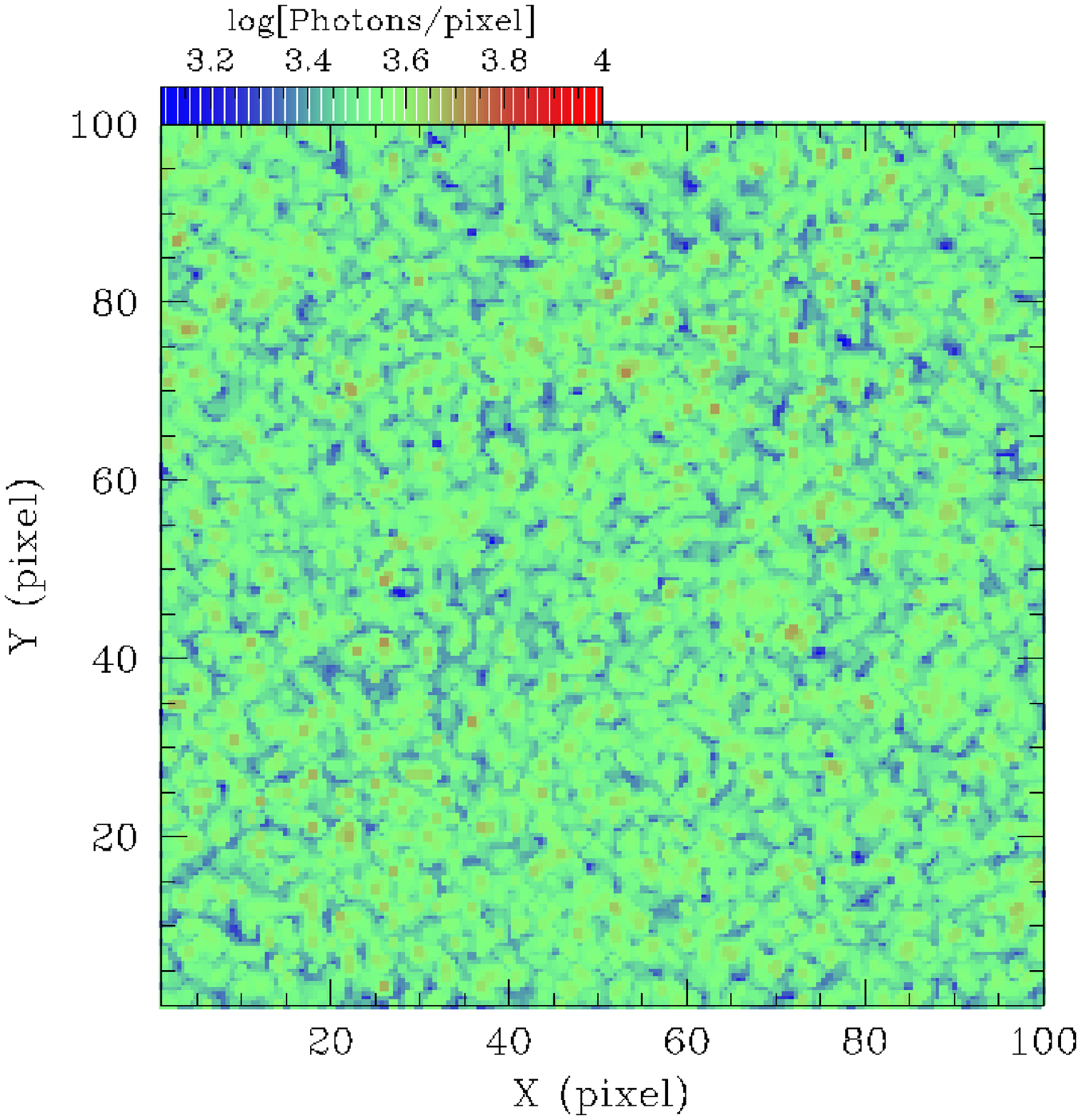,width=0.38\linewidth,clip=} & \epsfig{file=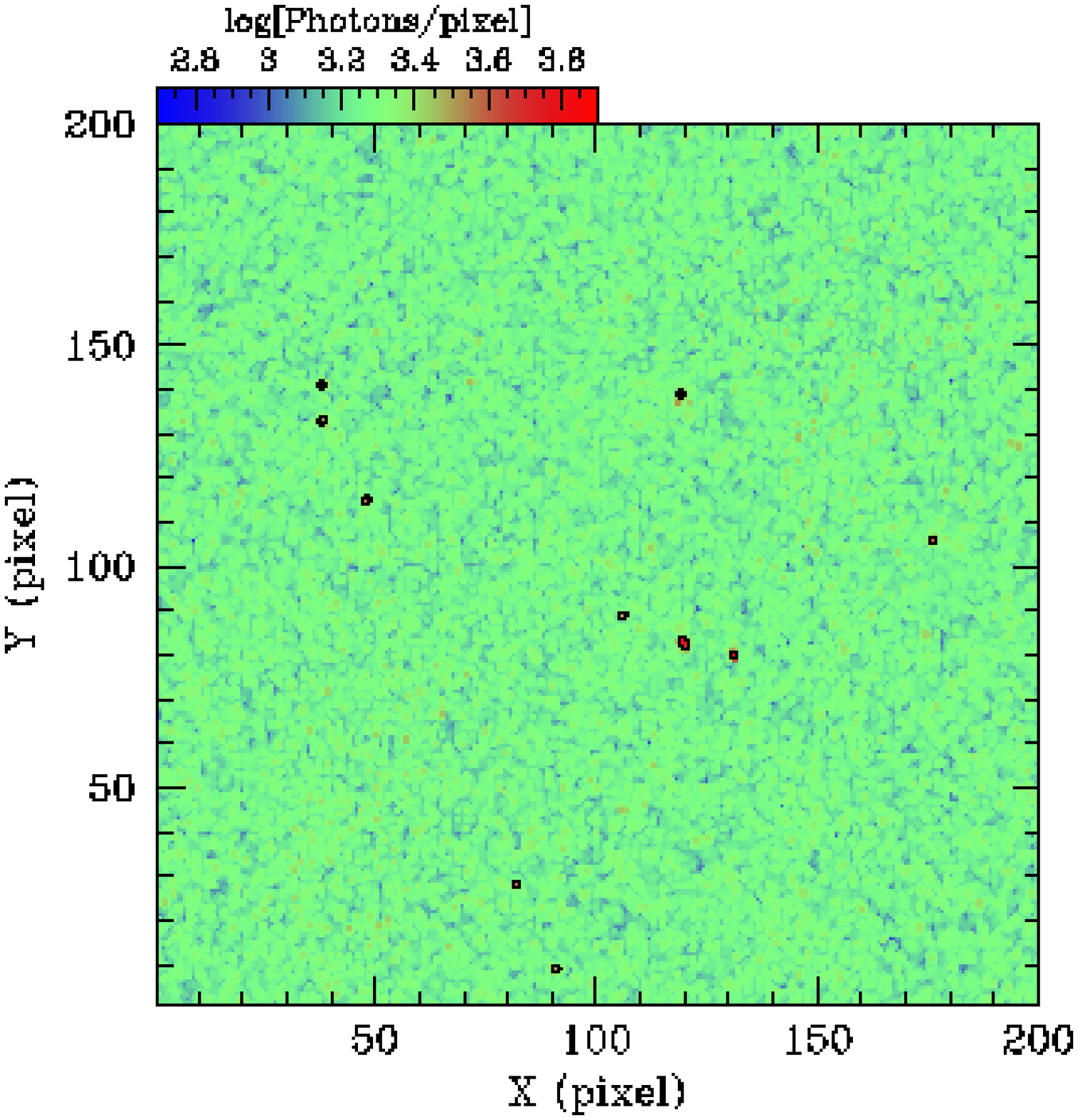,width=0.38\linewidth,clip=} \\
\epsfig{file=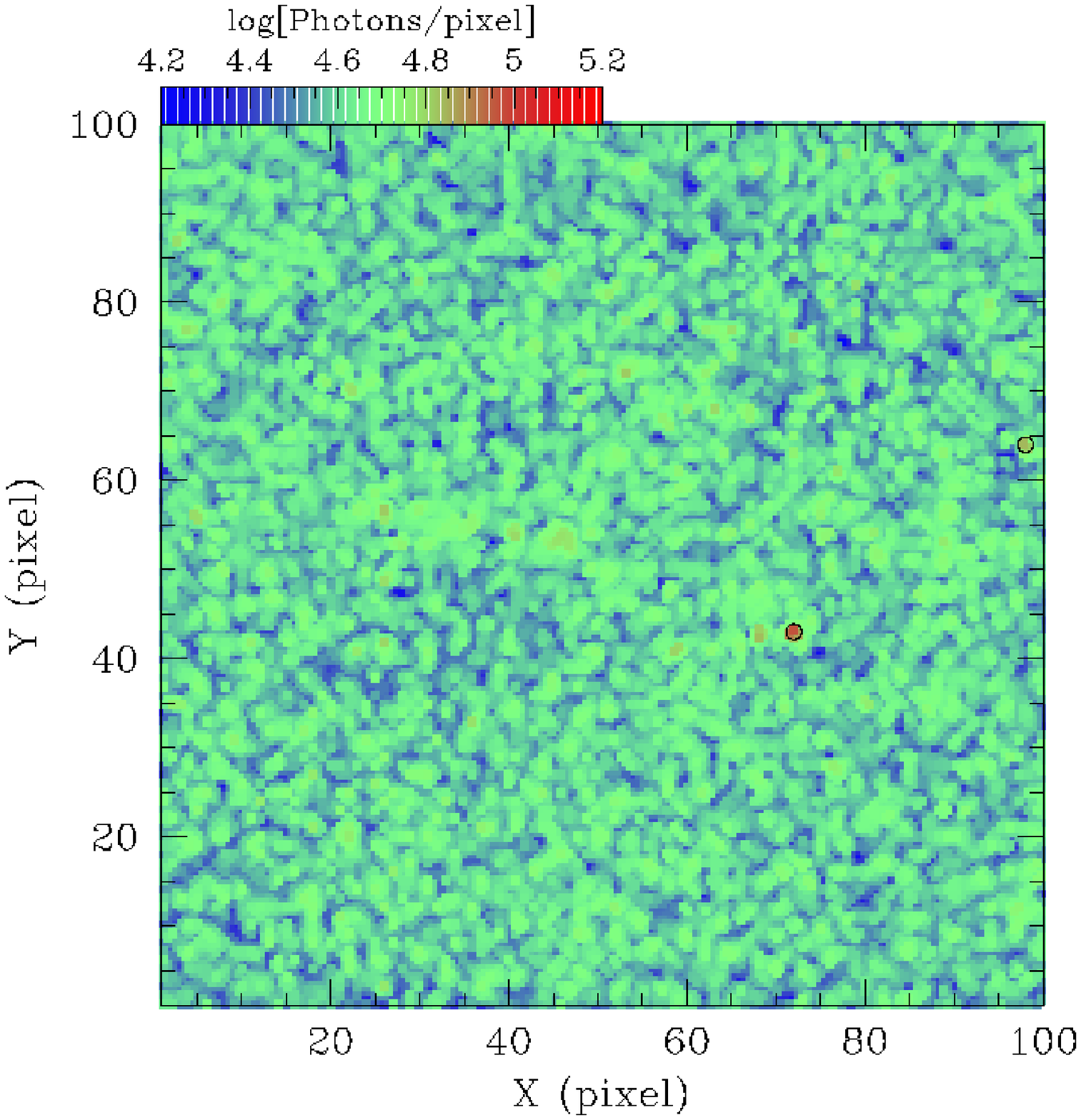,width=0.38\linewidth,clip=} & \epsfig{file=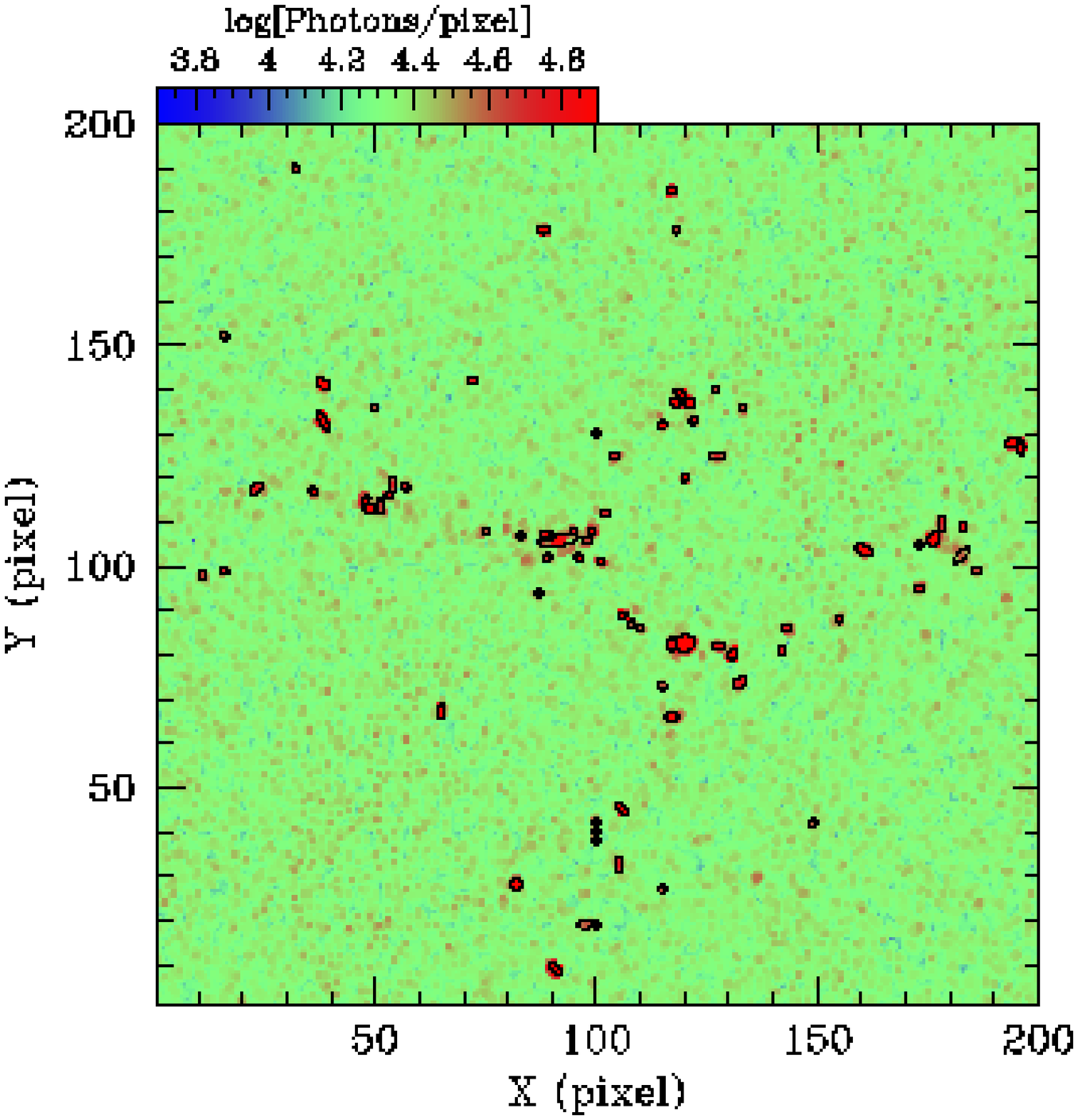,width=0.38\linewidth,clip=} \\
\epsfig{file=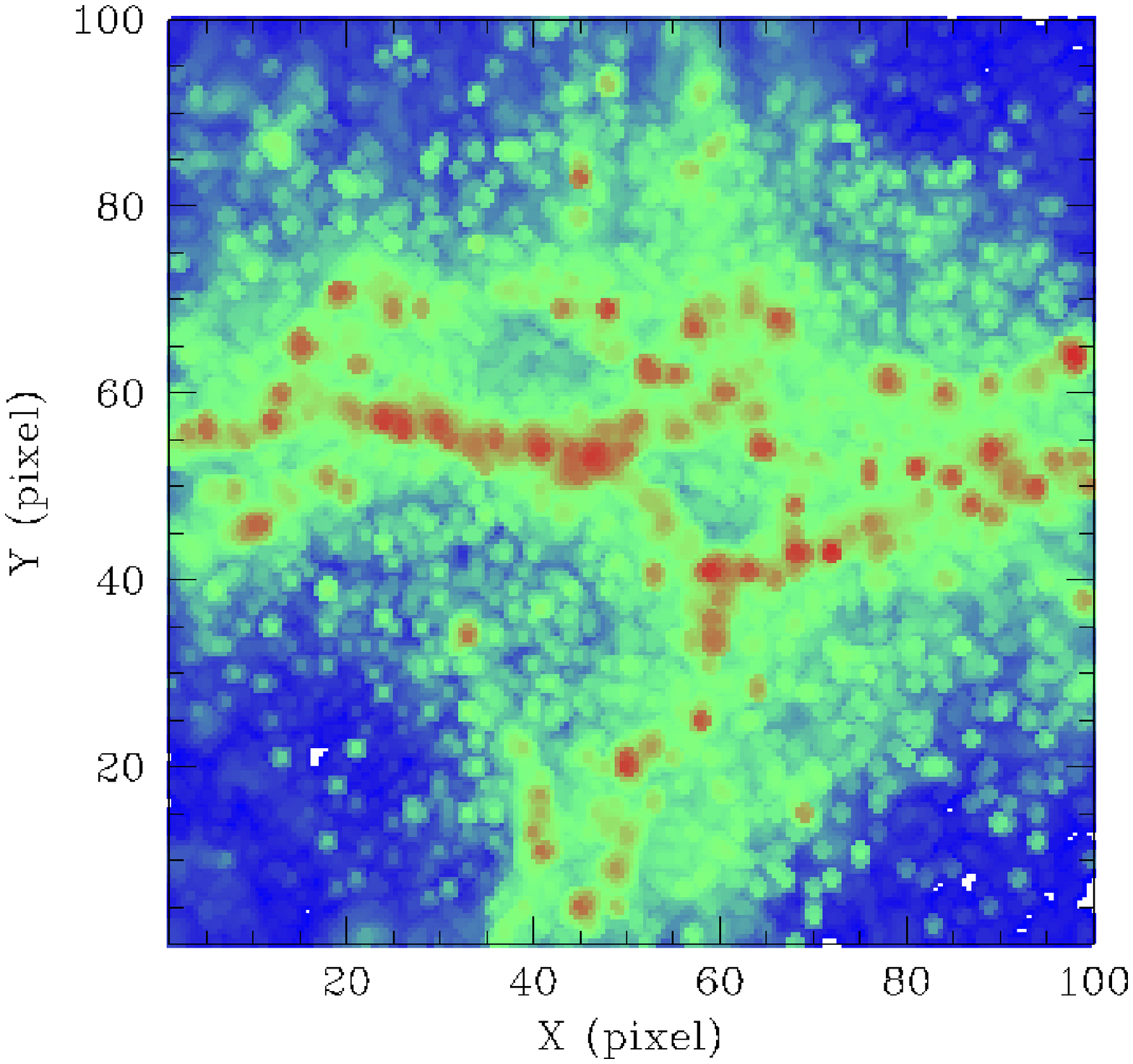,width=0.38\linewidth,clip=} & \epsfig{file=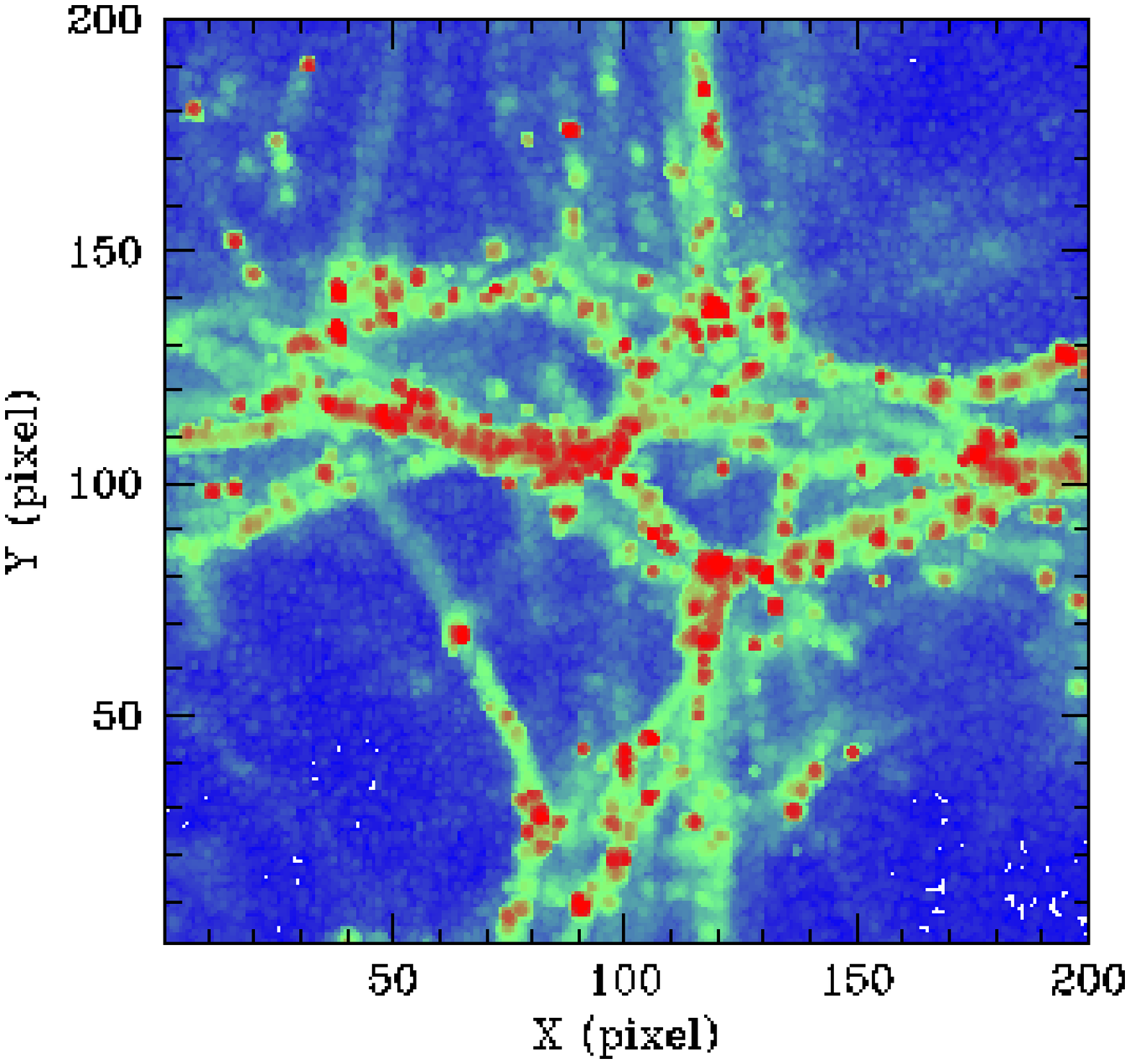,width=0.38\linewidth,clip=} \\
(a) & (b)
\end{tabular}
\caption{Simulated observed maps of \lya\ fluorescence in the L5 region at
$z=3$.
Top panels show a 10 hour observation using a 10\AA\, filter on a 10m telescope with 30\% efficiency.  
Middle panels show a 1500 hour integration.  
Bottom panels show the noiseless image to aid with identifying features in the
noisy maps.
Left panels are the case for fluorescence from the UVB only.
Right panels are for fluorescence boosted by the presence of a bright quasar.}
\label{fig:noisymapsL5} 
\end{figure}

In Figures~\ref{fig:noisymapsL22} and \ref{fig:noisymapsL5} we show
the maps created by this procedure for our simulated results at
redshifts 2 and 3 respectively.  The left panels in these figures show
the UVB only cases.  Right panels in the figures show the case of
UVB+QSO.  The difficulty of observing fluorescence is clear from the
top panels of these figures, which show the resulting maps after 10
hours of integration.  The middle panels show 1500 hour observations
and the bottom panels show the ``perfect'' case (equivalent to an
infinite exposure time) at the same resolution as the images to assist
with identifying the features.  While fluorescence from the UVB alone
is not observable in 10 hours, and only marginally detected after 1500
hours at $z=2$ and not at all at $z=3$, the quasar illuminated
structures glow brightly and with high significance after a single
night of observation.

With minimum high-resolution volumes of (5.555\hmpc)$^3$ we can begin
to examine the statistics of fluorescent sources in our models.  Such
statistics can be compared to observations from large \lya\ surveys
(e.g., \citealt{matsuda06, steidel00, cantalupo07, rauch08}).  We
leave a more complete statistical analysis of large simulation volumes
to future work, but we demonstrate here the types of measurements we
can make from our simulated data.  We first look at the distribution
of sources from the Source Extractor software applied to the observed
maps from the UVB+QSO case presented in Figures~\ref{fig:noisymapsL22}
and \ref{fig:noisymapsL5}.  Sources are defined within the SExtractor
software as being 5-$\sigma$ detections.

\begin{figure*}
\plotone{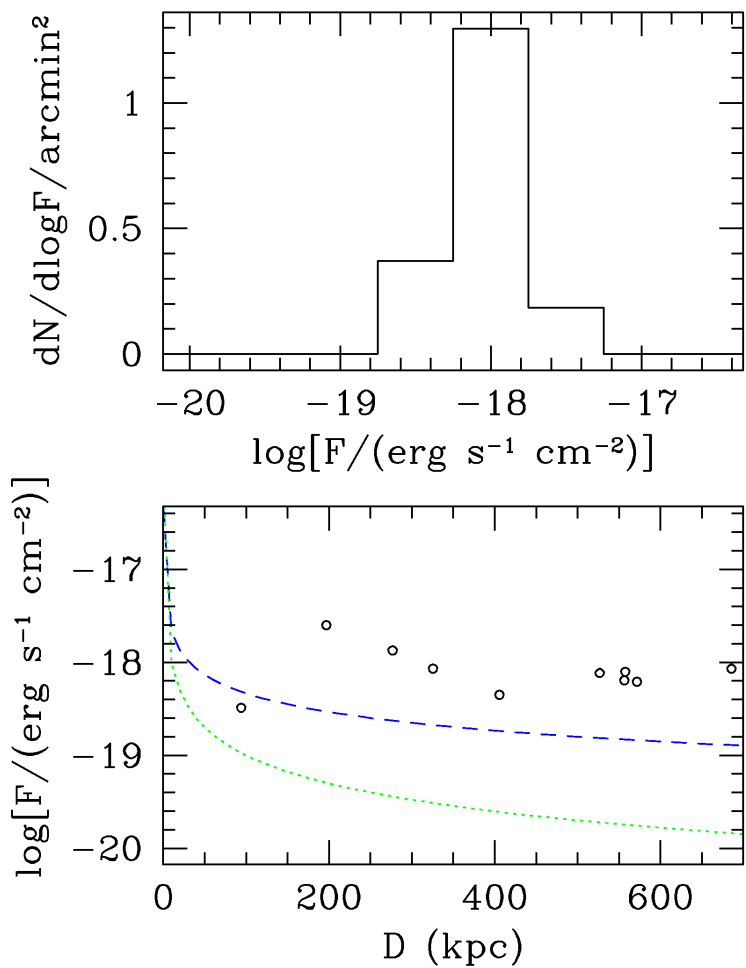}
\caption[Distribution of \lya\ sources/ in the presence of a QSO.]
{\singlespace Distribution of \lya\ sources for the UVB+QSO case
identified from simulated noisy images for a 10 hr exposure using the
Source Extractor software.  The top panel shows the
differential distribution in flux of identified sources.  The bottom panel
shows the source fluxes as a function of projected distance from the center
of the image, where the QSO is located.  The dashed curve in the lower panel shows a $d^{-2/3}$ decay,
which would be expected for a population of identical, self-shielded
isothermal spheres.  The dotted curve shows a $d^{-2}$ decay.
}
\label{fig:sexstatL5Q10hr} 
\end{figure*}

\begin{figure*}
\plotone{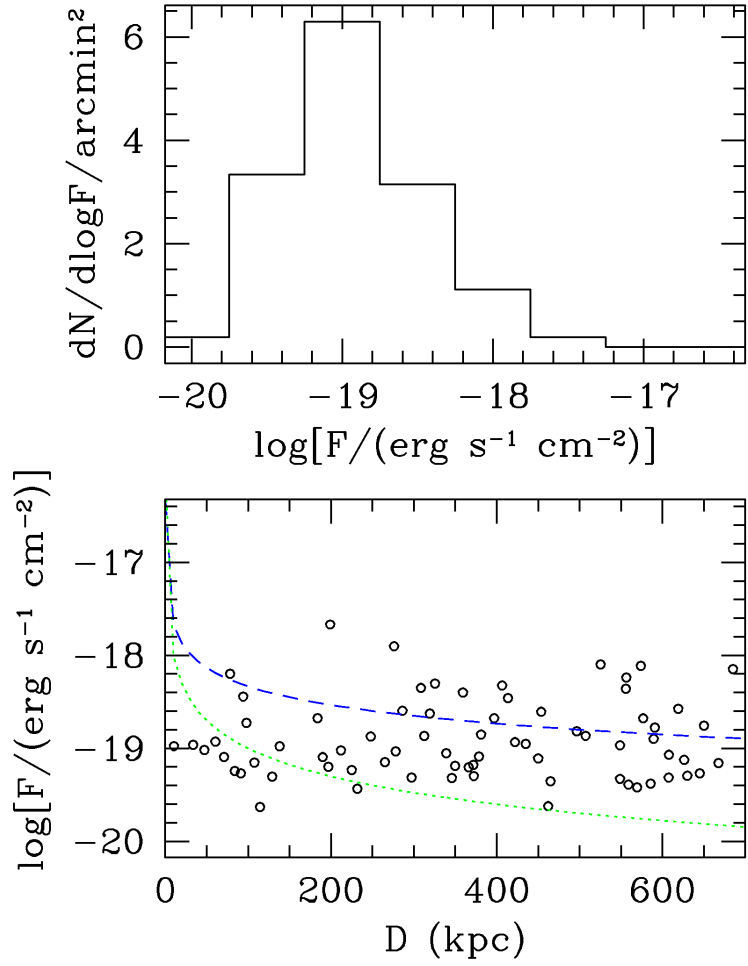}
\caption[Distribution of \lya\ sources in presence of QSO.]
{\singlespace The same as Fig.~\ref{fig:sexstatL5Q10hr} for a 1500 hour
exposure.  Again we detect a radial trend but not an inverse-square dependence.}
\label{fig:sexstatL5Q1500hr} 
\end{figure*}

Figures~\ref{fig:sexstatL5Q10hr} and \ref{fig:sexstatL5Q1500hr} 
present results of this analysis for the 1.5Mpc (physical)
subregion of the L5 simulation, for the 10 and 1500 hour cases,
respectively.  
The top panels in these figures show the differential
distribution of sources as a function of \lya\ flux.  The bottom
panels show the fluxes of identified sources as a function of radial
distance from the quasar, which is located at the center of box.  In
the ``quasar field'' the distribution of fluxes is substantially
skewed toward higher values, which results from the increased
photoionization of optically thick systems in the sub-region.

In the 10-hour field, 10 sources are detected, with fluxes in the range
$\sim 2\times 10^{-19} - 2\times 10^{-18} \fluxunits$.  There is only
a marginal trend of flux with distance from the quasar.  One might
naively expect a $d^{-2}$ falloff in flux, but sources further from
the QSO can remain self-shielded to a larger radius and therefore
present a larger reflecting area.  For a population of isothermal
spheres, like those in \S\ref{sec:results1}, one can show that the
expected falloff in flux is $d^{-2/3}$, shown by the dashed line,
which approximately describes the overall trend of points.  Because
of the smaller reflecting area of sources at smaller radii, we expect
that our fixed grid resolution smears out the brightest sources near
the quasar. Therefore the radial falloff may be even flatter than
$d^{-2/3}$.  Given the slowness of the radial trend, we expect
observable sources beyond the $750$\hkpc radius of the volume we have
analyzed.  The 1500-hour map contains 76 detected sources, down to
fluxes $\sim 3\times 10^{-20}\fluxunits$, and it again shows only a
weak trend of source flux with distance from the quasar.

\subsection{A Case Study \label{sec:d10}}
In this section, we compare a simple model of \lya\ fluorescence with
observations to investigate the origin of the emission. Our method has
already been applied to constrain the emission mechanism for emission
seen in a DLA absorption trough seen in close proximity to a
background quasar (a ``proximate DLA'' Hennawi et al. 2009) and here
we show the the application to a case with different geometry.

Adelberger et al.\ (2006) observed a DLA with column density $N_{\rm
  HI}=(2.5 \pm 0.5) \times10^{20} {\rm cm^{-2}}$ at $z=2.842$ in the
spectrum of a background quasar Q1549-D10.  The absorber is at a
projected angular separation $\theta_Q = 49 \asec$ (corresponding to
380 kpc) from the bright $(G\sim 16)$ quasar HS1549+1919, which has
the same redshift as the DLA.  Extended \lya\ emission with a
double-peaked spectrum is observed at $\theta_l=1.5\asec$
(corresponding to a physical size of $\sim 11 \kpc$ proper) offset
from the absorber.  The extended \lya\ emission region has an apparent
AB magnitude of $G=26.8 \pm 0.2$ mag, and the line has an equivalent
width of $\Delta\lambda_{EW}=275\pm 75$ \AA\ in the $G$ band, which is
$\sim 1000$ \AA\ wide.  The emission line has a peak separation of
$\sim 8$ \AA\ and a line flux of $2.1\times 10^{-17} {\rm erg\ s^{-1}\
  cm^{-2}}$, which yields an inferred surface brightness of $\sim 1
\times 10^{-16} {\rm erg\ s^{-1}\ cm^{-2}\ arcsec^{-2}}$, assuming
that the emission region has diameter of $0.5\asec$.  The AB magnitude
at $912$ \AA\ estimated for the foreground quasar is $m_{912}=16.7$,
which corresponds to a luminosity at the Lyman limit of
$L_{\nu_L}=1.34 \times 10^{32} \rm erg/s/Hz$.  From these observables
we can calculate the expected \lya\ flux owing to fluorescence induced
by the quasar's impinging radiation.  Recall that in the ``mirror''
approximation, a fraction $\eta=0.66$ of the ionizing photons
impinging on an optically thick cloud get re-radiated as \lya\
photons.  For a given background and quasar intensity, the observed
surface brightness should be given by the following expression:

\begin{equation}
\label{eq:mirror}
       \pi \times SB=\frac{h\nu_\alpha \eta}{(1+z)^4} \times 
                     \left[ \pi\int \frac{I_\nu d\nu}{h\nu}
                      +  \int \frac{L_\nu d\nu}{h\nu} \frac{ cos\theta}{4\pi d^2} \right ]
\end{equation}
where $\phi$ is the angle between the foreground and background
quasars such that $d=d_\perp/\sin\phi$, and $\theta$ is the angle
between the mirror normal and the line-of-sight. Substituting the
observed values into equation (\ref{eq:mirror}), we obtain a value for the surface brightness of $5.3\times 10^{-20}$ (background) + $2.7\times 10^{-16}$
$\cos\theta\sin^2\phi$ (quasar) erg/s/cm$^2$/arcsec$^2$.  This matches
the observed value for a plausible geometric factor of
$0.3=\cos\theta\sin^2\phi$.  Approximating the absorber as an
isothermal sphere, we can compare our more sophisticated numerical
results described in \S\ref{sec:iso_qso} directly to this system.

We can also compare our results described in \S 3.2 to this absorber,
modeled as an SIS. We assume the quasar radiates isotropically to
compute its luminosity from the observed flux. The highest predicted
\lya surface brightness in our image is $3.75\times10^{-17} {\rm erg\
  s^{-1}\ cm^{-2}\ arcsec^{-2}}$ --- a factor of $\sim
7\,\cos\theta\sin^2\phi$ below the analytic mirror prediction and
$\sim 2.2$ below the observed value of $0.84\times10^{-16} {\rm erg\
  s^{-1} cm^{-2}\ arcsec^{-2}}$.  For our system, while we know $\phi$
exactly, $\theta$ is uncertain, and we take an average value of
$\theta=60$ degrees for our system.  Our calculation is therefore a
factor of $\sim 3$ below the expectation for the mirror under these
assumptions.  This is compatible with the expected losses due to
geometry (e.g. \lya\ photons leak out of the system and are therefore
emitted over a solid angle larger than $2\pi$).  We show a comparison
of the surface brightness for both the analytic and numerical
predictions and the observations in Table 1. for reference.

\begin{figure*}
\plotone{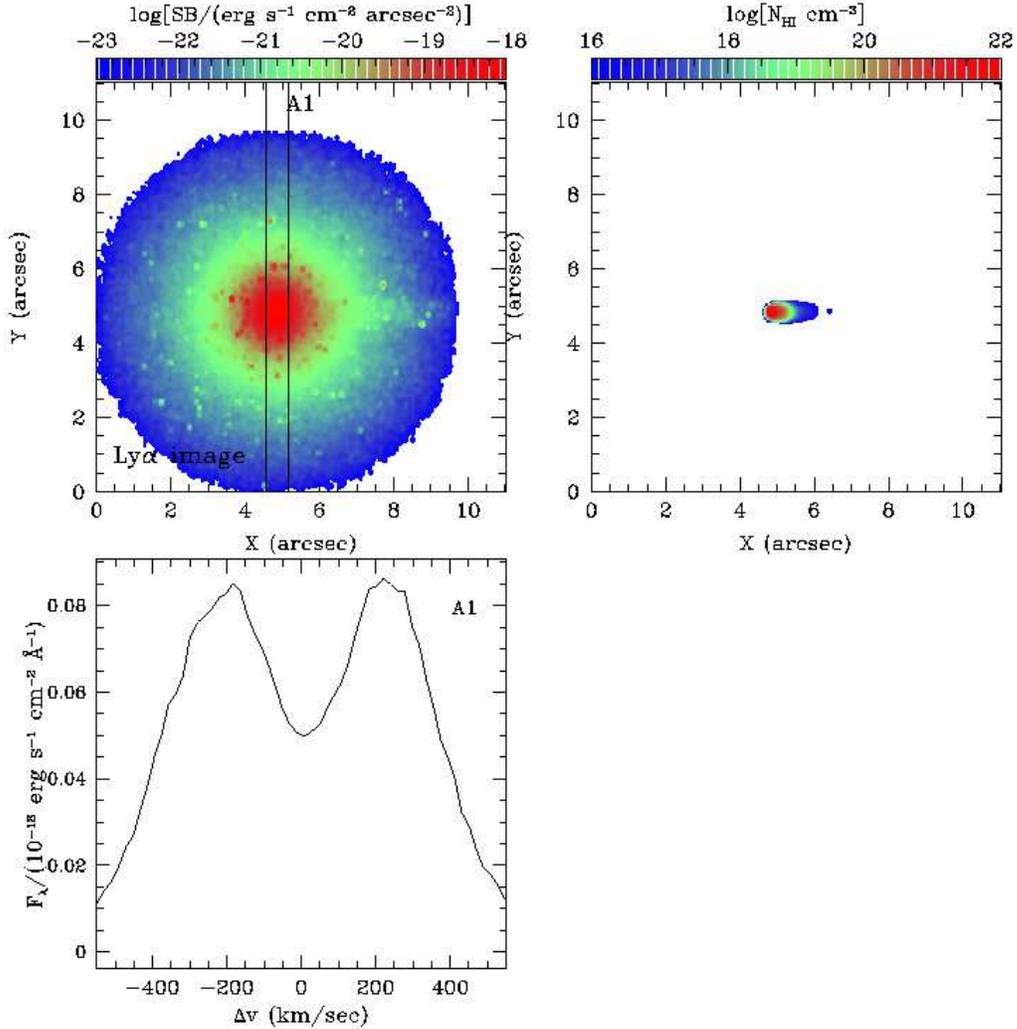}
\caption[A Case Study]
{\singlespace A detailed model of the system observed by 
Adelberger et al. (2005).  The cloud is modeled as a singular isothermal
sphere anisotropically illuminated by a quasar from the left.
The upper left panel shows the \lya\ image for this configuration.
Vertical lines in this panel show a long-slit with 0.7 arcsec slit width placed
along the edge of the cloud.  We plot the 1-d spectrum from this aperture 
in the lower left panel.
The upper right panel shows the neutral column density distribution from the
cloud.  }
\label{fig:a05_comp} 
\end{figure*}

For this high incident quasar flux, the optically thick regions of
the cloud are eroded by photoionization, resulting in a decreased
region of high enough column density to act as an efficient fluorescent
surface.  For the density profile we have adopted, the bright
emission region itself becomes nearly coincident with the high
column-density absorber and not offset from the absorber (as in the
lower flux half-moon illuminated cases).  This can be seen in
Figure~\ref{fig:a05_comp} where we show the \lya\ image, neutral
column density distribution and 1D spectrum for comparison with the
observations.  The most significant differences between our
predictions and the observations are the large absorber size and the
spatially offset high \lya\ surface brightness.  This places some
tension on our simple model for this system.  To obtain a large
surface brightness and maintain a high neutral fraction over 10 kpc
scales, a large, dense sheet rather than a centrally concentrated ball
may be required.

This individual system provides an exciting glimpse into what is
possible by combining detailed observations of \lya\
fluorescence and the type of predictions that are now
possible. Variations in density profile, temperature structure, and
velocity structure give rise to distinct signatures, with the
appropriate data allowing us to discriminate between different
physical mechanisms for powering the observed \lya\ emission.
Combining such modeling with larger samples of quasar-absorber pairs
(e.g., Hennawi et al. 2006a, Steidel et al. 2006), 
one should be able to
directly constrain the ensemble physical properties of the densest
regions of the IGM.

\subsection{Future Studies}

\begin{figure*}
\plotone{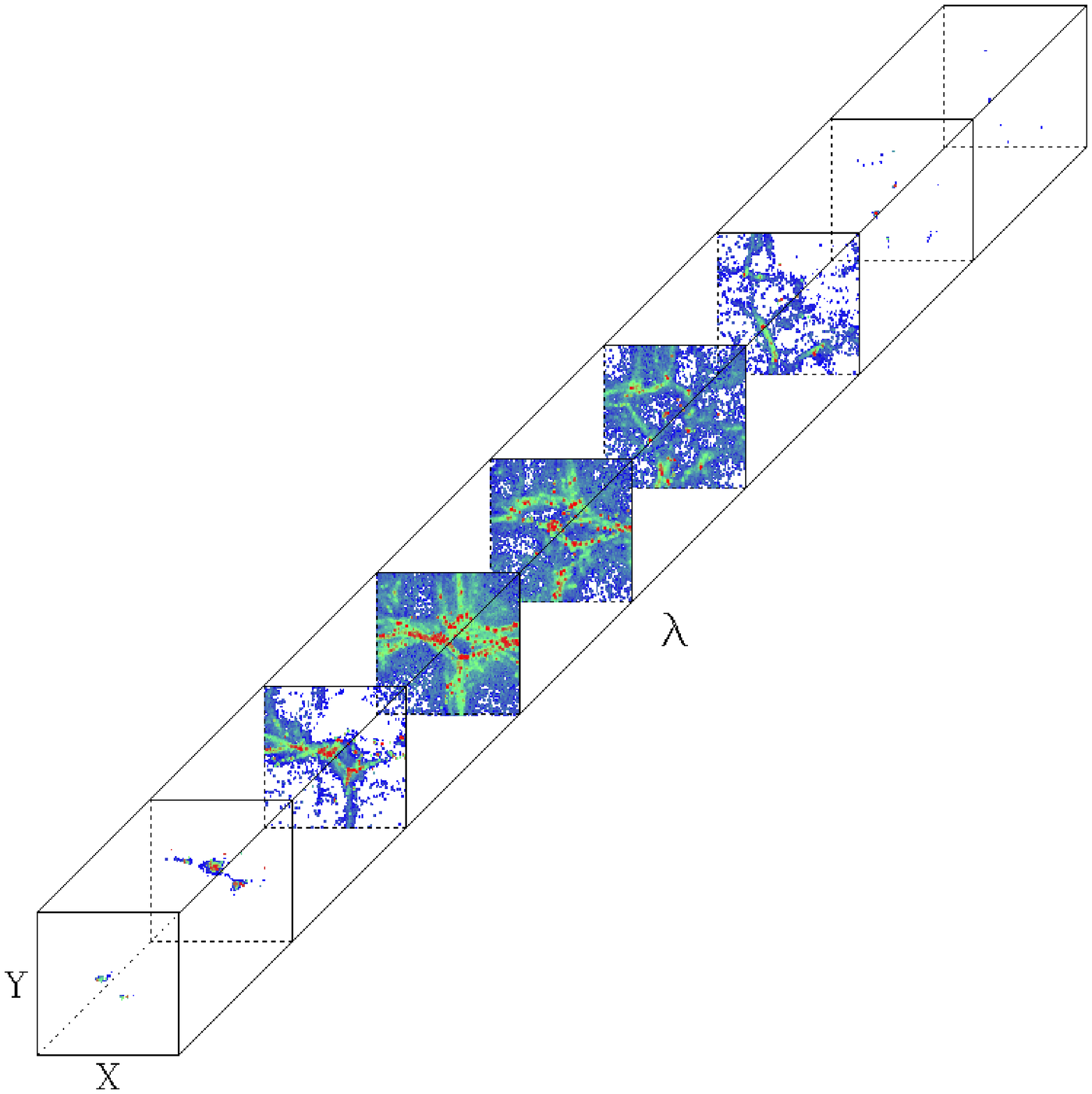}
\caption[Simulated channel map from cosmological simulation region]
{\singlespace Channel maps of \lya\ fluorescence around a bright quasar.
The channels are 0.33 \AA\ in width and spaced by 0.66 \AA.  An IFU on a large
telescope could in principle produce data for direct comparison with
maps such as these constructed from the simulations. }
\label{fig:channel_map} 
\end{figure*}

In order to make the best use of the predictions presented here, one
would like to exploit both spatial and spectral
information. Currently, one can already compare our \lya\ maps and 1D
spectra to observations obtained in deep long-slit spectroscopic or
narrow-band surveys with follow-up slit spectroscopy (e.g. Rauch et
al. 2008, Steidel et al. 2009, in prep).  Blue-sensitive integral
field units (or tunable narrow band filters) on large telescopes will
have the capability to make channel maps of \lya\ emission around
structures, identified in imaging or spectroscopic surveys, at a given
redshift that can be directly compared to the kinds of predictions
that we are making here.  We show such a configuration in
Figure~\ref{fig:channel_map}, which presents a series of
frequency-slices through our $z=3$ cosmological calculation of
fluorescence around a bright quasar.  As the filter is tuned past
\lya\ at the appropriate redshift, the ``forest'' of optically thick
absorbers comes into view and then fades away.

\section{Discussion and Conclusions \label{sec:discussion}}

We have presented our numerical methods for accurately obtaining the
fluorescent \lya\ emission signature from cosmological smoothed
particle hydrodynamic simulations using these simulations in
conjunction with a Monte Carlo scheme for radiative transfer of the
\lya\ line.  We apply a self-shielding correction to the cosmological
particle distribution to correct for the fact that the simulation is
initially run in the optically thin regime, which makes particles
hotter and more highly-ionized than they would have been if radiative
transfer was included included during the simulation. We find that
one must carefully treat the boundaries between optically thin and
optically thick structures to obtain accurate recovery of the neutral
fraction profile within simulated structures.  Failure to do this
results in a systematic underprediction of the \lya\ surface
brightness and an overprediction of the sizes of high-column density
structures.  We find that this is sensitive to the resolution of a
given simulation, and we develop techniques that are relatively robust
to changes in resolution.  Because SPH is notoriously problematic at
boundaries, this calculation is nontrivial.  However, because many
widely available codes (e.g. GADGET2; Springel (2005), GASOLINE;
Wadsley et al. (2004)) make use of the SPH technique, it is useful to
be able to make \lya\ predictions from this type of cosmological
calculation.  We further find that the treatment of temperatures in
these simulations can have a significant impact on the morphology and
luminosity of fluorescent \lya\ emission from SPH cosmological
simulations.  While this has obvious implications for cooling
radiation (which we will present in future work), it also has
substantial implications for fluorescence as we show in A3.  One
commonly used strategy, setting {\it all} gas to a constant
temperature $T \sim 10^4-2\times 10^4\,$K, can produce misleading
results by ascribing a large amount of \lya\ emission to high
temperature, shock heated gas, which should have very low neutral
fraction.

We find that in the absence of a strong ionizing continuum source, the
highest {\it fluorescent} surface brightnesses within our $z=3$
simulation box are of order $\sim 2\times 10^{-19} {\rm erg\  s^{-1}\ cm^{-2}\
  arcsec^{-2}} $.  To detect such sources will take 1500 hr
campaigns on the current generation of telescopes given a typical
narrow-band setup.  For practical applications, we show that the
fluorescent surface brightness can be substantially enhanced by the
presence of powerful ionizing sources. We show that fields with bright
quasars are significantly more fruitful regions to search for
fluorescent \lya\ emission from dense optically-thick structures at
present, yielding significant detections in mere hours on the current
generation of telescopes.  Such observations can be compared directly
to the calculations we present here.  We test our methods and present
results for the simple case of an anisotropically illuminated singular
isothermal sphere to demonstrate both our method and an application of
interest, and we compare these predictions to recent observations of
such a system \citep{adelberger06}. We find that the combination of
spectral shape, surface brightness, and absorber size give important
constraints on the fluorescent emission from such systems.  With this
machinery in place, opportunities abound for understanding \lya\
emission in the high-redshift universe.

\acknowledgements{We thank Chuck Steidel, Alice Shapley, Michael
  Rauch, and Kurt Adelberger for many stimulating discussions
  throughout the course of this work.  J.A.K. and Z.Z. were supported
  for portions of this work by NASA through Hubble Fellowship grants
  HF-01197 and HF-01181 awarded by the Space Telescope Science
  Institute, which is operated by the Association of Universities for
  Research in Astronomy, Inc., for NASA, under contract NAS
  5-26555. Z. Z. gratefully acknowledges support from the Institute
  for Advanced Study through a John Bahcall Fellowship.}

\newpage

\clearpage

\clearpage

\clearpage

\clearpage

\begin{deluxetable}{ll}
\tablenum{1}
\tablecolumns{2}
\tablecaption{Comparison of observed surface brightness for the
system seen in Adelberger et al. (2006) with two theoretical models for the
system. \label{tbl-1}}
\tablehead{
\colhead{System}&
\colhead{Surface Brightness ($ {\rm erg\ s^{-1}} {\rm cm^{-2}\  arcsec^{-2}})$}
}
\startdata
Observed Value & $0.84\times10^{-16}$ \\
Analytic Mirror Prediction  & $2.7\times 10^{-16} \cos\theta\sin^2\phi$ \\
SIS + Radiative Transfer & $0.4 \times 10^{-16}$ 
\enddata
\end{deluxetable}

\appendix
\section{Accurate Self-shielding Correction}

In the SPH technique, the smooth density field is represented by
discrete particles.  One must therefore always be cautious that there
are sufficient particles to adequately resolve structures of interest.
In this application, we are primarily concerned with resolving the
optically thick skins of dense gas clouds within our simulation
volume.  While the clouds themselves are well-resolved, usually with
several hundred to thousands of particles (in our L5 simulation), it
is the distribution of particles at the interface between optically
thick and optically thin regions that contribute the majority of the
\lya\ emission.  We, therefore, must pay close attention to the
accuracy of our self-shielding correction at these transition layers.
We develop a method to perform the self-shielding correction that
accounts for the effects of low resolution, and we test this method
using a series of SPH approximations to an isothermal sphere (for
which we have exact analytic results).

To perform the self-shielding correction for a given particle
distribution we determine the optical depth for photons to reach each
particle's position.  We do this by evaluating the attenuated
ionizing photon intensity along 6 directions (and the quasar direction
when present). In the SPH technique, each particle has a density
distribution defined by its mass and smoothing length. At a given particle's
position, the optical depth for ionizing photons towards a direction
can be straightforwardly computed by integrating the neutral density
profiles of those particles that contribute in this
direction. However, because of the finite size of particles and the
steep gradient in the neutral density profile near the self-shielding
layer, such a simple computation may lead to large errors in the
optical depth and thus an inaccurate self-shielding correction.  The
problem is analogous to computing the optical depth from a steep
density distribution by using finite rectangle bins and evaluating the
density at the center of each bin. To compute the optical depth to the
center of a bin, the contribution from that bin is evaluated as the
density at the bin center times the half-width of the bin. If the
gradient of the density profile is large, this obviously overestimates
the contribution to the optical depth in the direction of decreasing density. 
A better way of computing the contribution is to use
the trapezoidal rule in this bin with the shape of the trapezoid
determined by the gradient of the density distribution. We apply a
similar idea for computing the optical depth from the SPH particle
distribution.

In the SPH formalism, the density at a given position {\bf r$_0$} is given by:
\begin{equation}
\label{eqn:sphden}
\rho({\bf r_0}) = \sum_{i}^{N} m_i  W({\bf r_0};{\bf r_i}, h_i) = \sum_{i=0}^{N}  \frac{m_i}{(\sqrt{2\pi} h_i)^3} {\rm exp}\left( - \frac{ |{\bf r_0 - r_i}|^2}{2h_i^2} \right)
\end{equation}
where $W({\bf r_0};{\bf r_i}, h_i)$ is the 3D Gaussian equivalent of the SPH cubic spline kernel used in the simulation\footnote{The cubic spline kernel is well represented by a Gaussian with appropriate width.  For ease of computation, we adopt the Gaussian-equivalent form for our post-processing calculations.  }, $N$ is the number of particles that contribute to the density at
$\bf{r_0}$ having position, neutral mass, smoothing length
${\bf r_i}$, $m_i$, $h_i$.  
For each particle we evaluate the density gradient at the position of the particle, ${\bf r_0}$ as
\begin{equation}
\nabla \rho|_{0} = \sum_{i=0}^{N} m_i \nabla W({\bf r_0};{\bf r_i}, h_i) 
= -\sum_{i=0}^{N} \frac{{\bf r_0 - r_i}}{h_i^2} \frac{m_i}{(\sqrt{2\pi} h_i)^3} {\rm exp}\left( - \frac{ |{\bf r_0 - r_i}|^2}{2h_i^2} \right)
\end{equation}
Accounting for the density gradient, the density profile along an arbitrary 
direction, $\hat{n}$, from this particle is then given by
\begin{equation}
\label{eqn:gradden}
\rho(s)  = \rho |_0 +(\nabla \rho|_0 \cdot \hat{n}) s ,
\end{equation}
where $s= ({\bf r} - {\bf r_0}) \cdot \hat{n}$ and $\rho|_0$ is the density at the position of the particle [${\bf r_0}$; eq.~(\ref{eqn:sphden})].  The 
optical depth at the particle's position is computed by integrating the density 
profiles of contributing particles along the given direction. The correction to the optical depth caused by the gradient near the particle's position is 
obtained from integrating the gradient term in equation~(\ref{eqn:gradden}) 
of the density profile.  The correction to the optical depth is given by
\begin{equation}
\Delta \tau_{\hat{n}} = \int_0^{fh_0} (\nabla \rho|_0 \cdot \hat{n})/m_H \sigma l dl ,
\end{equation}
where $h_0$ is the smoothing length of the particle and $f$ is a
factor we can adjust to reflect where we truncate the integral.  We
take $f=2$ in our calculations, corresponding to truncating the
integral at twice the particle smoothing length.  Since we evaluate
the gradient from discrete particle distributions, there are
unavoidable numerical effects in the computed gradient, which can
sometimes lead to corrections that are large and negative relative to
the original optical depth.  For these cases, we limit the corrected
optical depth to be no less than 10\% of the total optical depth.
While these constitute only a small fraction of the total number of
particles, they cannot be simply ignored because they typically lie at
the transition region between optically thin and thick material.

\begin{figure*}
\resizebox{5.5cm}{!}{\includegraphics{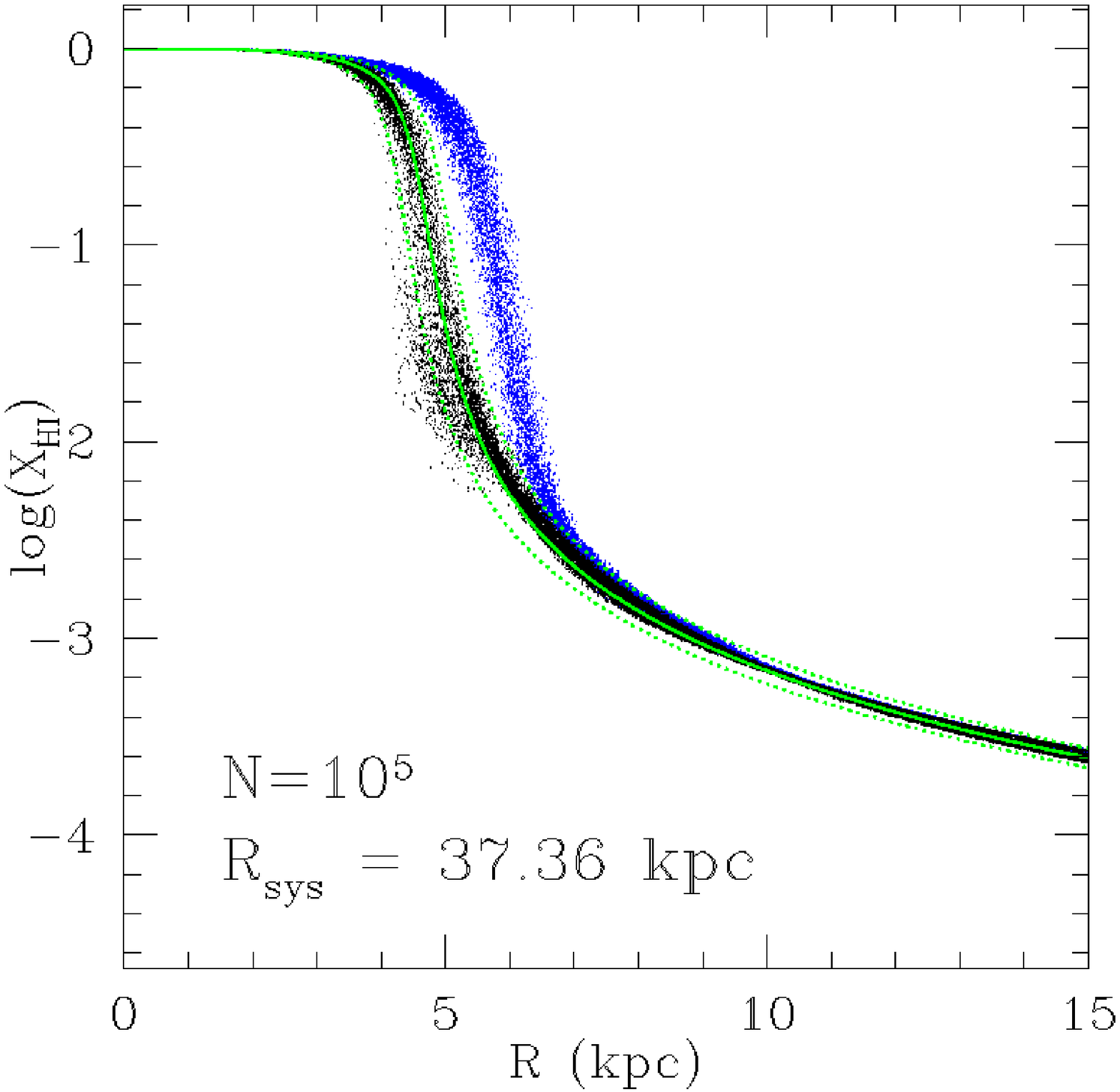}}\hspace*{0.1cm}%
\resizebox{5.5cm}{!}{\includegraphics{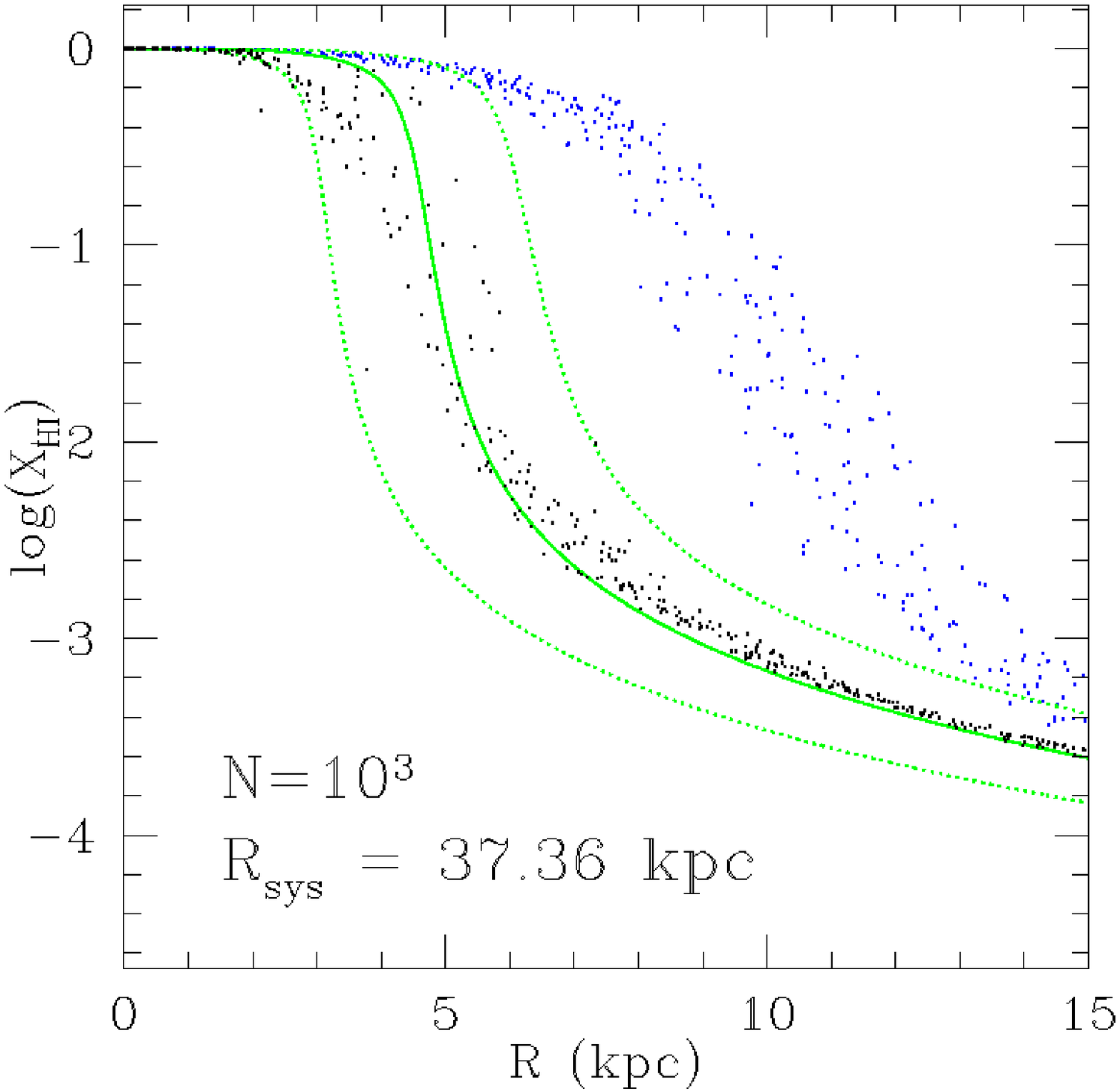}}\hspace*{0.1cm}%
\resizebox{5.5cm}{!}{\includegraphics{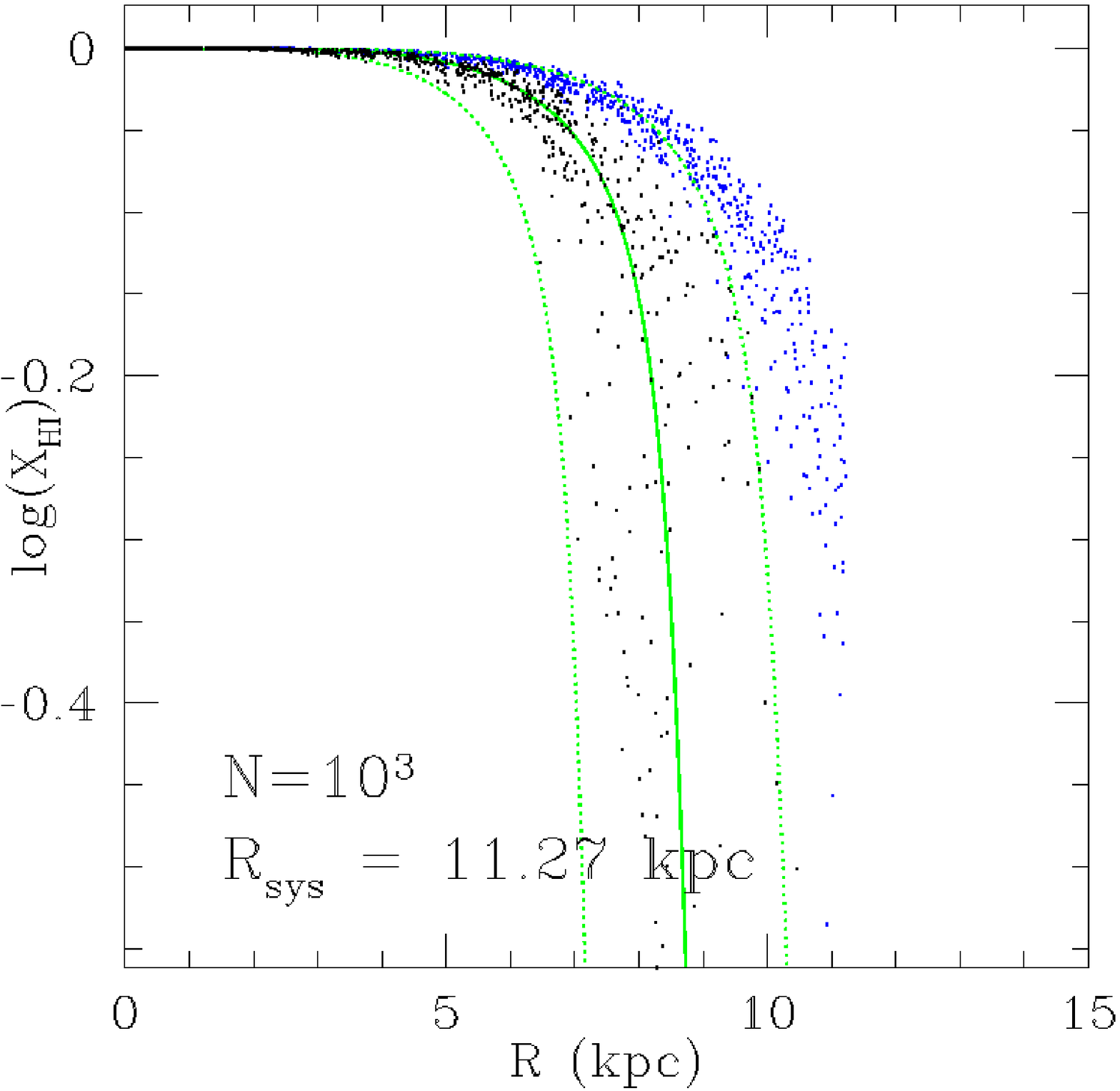}}\vspace*{0.1cm}\\
\resizebox{5.5cm}{!}{\includegraphics{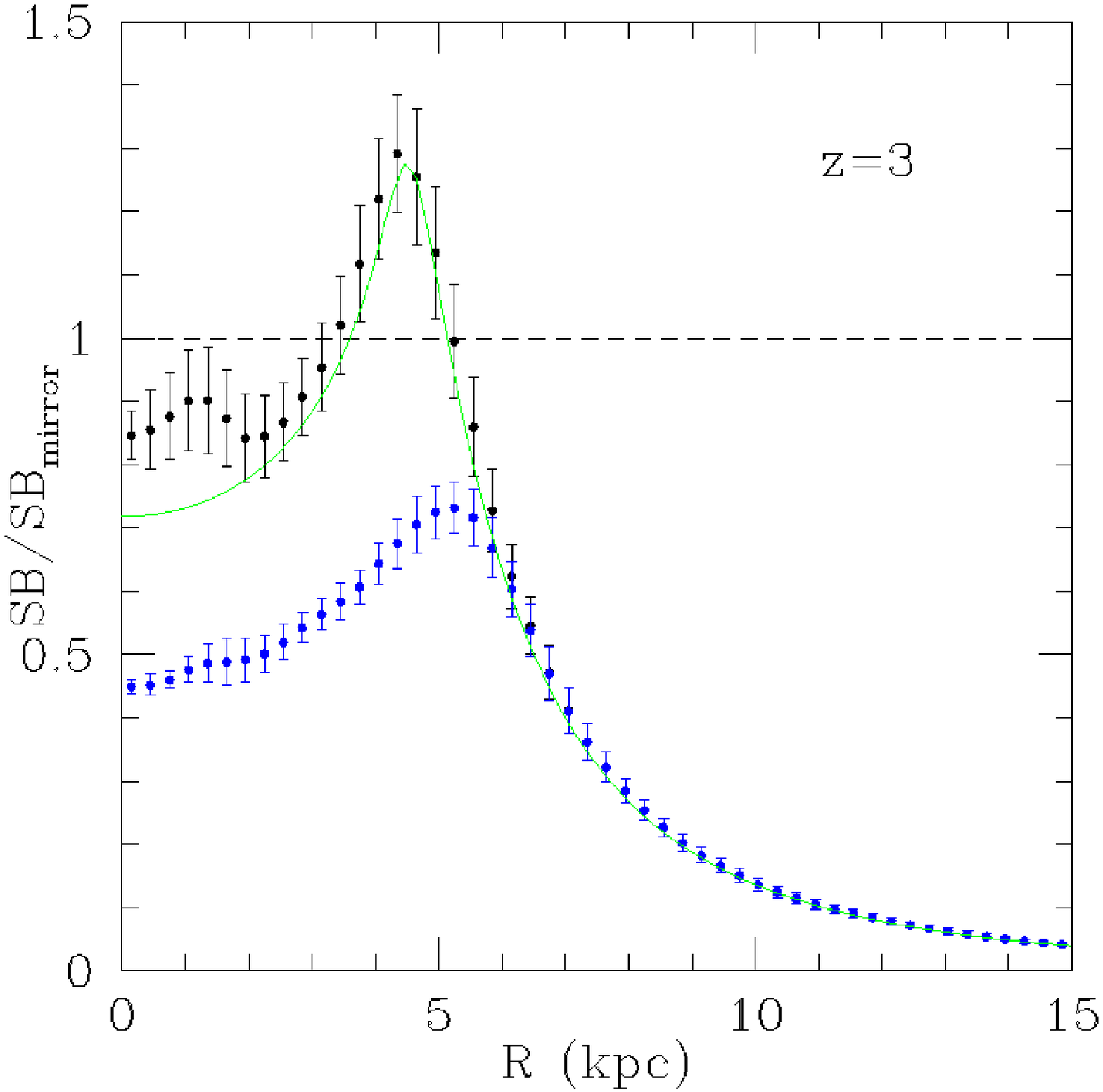}}\hspace*{0.1cm}%
\resizebox{5.5cm}{!}{\includegraphics{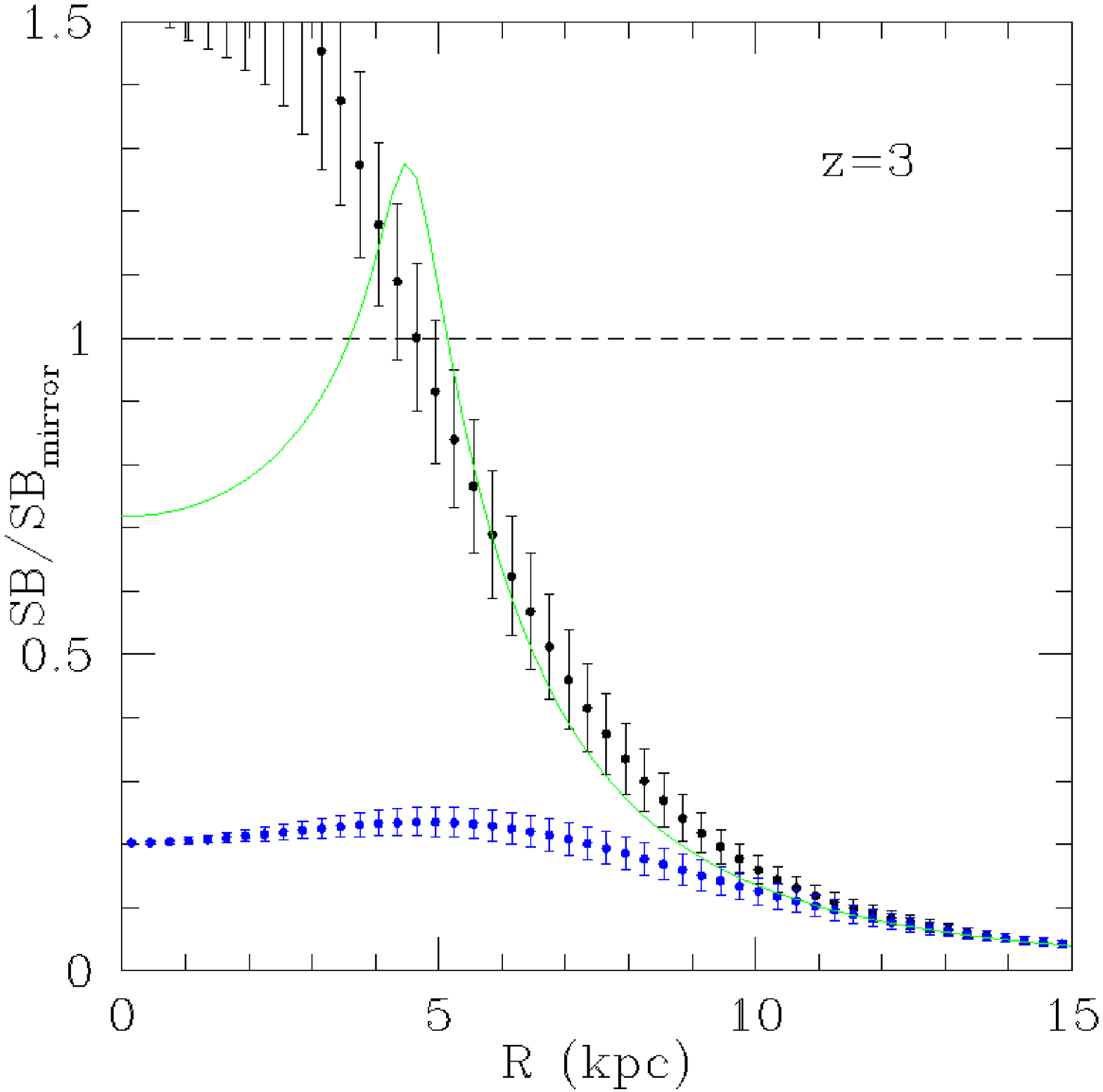}}\hspace*{0.1cm}%
\resizebox{5.5cm}{!}{\includegraphics{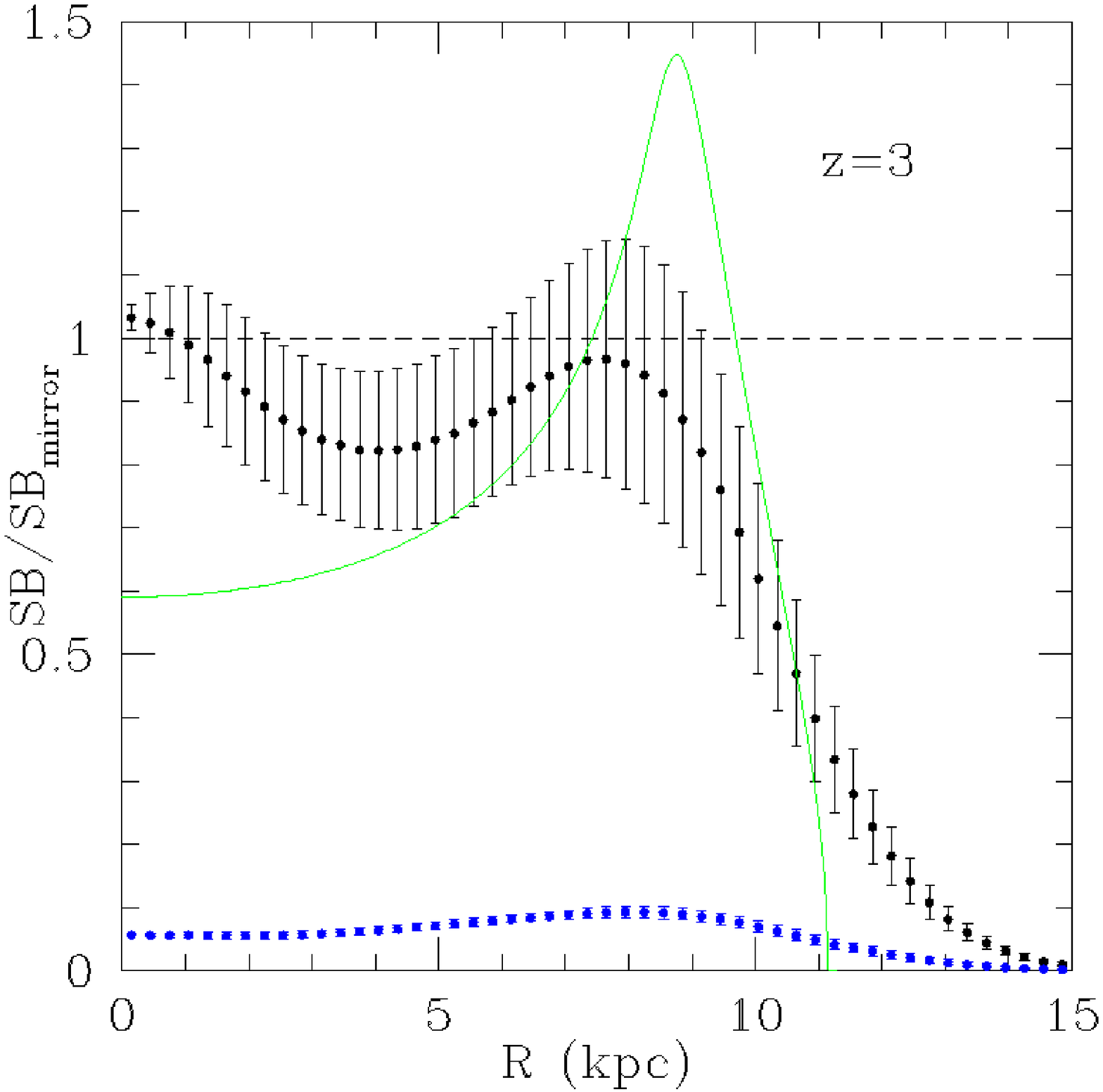}}\vspace*{0.1cm}\\
\caption{The effect of particle resolution and density gradient on the
  self-shielding solution.  Top panels show the neutral fraction
  profiles for an SPH sphere with 100k particles (left), 1k particles
  (middle), and 1k particles (right).  The left and middle panels have
  gas particles distributed within the entire virial radius of the
  halo in which the gas resides.  The right panel has the particles
  arranged such that they only occupy the inner 30\% of the virial
  radius.  Bottom panels show the same configurations, but plot the
  surface brightness profiles without scattering (i.e. the column
  emissivity).  Green solid lines in each panel show the exact
  solution, green dotted lines indicate the particle smoothing length,
  and black points show our solution.  Blue points show the solution
  when the density gradient is neglected.}
 \label{fig:a1_fig1}
\end{figure*}

We test our code on an SPH version of an isothermal sphere for varying
resolutions and spatial configurations.  We consider an isothermal
sphere with a halo mass of $10^{11}\msun$ (with a virial radius of
37.36 kpc).  We consider two configurations: either the gas particles
extend to the full virial radius or they only extend out to the inner
30\% of the virial radius.  The latter compact configuration may
represent a case more akin to what we predict in cosmological
hydrodynamic simulations.  For each configuration we perform tests
with three different mass resolutions by representing the gas with
$10^3, 10^4$ and $10^5$ particles, respectively.  We show the effect
of this gradient as a function of geometry and resolution in
Figure~\ref{fig:a1_fig1}.  The $10^5$ particle case in which the
particles are distributed to the full virial radius is shown in the
far left panels of Figure~\ref{fig:a1_fig1}.  This case is shown to
match the analytic predictions both for the neutral fraction profile
(upper panels) and the surface brightness profile (lower panels).  The
reference analytic solution is computed for the singular isothermal
sphere illuminated by the UVB by iteratively evaluating the attenuated
UV intensity and solving the photo-ionization equilibrium equation at
each radius \citep{ZM02b}. The radial bin size is set to be
sufficiently small to ensure an accurate solution.  The surface
brightness profile is obtained by griding the particle emissivities on
a regular $256\times 256$ grid.  This is therefore a projected surface
brightness profile, or a column emissivity as we discuss in the text.

Cosmological SPH simulations typically do not have many structures
resolved this sharply.  More commonly, structures will have one
thousand to several tens of thousands of particles.  To show the
effect of low resolution, we show the neutral fraction and surface
brightness profiles for this case represented by only 1000 particles
in the middle panels of Figure~\ref{fig:a1_fig1}.  We see that the
neutral fraction profile is reasonably well recovered when our
gradient correction is included even at this low resolution (top
middle panel).  The bottom middle panel of Figure~\ref{fig:a1_fig1}
shows the surface brightness profile for this case.  For the bulk of
the sphere, the surface brightness is well recovered.  However in the
very center of the sphere, our calculation overpredicts the emissivity
relative to the analytic case.  This owes to the fact that the
particle smoothing lengths are quite large in this case, and particles
with large emissivities, centered at the transition between optically
thick and thin gas, contribute emissivity formally in the center of
the cloud where the neutral fraction approaches unity (and hence the
emissivity approaches zero).  The right panels in
Figure~\ref{fig:a1_fig1} show the low-resolution ($10^3$ particles)
compact configuration case (particles distributed between the center
of the cloud and 30\% of the virial radius).  The neutral fraction
profile (top right) is recovered with large scatter.  The surface
brightness profile for this case is smoothed out relative to the
$10^5$ particle case.  Here again, high emissivity particles are
contributing flux at the center of the sphere owing to their large
smoothing lengths and, similarly, the emissivity is diluted in the
peak region owing to low emissivity particles.

The blue points in the top panels show the results of our calculation
when we neglect the density gradient. Ignoring the density gradient
results in significantly different neutral fraction profiles. While
for very high optical depth (at the Lyman limit) and for very low optical
depths the gradient is not important, at the transition region
($1<\tau_{LL}<10$) the density gradients are quite large and play an
important role.  Because this region also produces and radiates the bulk of
the \lya\ emission, it is critical to get this
region correct for fluorescence calculations.  If we did not correct
for the gradient, our peak surface brightness estimates would be in
error (too low) by factors of 2, 3 and 5 in the $10^5$, $10^4$, and
$10^3$ cases (for particles distributed to the virial radius).  This would
clearly have a major impact on our predictions.  Therefore, even with
the over-correction at the very center of these structures, it is far
superior to the uncorrected case.

\begin{figure*}
\resizebox{5.5cm}{!}{\includegraphics{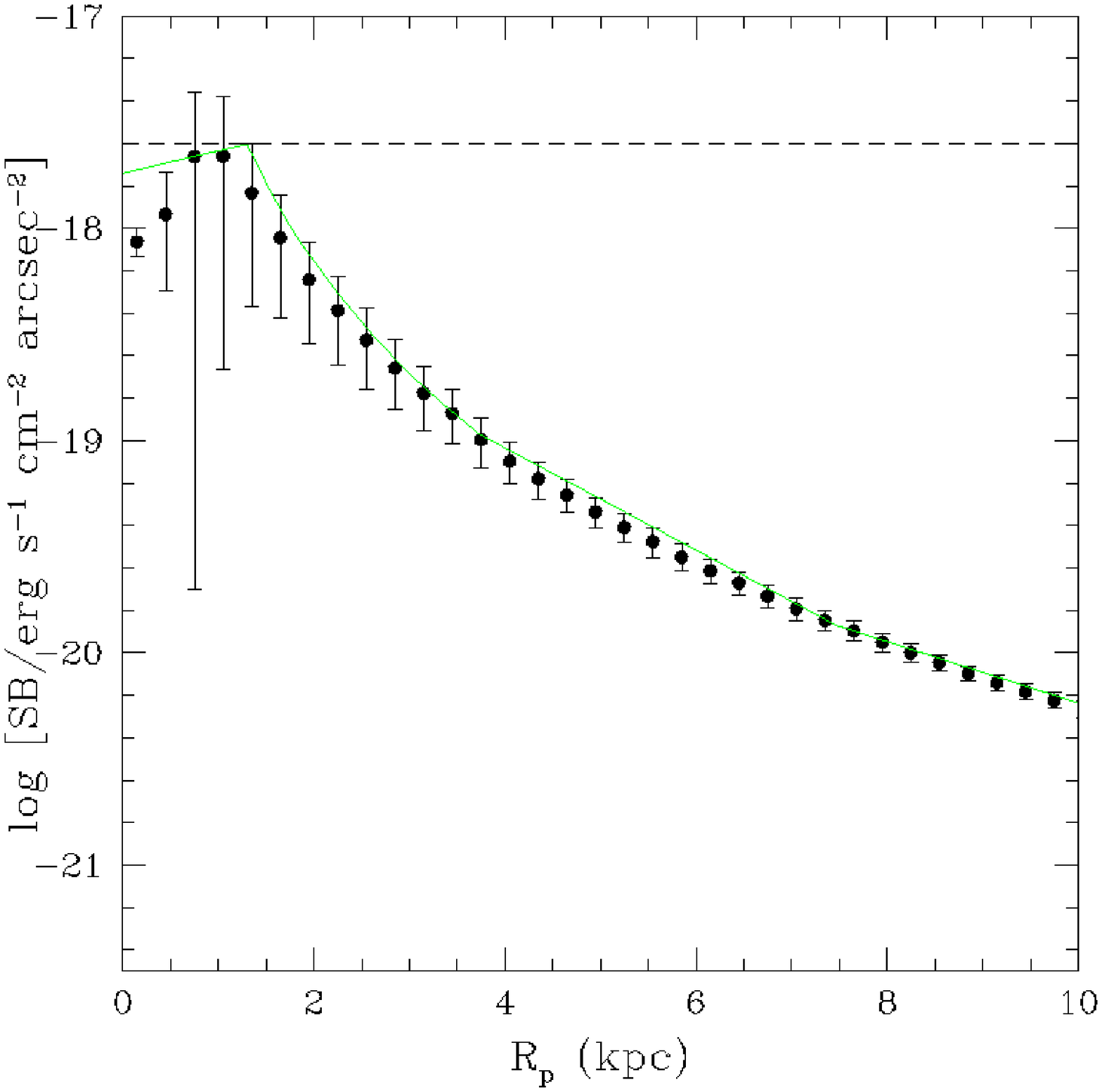}}\hspace*{0.1cm}%
\resizebox{5.5cm}{!}{\includegraphics{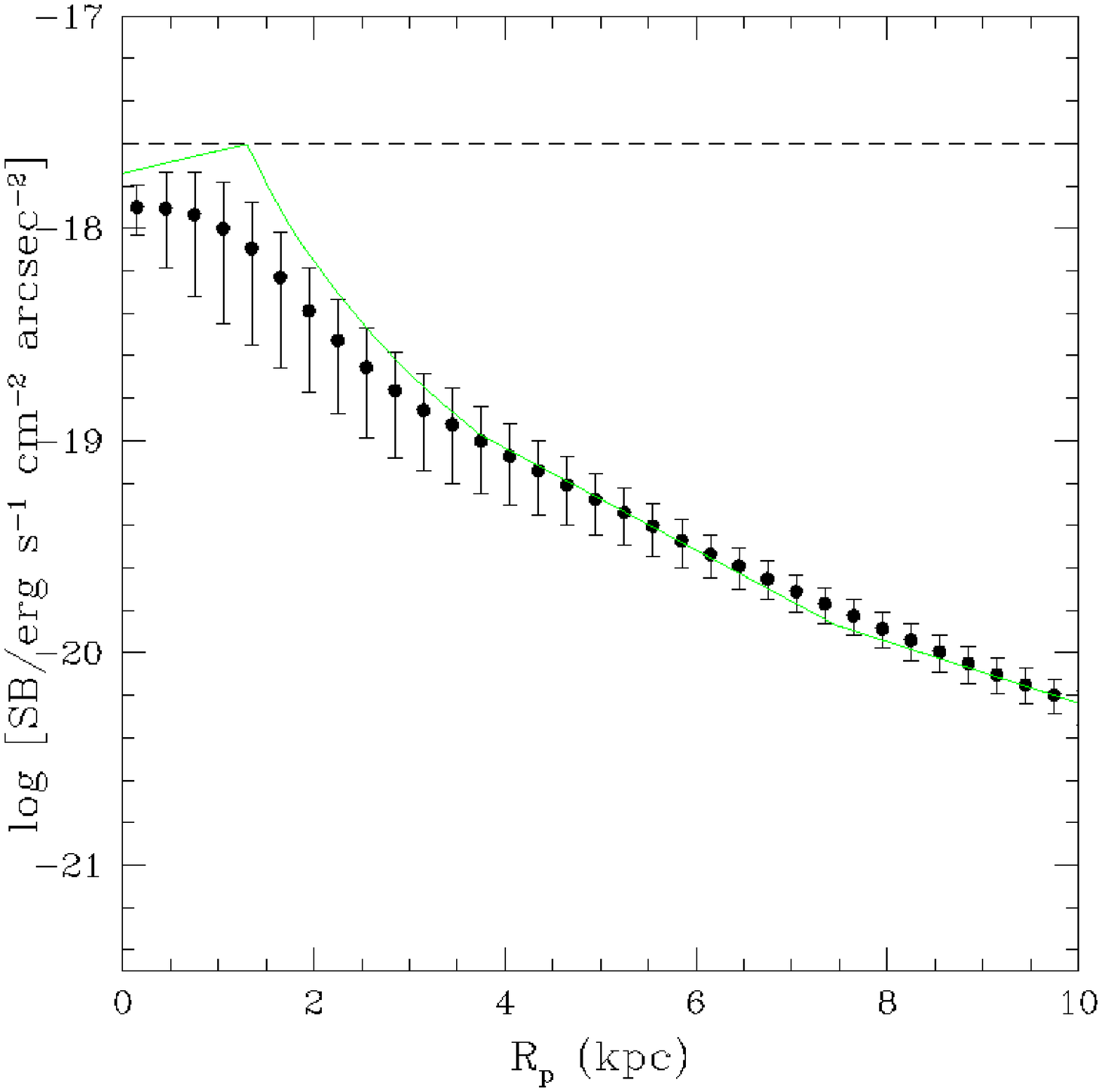}}\hspace*{0.1cm}%
\resizebox{5.5cm}{!}{\includegraphics{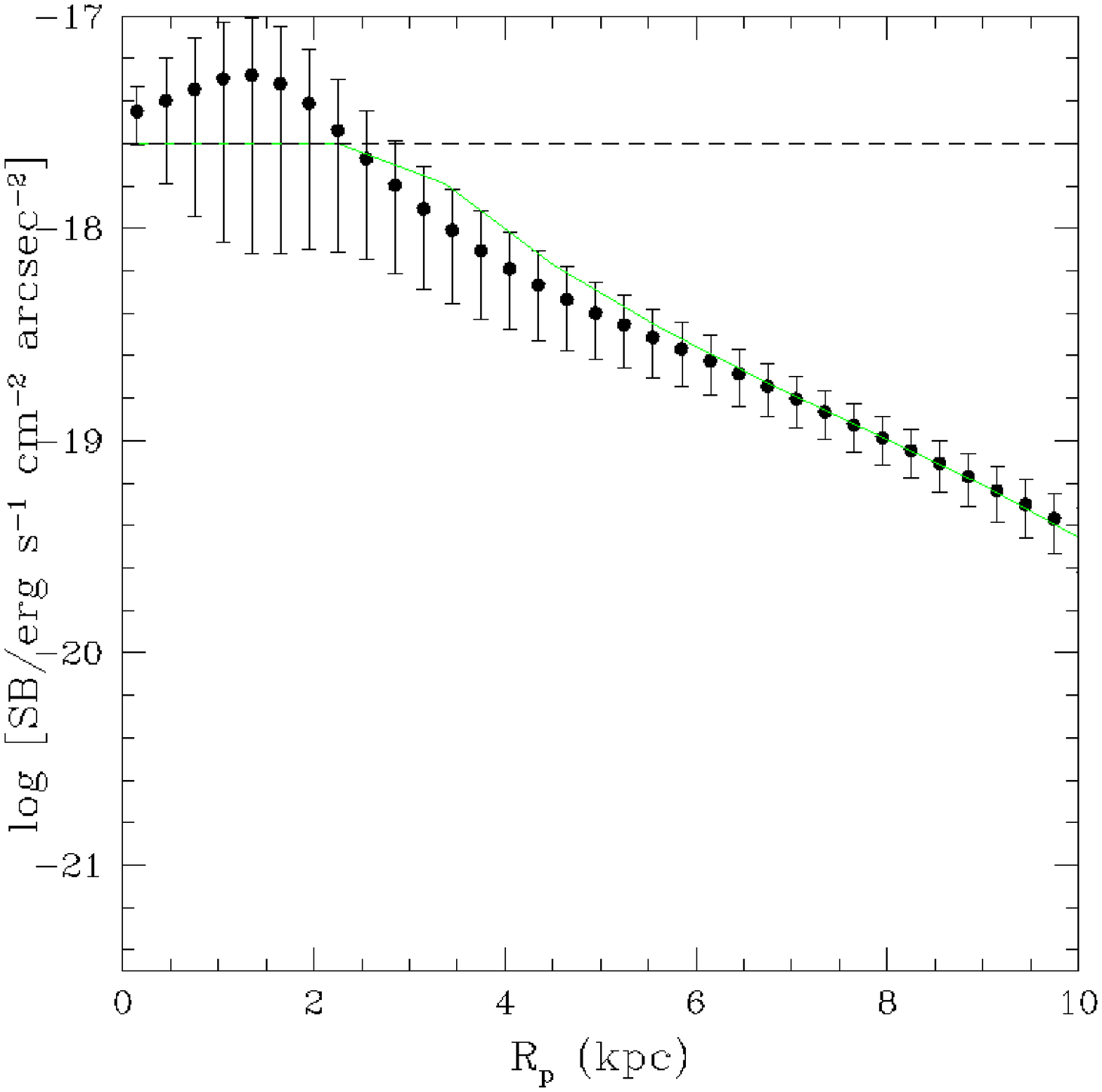}}\vspace*{0.1cm}\\
\caption[QSO Surface Brightness Test] {\singlespace The recovery of a
  quasar only radial surface brightness profile for the three cloud
  configurations and resolutions shown in the lower panels of
  Figure~\ref{fig:a1_fig1}.  Blue points show the analytic case, black
  points show the dispersion of surface brightness pixels in a 2D map
  of the \lya\ column emissivity and green points show the mean value
  of the black points.  The dashed horizontal line shows the ``mirror''
  expectation for this case.  The gradient-corrected quasar case
  recovers the true surface brightness distribution to better than a
  factor of three throughout even at low resolution.  }
\label{fig:a1_fig2} 
\end{figure*}

We further test our gradient correction in the presence of a bright
ionizing source.  While it is not feasible to analytically compute the
UVB+QSO case, we can compute a ``quasar only'' case with our SPH
sphere and compare the resultant surface brightnesses.  For such a
case, at each projected radius along the quasar-cloud direction, the
calculation is reduced to a 1D problem and we use a method similar to
\citet{ZM02b} to obtain solution numerically.  We use the same
isothermal sphere configurations as in Figure~\ref{fig:a1_fig1}, but
we now irradiate these structures by our fiducial quasar. We show the
results of this in Figure~\ref{fig:a1_fig2}.  Our gradient-corrected
neutral density profile correctly recovers the surface brightness in
this quasar illuminated case to better than a factor of two throughout
the profile and particularly over the peak for these test configurations.

The SPH technique has natural limitations at boundaries with large
density gradients and, for the purpose of fluorescence calculations,
these boundaries are critical.  We put forth a technique to address
this issue here that works with good but not perfect accuracy in the
cases that we test.  We adopt this method throughout the paper.

\section{Grid Convergence}

To further assess the effects of grid resolution, we analyze a small
sub-volume of the L5 simulation corresponding to just $65\,$kpc on a
side.  For the smallest smoothing lengths (0.07 kpc) in this
simulation, this sub-volume resolves the minimum smoothing length with
a grid of $1024^3$, which is manageable.

\begin{figure*}
\plottwo{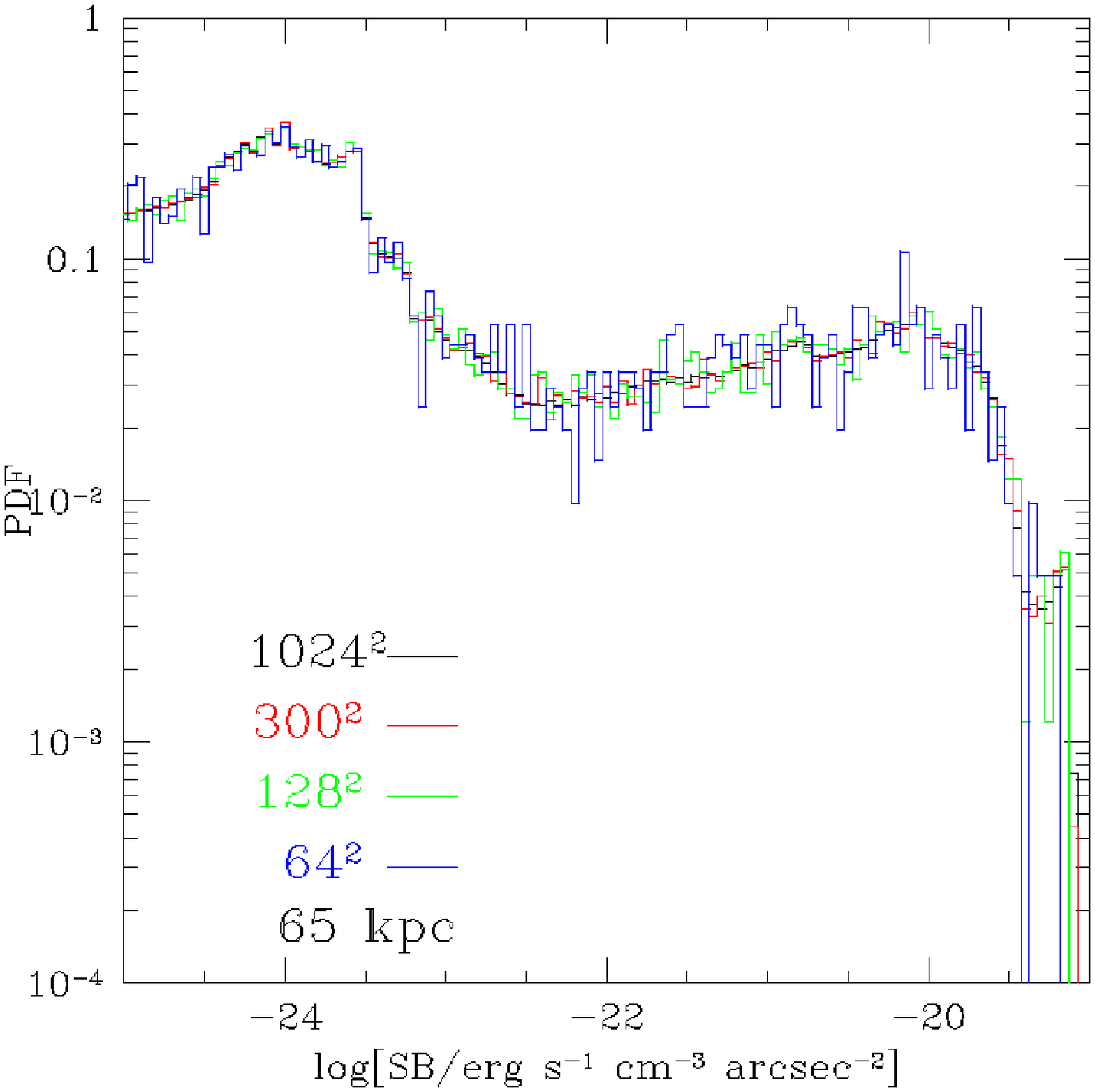}{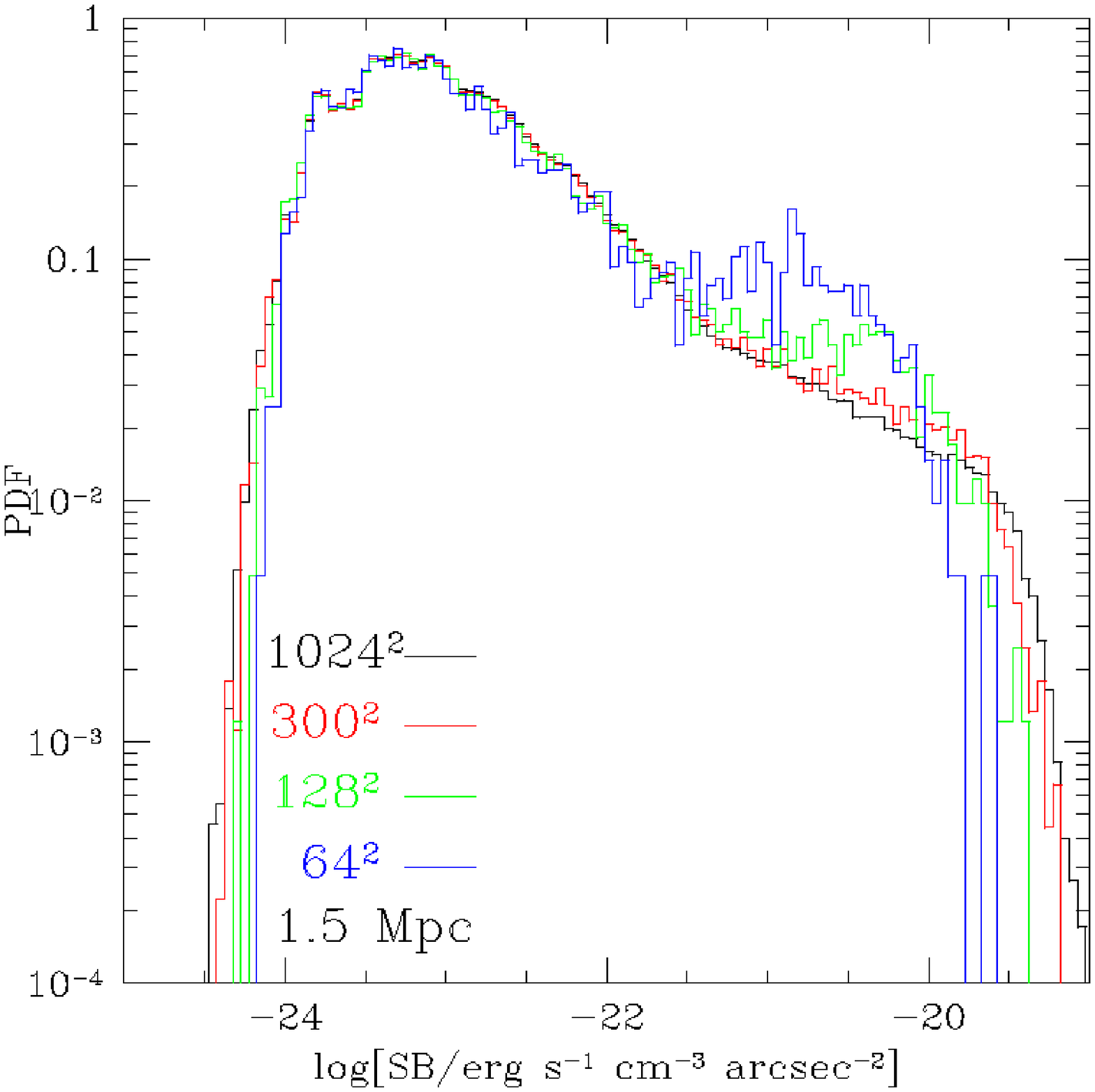}
\caption[Convergence of projected emissivity with grid resolution]
{\singlespace Distribution of projected emissivity of pixels as a function of
grid resolution.  The left panel shows convergence for a cubic sub-region 
simulation of 65 kpc (physical) on a side at $z=3$.  
The projected emissivity distribution converges even with a $64^2$ grid,
which corresponds to a pixel size of $\sim$ 1 kpc on a side.
The right panel shows the distribution of 2D emissivity for the fiducial
1.5 Mpc (physical) region.  
The $300^2$ grid (corresponding to a cell size of $\sim$ 5 kpc on a side)
reaches convergence, and we therefore adopt this for 
our \lya\ radiative transfer calculations.}
\label{fig:emis_pixels} 
\end{figure*}

We first examine the effect of grid resolution on the projected
emissivity distribution, i.e., the expected \lya\ surface brightness
if \lya\ photons escaped without scattering.  We note that the 2D
resolution used for our \lya\ and column emissivity images is always
matched to the underlying 3D grid resolution we use to perform the
scattering calculation.  In the left panel of
Figure~\ref{fig:emis_pixels} we plot the distribution of pixel
emissivities projected on a $64^2$, $128^2$, $300^2$, and $1024^2$ 2D
grid, which shows that our projected emissivity distribution is not
sensitive to grid resolution and that the projected emissivity
distribution converges for the $64^2$ grid in this case.  This
suggests that typical sources are generally larger than 1kpc in size.
To demonstrate convergence of our results for the pre-transfer
emission for the full L5 region (1.5Mpc on a side), we show in the
right hand panel of Figure~\ref{fig:emis_pixels} the pixel statistics
for 2D \lya\ emissivity with the $64^2$, $128^2$, $300^2$, and
$1024^2$ grid resolutions for the full region.  The $300^2$ grid
converges with the $1024^2$ grid, particularly at the high
surface-brightness end, indicating a typical source size of $\gtrsim
$5 kpc and we, therefore, feel comfortable adopting this resolution
for our UVB-only computations.  For the quasar-illuminated cases, the
structures get smaller and we therefore go to a $600^2$ grid to
adequately recover the surface brightness distributions in this case.

\begin{figure*}
\plotone{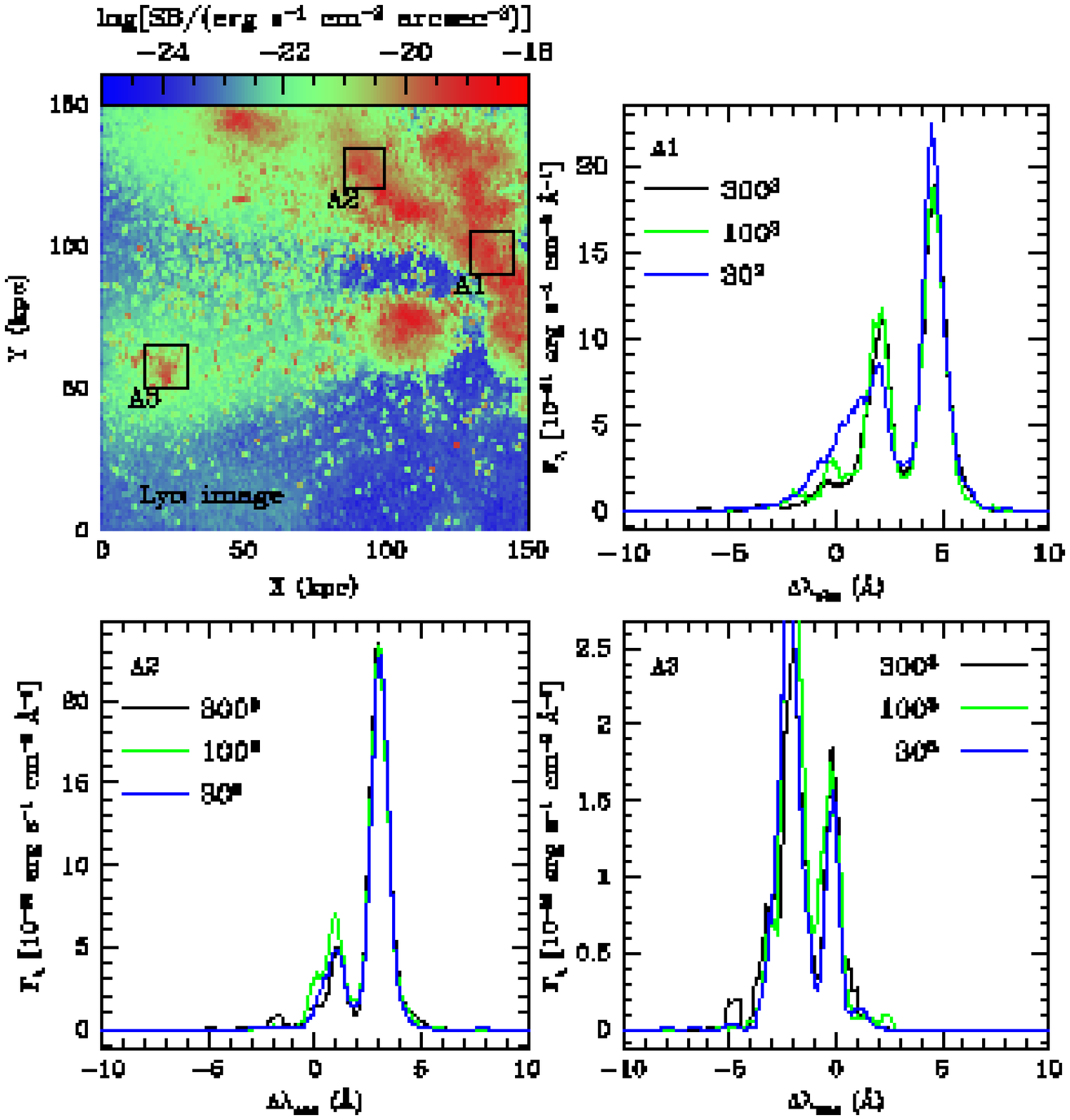}
\caption[Apertures for convergence test]
{\singlespace Convergence of the 1D spectra in a 150 kpc region of the 
L5 simulations.  
The upper left panel shows the apertures overlaid on the \lya\ image.
Remaining panels show the 1D spectra from these apertures as a function of
grid resolution.  From the figure, we see that the $30^3$ grid
recovers the true 1D spectrum well from this small sub-region of
the simulation.  This is analogous to using a $300^3$ grid on the
larger portion of the L5 simulation from which we obtain
the results presented in this paper. }
\label{fig:resolution_apertures} 
\end{figure*}

The scattering process itself depends on how accurately the density,
velocity and temperature distributions are rendered.  To examine the
robustness of our resulting \lya\ images and spectra, we perform
radiative transfer calculations for a region with a physical size of
$150\,$ kpc on a side gridded to a resolution of $30^3$, $100^3$ and
$300^3$ cells.  Even the $300^3$ grid does not resolve the smallest
SPH smoothing length in this region.  However, because gravitational
forces are softened on scales of 0.48 kpc (for the spline kernel), the
$300^3$ grid should be sufficient to accurately capture the physical
structure of the gas.  Adopting a $30^3$ grid for this region is
equivalently coarse to using $300^3$ for our main region of the L5
simulation of $1.5\,$Mpc.  In the upper left panel of
Figure~\ref{fig:resolution_apertures} we show this region and overlay
several representative apertures from which we extract spectra.  We
compare the spectra for these apertures between the three resolutions
in the remaining panels of Figure~\ref{fig:resolution_apertures}.  We
see from this figure that grid resolution does not play a large role
in the resulting spectra as long as the cell size is $\la 5\kpc$;
there are modest variations with resolution, but they do not
systematically change the flux or spectral features.  Similar
conclusions hold for \lya\ images (not shown).  Based on this test, we
adopt a grid resolution of $300^3$ for the L5 simulations throughout
our work for the UVB-only case.  Since our L22 simulation has much coarser resolution (with
a minimum spline-kernel softening length of 2.58 kpc), we
conservatively adopt a grid resolution of $300^3$ for calculations
with this simulation.  In the presence of a bright quasar, we adopt a $600^3$ grid to ensure that grid resolution does not become an issue for the relatively smaller sources.  We have tested that grid resolution higher than this leads to little change in the projected emissivity distribution for the high quasar luminosity we adopt.

\section{The Effect of Temperatures}

A limitation for computing the \lya\ emission from the simulations
here is that the simulations are run assuming an omnipresent ionizing
background (the optically thin approximation). Just as this background
creates high ionization fractions in regions that would be
self-shielded, those regions also have unrealistically high
temperatures because of photoionization heating.  The temperature
differences themselves are moderate (1.5--$2 \times 10^4$K
v.s. $10^4$K), but they have an impact on collisional ionization rates
and recombination rates, and hence on neutral fractions, and a large
impact on collisional excitation rates, and hence on \lya\ emissivity
from cooling radiation.  We present our method for correcting the
simulation temperatures and the full prediction for cooling radiation
(including collisional excitation as well as collisional ionization)
in Paper II.  The effect of the increased temperature is not only to
change the emissivity of the gas in some regions, but also to decrease
the neutral fractions in the gas.  In this Appendix, we perform
several tests to obtain a general idea about the effect of temperature
change and collisional ionization on the {\it fluorescence} signature.

\begin{figure*}
\epsscale{0.85}
\plotone{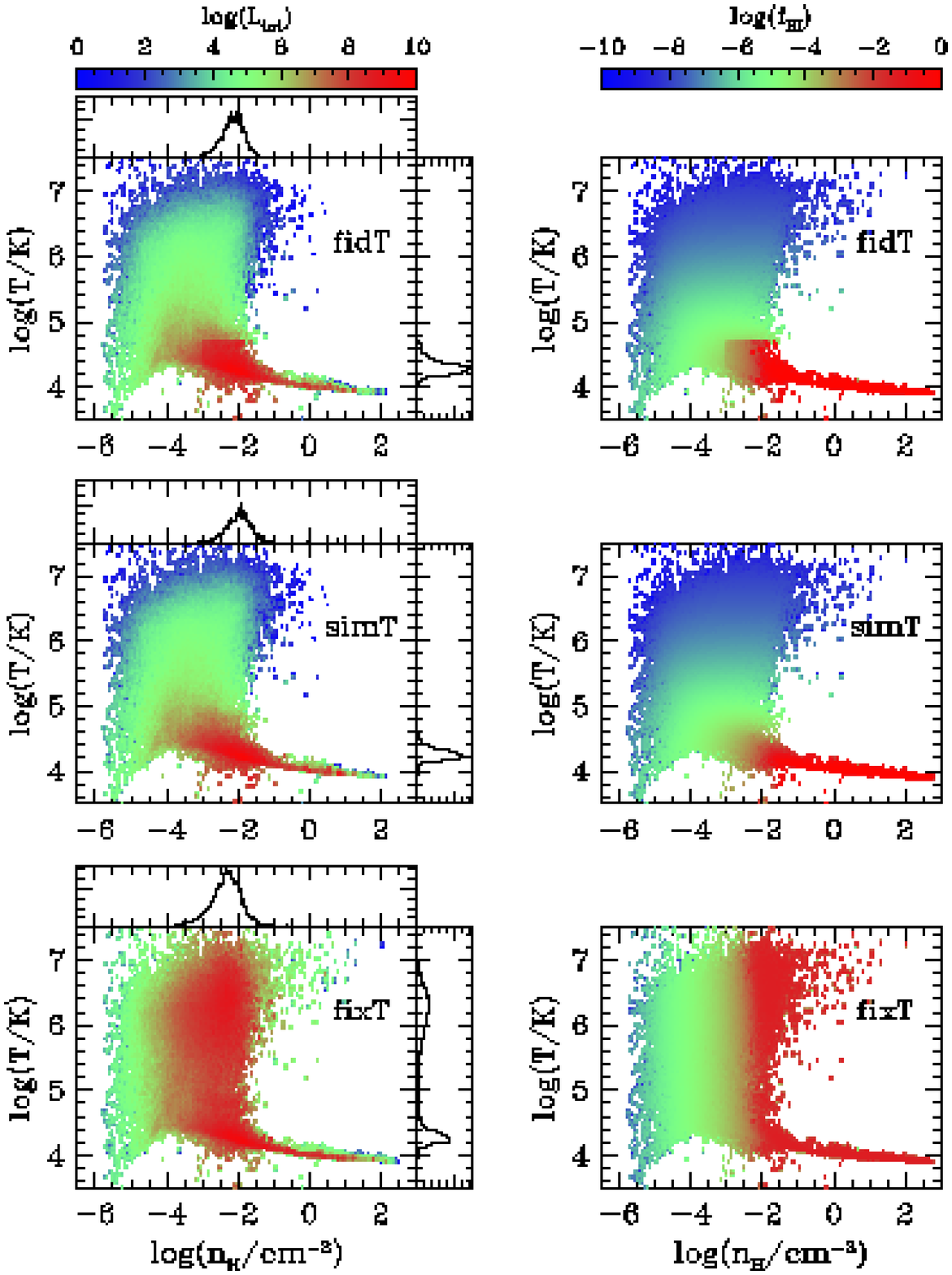}
\caption{\singlespace Distribution of particle luminosity ({\it left}) and neutral 
fraction ({\it right}) in the temperature-density plane for particles
in the sub-region of the L5 simulation.  Top panels show our fiducial
case (fidT).  Middle panels show the case in which simulation
temperatures are adopted directly (simT).  Bottom panels show the case
in which particles are set to a constant temperature of $T=2\times
10^4$K (fixT).  The histograms associated with the left panels show
the luminosity distribution as a function of particle density (top
histogram) and temperature (right histogram). }

\label{fig:rhot}
\end{figure*}

In general, the distribution of particles in the plane defined by the
hydrogen number density and temperature in the simulation (the $n_{\rm
H}$--$T$ plane) has three components (see KWH96 and
Fig.~\ref{fig:rhot}): low-density gas that has been adiabatically
cooled by cosmic expansion; overdense, shock-heated gas; and extremely
overdense, radiatively cooled gas around $10^4$K. The low-density gas
is likely to be exposed to the full ionizing background, so the
optically thin approximation for background ionizing photons in the
simulation is reasonable for this component. The temperatures of shock
heated gas particles are not artificially high owing to the lack of
self-shielding in the simulation, since photoionization alone cannot
heat this gas to such high temperatures.  In the $n_{\rm H}$--$T$
plane, the particles most affected by the optically thin approximation
are those with moderately high density and temperatures $\sim 10^4\rm K -
3\times 10^4\rm K$. If self-shielding were correctly done in the
simulation, these particles would be able to radiatively cool to
$10^4$K.

Based on the above arguments, we define our fiducial model by applying
a crude correction for the particle temperature: the temperatures of gas 
particles with high density ($n_{\rm H} > 10^{-3} {\rm cm}^{-3}$) and low 
simulation temperature ($T<5\times 10^4$K) are set to be $10^4$K, and the
temperatures of all other particles are unaltered. The calculations 
presented in \S 4 and \S 5 are performed using this fiducial model.

To investigate the effect of temperature, we compare the results
from the fiducial model (denoted as the ``fidT'' case) to those from 
two test cases. In the first test case (``simT'' case), we simply 
adopt the particle temperatures as given by the simulation. 
In the other test case (``fixT'' case), we set the temperatures of all
particles to $2\times 10^4$K as has been adopted by other
authors (e.g., Cantalupo et al. 2005). In {\it all} cases, the neutral hydrogen 
fractions are computed assuming equilibrium between recombination and 
the sum of photoionization and collisional ionization, and the 
\lya\ emissivity is computed as 66\% of the photoionization rate (i.e., 
we are calculating only the fluorescent \lya\ emission, not the \lya\ 
cooling radiation). We choose the sub-region of the L5 simulation 
as in \S 4 to perform the comparison.

Figure~\ref{fig:rhot} shows the probability distribution of particles
in the $n_{\rm H}$--$T$ plane for the three cases. We divide the $\log
n_{\rm H}$--$\log T$ space into a uniform grid. For each grid cell, we
compute the total \lya\ luminosity and the median neutral fraction
from particles in that cell. The left panels show the luminosity
distribution (the luminosity is arbitrarily normalized, but the
normalization is the same for all cases), and the right panels show
the neutral fraction distribution. The component of adiabatically
cooled gas is not prominent in the plot, since we are zooming in on an
overdense region. The top and middle panels compare the fidT and the
simT cases. They look reassuringly similar. The fidT case lowers the
temperature of particles that are likely to be artificially heated in
the simulation. This change of temperature leads to an increase in the
neutral fraction of these particles, as can be seen by comparing the
right panels of the fidT and simT cases. Consequently, more of the gas
can be self-shielded, which increases the effective area for
intercepting ionizing photons and ``reflecting'' them back as \lya\
photons.  That is, the total fluorescent \lya\ luminosity
increases. However, the increase in the \lya\ luminosity is small,
which can be seen clearly from the comparison of the top and right
histograms associated with the luminosity distribution panels for fidT
and simT cases. The histograms show the \lya\ luminosity distribution
as a function of density and temperature.  We also compare the
post-transfer results for the two cases and again there is no large
difference in \lya\ images and spectra. Therefore, our fiducial case
and the case adopting the simulation temperature are similar to each
other for fluorescent \lya\ emission.

A comparison between the fidT (top panels) and the fixT (bottom
panels) cases shows that adopting a fixed particle temperature of
$2\times10^4$K has a dramatic impact on both the neutral fraction and
luminosity distributions of particles.  The effect is primarily on the
shock-heated gas. While this diffuse gas is largely optically thin and
contributes little to the \lya\ emissivity in the fidT case, reducing
the temperature as in the fixT case leads to significant shielding
effects for some fraction of this gas in dense regions. This can be
clearly seen in Figure~\ref{fig:rhot} by comparing the neutral
fraction distributions of the fidT and fixT cases.  Because of the
large spatial extent of the shock-heated gas, the artificial shielding
caused by lowering the temperatures greatly increases the effective
area for ``reflecting'' ionizing photons. Therefore, we see a
substantial increase in the \lya\ luminosity caused by shock-heated
gas (see the histograms in the bottom-left panel), which is physically
implausible.

\begin{figure*}
\plotone{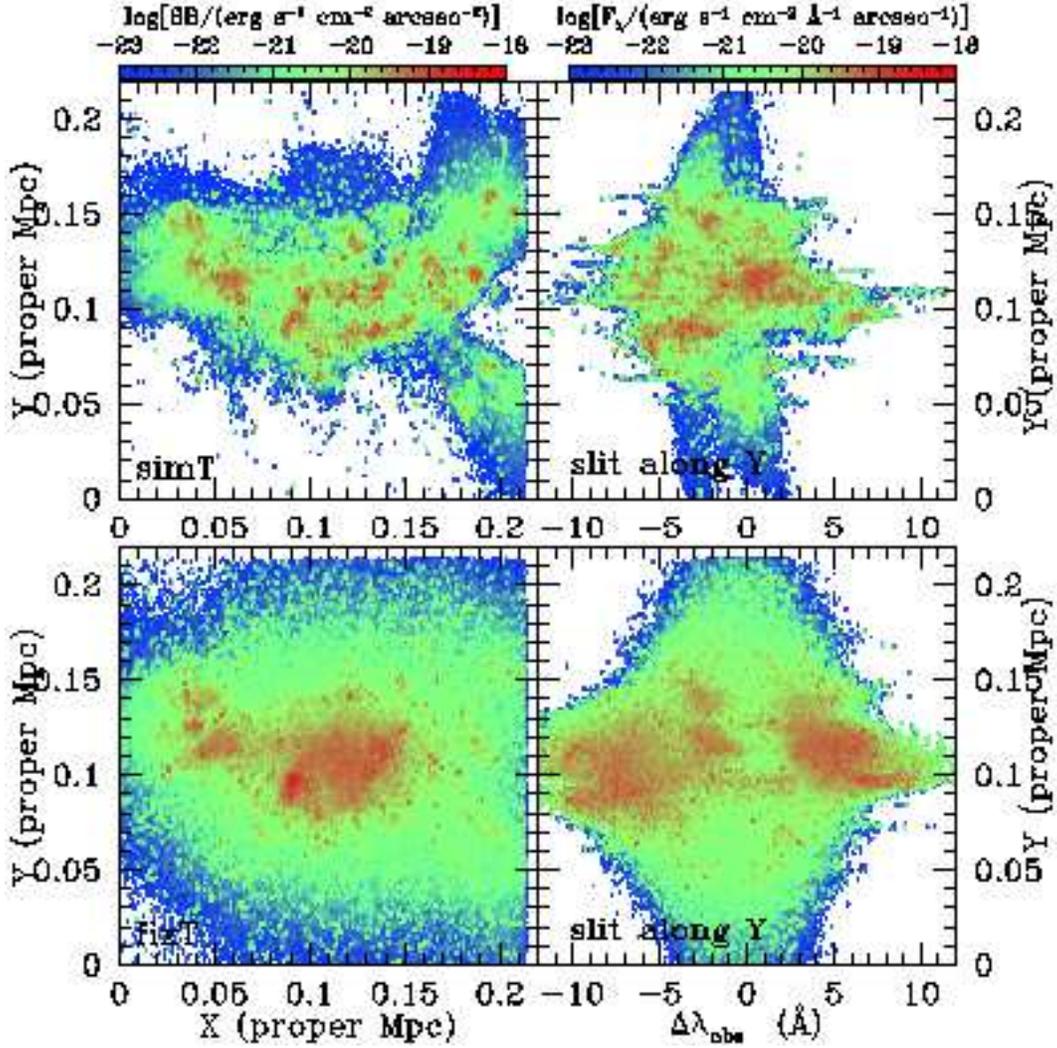}
\caption[\lya\ fluorescence from 200kpc subregion of simulation L5.]
{\singlespace Comparison of \lya\
fluorescence from the sub-region of the L5 simulation when the
simulation temperatures are adopted (simT, top panels)) and when the gas
temperature is fixed at $2\times 10^4$K (fixT, bottom panels).
The results of the simT
case are similar to our fiducial model shown in Figure~\ref{fig:lya_L5zoom}.
The striking differences in morphology and emissivity directly result from
the false shielding effect of shocked gas when the temperatures are
artificially lowered to $2\times 10^4$K.}
\label{fig:lya_L5ttest}
\end{figure*}

As a consequence of the differences in the \lya\ luminosity and
neutral hydrogen fraction distributions, the \lya\ images and spectra
from the fixT case and the fidT (or simT) case are dramatically
different as shown in Figure~\ref{fig:lya_L5ttest}. Fixing the
temperature to $2\times 10^4$K significantly alters the morphology of
the \lya\ emission. The image from the fixT case shows far more
extended \lya\ emission, giving the impression of a single large
structure of emitting gas.  Adopting more realistic temperatures
correctly removes the contribution of moderately dense but
shock-heated gas from the emission signal.  As a result, we are left
with emission from denser, compact knots of material, seen in the
image of the simT case. The 2D spectra in the right panels also
reflect this morphological change -- the spectra are far more diffuse
in the fixT case, as the fixed (low) temperatures increase the neutral
column densities of structures with respect to the fidT/simT
cases. This highlights the necessity of having accurate simulation
temperatures when computing \lya\ emission for comparison with future
observations. Since simulations with a fully self-consistent self-shielding
correction are not available, our approach is acceptable in that we
compute the neutral fractions by making reasonable corrections to gas
temperatures rather than adopting either simulation temperatures or
fixing the temperature to a constant value.  We note however, that total
\lya\ emission
(including cooling radiation) is much more sensitive to the differences
between SimT and FidT than is the fluorescent emission.  Hence, the temperature
treatment is extremely important when predicting total \lya\ fluxes for
comparison with observations.

\end{document}